\documentclass[aps,twocolumn,nofootinbib,groupedaddress,superscriptaddress,longbibliography]{revtex4-1}

\usepackage{graphicx}
\usepackage{xcolor}
\usepackage{bm}
\usepackage{times}
\usepackage[english]{babel}
\makeatletter
\adddialect\l@en\l@english
\makeatother
\usepackage{amsmath,amssymb,amsfonts}
\usepackage{mathrsfs}
\usepackage[colorlinks=true,linkcolor=blue,citecolor=blue,urlcolor=blue,hypertexnames=false]{hyperref}
\usepackage{lineno}
\usepackage{xr-hyper}
\newcommand{\op}[1]{\operatorname{#1}}
\newcommand\bb[1]{\mathbb{#1}}
\newcommand\bk{\mathbf{k}}
\newcommand\br{\mathbf{r}}
\newcommand\ket[1]{|#1\rangle}
\newcommand\bra[1]{\langle#1|}
\newcommand\braket[2]{\langle#1|#2\rangle}
\newcommand\Tr{\operatorname{Tr}}
\newcommand\bz[1]{\mathbb{T}_{\mathrm{BZ}}^{#1}}
\newcommand\eq[1]{Eq.~\eqref{#1}}
\newcommand\dd{\mathrm{d}}
\newcommand\set[2]{\{#1\,|\,#2\}}
\newcommand\real{\operatorname{Re}}
\newcommand\hp{\mathbb{H}\mathrm{P}}
\newcommand\cp{\mathbb{C}\mathrm{P}}

\newcommand{\PropMetric}{S1}
\newcommand{\PropSym}{S2}
\newcommand{\PropQQGT}{S3}
\newcommand{\PropNN}{S4}
\newcommand{\ThmMain}{S1}
\newcommand{\PropVortex}{S5}
\newcommand{\PropFour}{S6}
\newcommand{\PropRigid}{S7}
\newcommand{\PropLowdim}{S8}
\newcommand{\PropTwo}{S9}
\newcommand{\PropThree}{S10}
\newcommand{\EqOmegaThree}{S216}

\newcommand{\MainQQGTEq}{13}

\newtheorem{definition}{Definition}
\newtheorem{proposition}{Proposition}
\newtheorem{theorem}{Theorem}
\newtheorem{corollary}{Corollary}
\newtheorem{lemma}{Lemma}

\begin{document}


\newcommand{\scititle}{Quaternion-K\"ahler geometry of time reversal symmetric crystals}
\title{\scititle}

\author{Hyeongmuk \surname{Lim}}
\affiliation{Department of Physics and Astronomy, Seoul National University, Seoul \& 08826, Korea.}
\affiliation{Center for Theoretical Physics (CTP), Seoul National University, Seoul \& 08826, Korea.}
\affiliation{Institute of Applied Physics, Seoul National University, Seoul \& 08826, Korea.}

\author{Junseo \surname{Jung}}
\affiliation{Department of Physics and Astronomy, Seoul National University, Seoul \& 08826, Korea.}
\affiliation{Center for Theoretical Physics (CTP), Seoul National University, Seoul \& 08826, Korea.}
\affiliation{Institute of Applied Physics, Seoul National University, Seoul \& 08826, Korea.}

\author{Yuting \surname{Qian}}
\affiliation{Department of Physics and Astronomy, Seoul National University, Seoul \& 08826, Korea.}
\affiliation{Center for Theoretical Physics (CTP), Seoul National University, Seoul \& 08826, Korea.}
\affiliation{Institute of Applied Physics, Seoul National University, Seoul \& 08826, Korea.}

\author{Bohm-Jung \surname{Yang}}
\thanks{Corresponding author. Email: bjyang@snu.ac.kr}
\affiliation{Department of Physics and Astronomy, Seoul National University, Seoul \& 08826, Korea.}
\affiliation{Center for Theoretical Physics (CTP), Seoul National University, Seoul \& 08826, Korea.}
\affiliation{Institute of Applied Physics, Seoul National University, Seoul \& 08826, Korea.}

\begin{abstract}
Quantum geometry reveals how the shape of Bloch wave functions governs correlated quantum phenomena. Its standard formulation describes isolated complex bands, where Berry curvature is Abelian and ideal geometry is K\"ahler. 
However, time reversal symmetric crystals with spin require a different language since Kramers degeneracy pairs Bloch states and turns Berry curvature into a non-Abelian $\mathrm{SU}(2)$ field.
Here we show that Kramers pair band geometry is quaternionic. A minimal Kramers pair defines a map into quaternion projective space $\hp^n$, and its quaternionic quantum geometric tensor unifies the quantum metric with the three $\mathrm{SU}(2)$ Berry curvature components. The non-negativity of this tensor imposes local metric–curvature inequalities, whose saturation defines the non-Abelian counterpart of ideal Chern bands. In four dimensions, the ideal limit further yields an algebraic structure related to the four-dimensional quantum Hall effect. Our results promote ideal quantum geometry from the Abelian geometry of Chern bands to the quaternionic, non-Abelian geometry of time reversal symmetric quantum matter.
\end{abstract}

\maketitle

\noindent
\textbf{\large Introduction}

\noindent
Quantum geometry is now recognized as a physical property of matter rather than merely a mathematical description of wave functions. When Bloch states vary across momentum space, they carry a quantum metric, which measures the distance between nearby quantum states, and a Berry curvature, which acts as an effective magnetic field in momentum space~\cite{Simon1983,Berry1984,WilczekZee1984,Zak1989,Provost1980}. These quantities enter electric polarization~\cite{KingSmithVanderbilt1993,Resta1994}, orbital magnetization and electromagnetic responses~\cite{Thonhauser2005,Ceresoli2006,EssinMooreVanderbilt2009,GaoYangNiu2015}, nonlinear transport~\cite{GaoYangNiu2014,SodemannFu2015,Das2023,Wang2023QuantumMetricTransport,Gao2023QuantumMetricNLHE}, superfluid weight~\cite{PeottaTorma2015,Julku2016,Liang2017}, Landau-level spectra~\cite{Rhim2020,Hwang2021,Jung2024}, and correlated flat-band phases~\cite{JacksonMollerRoy2015,Roy2014,Bauer2016,Yu2025QuantumGeometry}. 
Recent spectroscopic studies have also made the quantum geometric tensor experimentally accessible in solids~\cite{Kang2024,Kim2025}. 
Thus, the arrangement of wave functions in Hilbert space can be as consequential as the energy spectrum itself~\cite{Yang2026}.

The Chern-band paradigm provides the starting point. There, ideal quantum geometry is controlled by a compatibility between the quantum metric and the Abelian Berry curvature, which is the band-theoretic manifestation of K\"ahler geometry. It explains why the lowest Landau level (LLL) is an ideal flat band and why lattice Chern bands satisfying analogous ideality conditions can support fractional Chern insulating phases~\cite{Roy2014,JacksonMollerRoy2015,Ledwith2020,Mera2021kahler,MeraOzawa2021Engineering,Wang2021exact,Wang2022hierarchy,Wang2023origin,Ledwith2022}.
K\"ahler geometry is thus the hidden structure behind a broad class of time reversal breaking topological flat bands.

However, spinful time reversal symmetric crystals are different. Kramers degeneracy binds two Bloch states into a pair that cannot be split without breaking symmetry. The pair carries an intrinsic $\mathrm{Sp}(1)\simeq\mathrm{SU}(2)$ gauge freedom, and its Berry curvature is non-Abelian. Thus the central problem is not to add matrix indices to the Abelian theory, but to identify the geometry that replaces K\"ahler geometry when the elementary object is a Kramers pair.

Here we show that this replacement is quaternionic geometry. A Kramers pair is a quaternionic line, and its natural description is provided by quaternion projective space $\hp^n$. The corresponding quaternionic quantum geometric tensor (QQGT) unifies the quantum metric and the three components of the $\mathrm{SU}(2)$ Berry curvature. In this way, the non-Abelian band geometry of time reversal symmetric crystals is identified with the canonical quaternion-K\"ahler geometry of $\hp^n$~\cite{Hatsugai2010,Swann1991,Freed1995}.

The non-negativity of QQGT imposes local metric--curvature bounds that generalize the determinant inequality of Chern bands. Saturation of the bound defines an ideal time reversal symmetric band: a non-Abelian analogue of ideal Chern band, in which the Brillouin-zone geometry inherits a quaternion-K\"ahler structure, and Kramers pairs obey algebraic closure relations that parallel the vortexability of quaternionic Landau levels~\cite{Zhang2001,Elvang2003,LiWu2013,LiWu2013top}.
In four dimensions, the local bound integrates to a quantum-volume bound controlled by the second Chern number. We further show that the bound-saturating local geometry reduces to a minimal four-band time reversal symmetric band system, and that lower-dimensional systems inherit local geometric bounds relevant to $\mathbb Z_2$ topological phases.

\mbox{}\\
\textbf{\large Results}

\noindent
\textbf{K\"ahler geometry of Chern bands.|}
We first recall why K\"ahler geometry is the natural language for ideal Chern bands. Let $|\psi(\bk)\rangle$ be the cell-periodic part of a normalized Bloch eigenstate, and let $P(\bk)=|\psi(\bk)\rangle\langle\psi(\bk)|$ be the projector onto an isolated band. With $\partial_a\equiv\partial/\partial k_a$, the quantum geometric tensor is
\begin{equation}
    Q_{ab}(\bk)=\mathrm{Tr}[P\partial_aP\partial_bP]
    =\langle\partial_a\psi(\bk)|(1-P)|\partial_b\psi(\bk)\rangle .
    \label{eq:qgt_chern_science}
\end{equation}
It is invariant under the local phase rotation $|\psi(\bk)\rangle\mapsto e^{i\phi(\bk)}|\psi(\bk)\rangle$ and is positive semidefinite. Its real and imaginary parts define the quantum metric and Berry curvature,
\begin{equation}
    g_{ab} = \mathrm{Re}\, Q_{ab},\qquad F_{ab} = 2\,\mathrm{Im}\, Q_{ab} .
\end{equation}
Thus the metric and curvature are not independent additions to band theory; they are two faces of the same gauge-invariant tensor.

In two dimensions, non-negativity of $Q_{ab}$ gives the pointwise bound
\begin{equation}
    \det g(\bk)\geq \frac14 F_{xy}(\bk)^2.
    \label{eq:chern_bound}
\end{equation}
The equality of \eq{eq:chern_bound} defines the ideal band condition. It is satisfied globally in the LLL as well as in certain band systems, and expresses the compatibility of a Riemannian metric with a symplectic form through a complex structure. 

More explicitly, the physical state of a single complex band is not a vector but a complex line, because $|\psi\rangle$ and $e^{i\phi}|\psi\rangle$ describe the same state. The space of such lines is complex projective space $\cp^n$, which carries the Fubini--Study metric $g_{\rm FS}$, the Fubini--Study two-form $\omega_{\rm FS}$, and a compatible complex structure $\mathbb J$ satisfying
\begin{equation}
    \omega_{\rm FS}(X,Y)=g_{\rm FS}(X,\mathbb JY).
\label{eq:kahler_compat_science}
\end{equation}
The quantum metric and Berry curvature are the pullbacks of these objects by the projector map: $g=P^*g_{\rm FS}$ and $F=2P^*\omega_{\rm FS}$. The determinant inequality in Eq.~\eqref{eq:chern_bound} becomes an equality precisely when the complex structure can be pulled back to the Brillouin zone so that the induced geometry is K\"ahler.

This geometric compatibility explains why ideal Chern bands inherit many Landau-level-like properties. In LLL, holomorphicity locks the metric and curvature and gives rise to special projected density algebras~\cite{gmp1986}.
When the complex structure saturating \eq{eq:chern_bound} can be chosen momentum independent, the same structure persists exactly in lattice bands: their Bloch functions admit an exact Landau-level representation, and the band-projected density operators close into a GMP-type algebra in a suitably renormalized Hilbert space even when the Berry curvature is nonuniform~\cite{Wang2021exact,Wang2023origin}.
Lattice Chern bands that approximate the same K\"ahler condition can therefore mimic aspects of Landau-level physics and provide favorable settings for fractional Chern insulating phases~\cite{Roy2014,JacksonMollerRoy2015,Lee2017,Ledwith2020,Mera2021kahler,MeraOzawa2021Engineering,Wang2022hierarchy,Ledwith2022}. Importantly, ideality is not merely the flattening of the energy dispersion. It is a constraint on how the occupied projector moves through projective Hilbert space.

This review also identifies the conceptual gap addressed below. K\"ahler geometry explains the ideality of Abelian Chern bands, whose states are complex lines and whose Berry curvature is $\mathrm{U}(1)$ valued. A Kramers-degenerate band is instead a quaternionic line: one-dimensional over $\mathbb H$, or equivalently two-dimensional over $\mathbb C$. Its symmetry-compatible gauge freedom is $\mathrm{Sp}(1)\simeq\mathrm{SU}(2)$, and its Berry curvature has three non-Abelian components. The corresponding ideal geometry must therefore be quaternionic rather than complex.

\begin{figure*}
    \centering
    \includegraphics [width=0.9\linewidth]{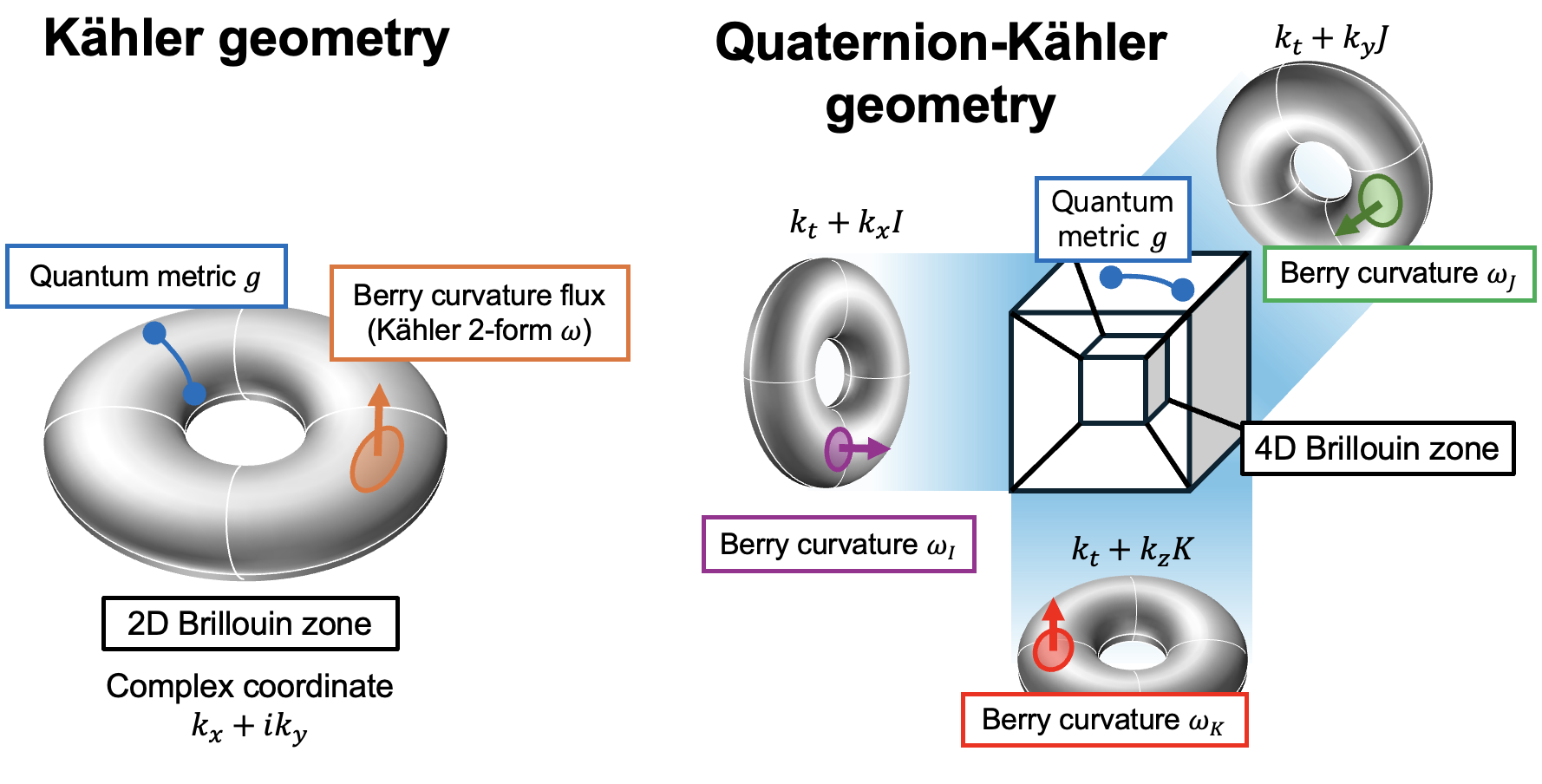}
\begin{tabular}{c|c|c}
\hline
    & K\"ahler (2D) & Quaternion-K\"ahler (4D)
\\ \hline\hline
    Riemannian metric & Quantum metric $g$ & Quantum metric $g$
\\
    Complex coordinates & $k_x+ik_y$ & $k_t+k_xI,\quad k_t+k_yJ,\quad k_t+k_zK$
\\
    Symplectic forms & $\mathrm{U}(1)$ Berry curvature $\omega$ & $\mathrm{SU}(2)$ Berry curvature $\omega_{A=I,J,K}$
\\
    Compatibility & $\omega(X,Y)=g(X,\bb{J}Y)$ & $\omega_A(X,Y)=g(X,\bb{A}Y)\;(\bb{A} = \bb{I},\bb{J},\bb{K})$
\\
    Metric--curvature inequality & $\sqrt{\det g}\geq|\omega|$ & $\sqrt{\det g}\geq(1/6) \left|\sum_{A=I,J,K}\omega_A\wedge\omega_A \right|$
\\ \hline
\end{tabular}

\caption{\textbf{Schematic comparison of K\"ahler and quaternion-K\"ahler band geometry.}
In an ideal two-dimensional Chern band, the quantum metric and the Abelian
Berry curvature are related by a complex structure: the metric measures
distances between nearby Bloch states, while the Berry curvature measures the
phase rotation accumulated around an infinitesimal momentum space loop.
For a time reversal symmetric Kramers pair in four dimensions, the Abelian
curvature is replaced by an $\mathrm{SU}(2)$ curvature with three local
components, $\mathscr{F}^I$, $\mathscr{F}^J$, and $\mathscr{F}^K$. These
components rotate into one another under a change of Kramers frame, while the
metric $g$ and the canonical four-form $\Omega=\sum_{A=I,J,K}\omega_A\wedge \omega_A$ remain gauge invariant.}
  \label{fig:kahler}
\end{figure*}

\mbox{}\\
\textbf{Quaternion-K\"ahler geometry of Kramers pairs.|}
We first consider systems with combined inversion $\mathcal{P}$ and time reversal $\mathcal{T}$ symmetry, $\Theta=\mathcal{PT}$, satisfying $\Theta^2=-1$. Because $\Theta$ leaves momentum invariant, every momentum carries a Kramers pair $\{\psi^1(\bk), \psi^2(\bk)= \Theta\psi^1(\bk)\}$. 
Because $\Theta$ preserves the distances between states, the two partners carry the same quantum metric
\begin{align}
    g_{ab} &\equiv \frac{1}{2} \Big[\bra{\partial_{a}\psi^1(\bk)} \,(1-P)\, \ket{\partial_{b}\psi^1(\bk)} + (a\leftrightarrow b) \Big],
  \label{isometry}
\end{align}
where $P=\ket{\psi^1} \bra{\psi^1} + \ket{\psi^2}\bra{\psi^2}$.
By contrast, the curvature part is intrinsically non-Abelian. The Berry connection
\begin{equation}
    A_a^{ij} = \langle\psi^i|\partial_a\psi^j\rangle
\end{equation}
is constrained by the Kramers relation to be $\mathfrak{su}(2)$-valued, and its curvature is
\begin{equation}
    F_{ab}=\partial_aA_b-\partial_bA_a+[A_a,A_b]
    =F^x_{ab}\sigma_x+F^y_{ab}\sigma_y+F^z_{ab}\sigma_z.
  \label{nhbc}
\end{equation}
In four dimensions, this curvature defines the second Chern number,
\begin{equation}
    \mathcal C_2=\frac{-1}{32\pi^2}\int_{\mathrm{BZ}}\mathrm{Tr}\,F\wedge F .
\end{equation}

The quaternionic structure of a Kramers pair is exposed by combining the two complex states into
\begin{equation}
\begin{split}
    |\Psi\rangle
    &=\frac{1}{\sqrt2}\left(|\psi^1\rangle+|\psi^2\rangle J\right) \\
    &=\frac{1}{\sqrt2}\left(\real|\psi^1\rangle+\op{Im}\ket{\psi^1}I +\real|\psi^2\rangle J+\op{Im}\ket{\psi^2}K\right).
\end{split}
\end{equation}
Here $I,J,K$ are the imaginary quaternion units satisfying $I^2=J^2=K^2=IJK=-1$. A generic quaternion has the form $q=a+bI+cJ+dK$. A symmetry-preserving rotation of the Kramers pair,
\begin{equation}
\begin{gathered}
    \psi^1\mapsto \alpha\psi^1+\beta\psi^2,\qquad
    \psi^2\mapsto -\beta^*\psi^1+\alpha^*\psi^2,\\
    |\alpha|^2+|\beta|^2=1,
\end{gathered}
\end{equation}
becomes right multiplication of $|\Psi\rangle$ by a unit quaternion $u=\alpha-J\beta$. The two states $\ket{\Psi}$ and $\ket{\Psi}u$ are therefore identified, just as $|\psi\rangle$ and $e^{i\phi}|\psi\rangle$ are identified in a single complex band. However, the distinction from a generic two-band non-Abelian QGT is important: a doubly degenerate complex band has a $\mathrm{U}(2)$ gauge freedom. A Kramers pair has the smaller freedom $\mathrm{Sp}(1)$, which is exactly the group of unit quaternions.

Thus, $|\Psi\rangle$ defines a quaternionic line, and the projector $\mathscr P=|\Psi\rangle\langle\Psi|$ invariant under $\ket{\Psi}\to\ket{\Psi}u$ defines a map from the Brillouin zone to
\begin{equation}
    \hp^{n}=(\mathbb H^{n+1}\setminus\{0\})/\mathbb H^\times,
\end{equation}
where $\bb{H}^{n+1}$ is the vector space of $(n+1)$-tuples of quaternions and $\bb{H}^{\times}$ is the group of nonzero quaternions acting on $\bb{H}^{n+1}$ by right multiplication. Thus $\hp^n$ plays for Kramers pairs the role that $\cp^n$ plays for single complex bands.

Quaternion projective space carries a Fubini--Study metric. Instead of one complex structure, as in K\"ahler geometry, a quaternion-K\"ahler manifold has a local triple $\mathbb I,\mathbb J,\mathbb K$ satisfying the quaternion algebra (Supplementary Text). The associated two-forms obey
\begin{equation}
\begin{aligned}
    \omega_I(X,Y) &= g(X,\mathbb IY),
\\
    \omega_J(X,Y) &= g(X,\mathbb JY),
\\
    \omega_K(X,Y) &= g(X,\mathbb KY),
\end{aligned}
  \label{ass_twoforms}
\end{equation}
and rotate into one another under changes of local quaternionic frame. 
Pulling back these Fubini--Study data by $\mathscr P(\bk)$ gives the QQGT (Proposition~\PropQQGT)
\begin{equation}
\begin{split}
    \mathscr Q_{ab}&=\langle\partial_a\Psi|(1-\mathscr P)|\partial_b\Psi\rangle\\
    &=g_{ab}+\frac{1}{2}\left(\mathscr{F}^I_{ab}I+\mathscr{F}^J_{ab}J+\mathscr{F}^K_{ab}K\right).
\end{split}
    \label{eq:qqgt_expanded}
\end{equation}
The real part is the quantum metric, and the three imaginary quaternion components are related to the non-Abelian Berry curvature in \eq{nhbc} by $\mathscr{F}^I=-iF^z,\, \mathscr{F}^J=-iF^y,\, \mathscr{F}^K=-iF^x$.
In complex notation, Eq.~\eqref{eq:qqgt_expanded} is equivalent to the matrix-valued QGT of the Kramers pair, whose antisymmetric part gives the $\mathrm{SU}(2)$ Berry curvature and whose symmetric part is the quantum metric multiplied by $\delta^{ij}$ (Propositions~\PropMetric,~\PropSym).

Reflecting the local rotation of the two-forms in \eq{ass_twoforms}, the QQGT is gauge covariant rather than gauge invariant. Under a change of Kramers frame $\Psi\mapsto\Psi u(\bk)$, with $u$ a unit quaternion, it transforms as $\mathscr Q_{ab}\mapsto \bar u\mathscr Q_{ab}u$. The three curvature forms therefore rotate by an $\mathrm{SO}(3)$ matrix, whereas the metric and the four-form
\begin{equation}
    \Omega = \frac{1}{4}\sum_{A=I,J,K}\mathscr{F}^A\wedge\mathscr{F}^A
  \label{can_4}
\end{equation}
are gauge invariant. Due to its invariance under the local $\mathrm{SU}(2)$ rotation of Kramers basis, we dub $\Omega$ as the canonical four-form. 
Equation~\eqref{eq:qqgt_expanded} is therefore not just a notation for a two-band QGT; it is the band-theoretic expression of the quaternion-K\"ahler geometry of $\hp^{n}$.

\mbox{}\\
\textbf{Metric--curvature bound and ideality.|}
The physical consequence of the QQGT is that the quantum metric and non-Abelian curvature cannot vary independently. \eq{eq:qqgt_expanded} satisfies non-negativity, which means
\begin{equation}
    \sum_{a,b}\bar q^a\mathscr Q_{ab}q^b\geq0
\end{equation}
for any quaternion vector $(q^a)$ (Proposition~\PropNN). At each point of a four-dimensional Brillouin zone, this algebraic condition implies (Theorem~\ThmMain)
\begin{equation}
    \sqrt{\det g}\geq
    \frac{1}{24}\left|\mathscr{F}^I\wedge\mathscr{F}^I+\mathscr{F}^J\wedge\mathscr{F}^J+\mathscr{F}^K\wedge\mathscr{F}^K\right| .
    \label{eq:q_bound_expanded}
\end{equation}
This is the quaternionic counterpart of the determinant inequality for a Chern band. It says that a Kramers pair carrying nontrivial $\mathrm{SU}(2)$ curvature must also carry a minimum quantum volume. 
The equality of \eq{eq:q_bound_expanded} was previously derived for Dirac Hamiltonians~\cite{zhang2022relating}; the present derivation gives the general inequality for spinful $\mathcal{PT}$-symmetric band Hamiltonians.
Integrating Eq.~\eqref{eq:q_bound_expanded} gives
\begin{equation}
    \operatorname{Vol}_{4D}(g)\geq \frac{2\pi^2}{3}|\mathcal C_2|,
\label{eq:volume_bound_expanded}
\end{equation}
where $\operatorname{Vol}_{4D}(g)$ is the 4D quantum volume and $\mathcal C_2$ is the second Chern number.

Saturation of Eq.~\eqref{eq:q_bound_expanded} defines an ideal time reversal symmetric band. Geometrically, when the equality holds, the tangent image of the Brillouin zone inside $\hp^{n}$ becomes closed under the local complex structures $\bb{I,J,K}$. Equivalently, on the regular set where $\det g>0$, the pulled-back structures $A_{\mathbb I},A_{\mathbb J},A_{\mathbb K}$ obey the quaternion algebra on the Brillouin-zone tangent space (Theorem~\ThmMain). This is the non-Abelian analogue of the K\"ahler condition for ideal Chern bands.

\mbox{}\\
\textbf{Vortexability and 4D quantum Hall physics.|}
The class AII ideal condition is further motivated by the 4D quantum Hall effect~\cite{Zhang2001}. It was first formulated on the curved space $\bb{S}^4$ for a particle with arbitrary spin, and later analyzed in the flat-space limit $\bb{R}^4$~\cite{Zhang2001,Elvang2003}. 
Four-dimensional quantum Hall states have also been formulated on K\"ahler manifolds such as $\cp^2$, where a $\mathrm{U}(1)$ background yields a holomorphic lowest Landau level~\cite{karabali2022}, and model Laughlin-type states annihilated by two- and three-body interactions have been constructed in that level~\cite{Wang2023cp2}. With an Abelian background, these states belong to the $\mathrm{U}(1)$/K\"ahler class: their lowest Landau level is the four-dimensional analogue of the holomorphic lowest Landau level that ideal Chern bands emulate. The $\bb{S}^4$ construction with its $\mathrm{SU}(2)$ monopole belongs instead to the $\mathrm{SU}(2)$/quaternionic class, and it is the one relevant for Kramers pairs.

For spin-1/2 particles, analogous three- and four-dimensional systems were subsequently studied as continuum Landau-level models with a nontrivial three-dimensional Fu--Kane--Mele $\bb{Z}_2$ invariant. The Hamiltonians considered in Refs.~\cite{Zhang2001,Elvang2003,LiWu2013} support quaternion-analytic wave functions, as demonstrated in Ref.~\cite{LiWu2013}. 
A related variant of these Hamiltonians was further shown to realize the 4D quantum Hall effect~\cite{LiWu2013top}.

Saturation of \eq{eq:q_bound_expanded} is associated with null quaternion directions of the QQGT,
\begin{equation}
    (1-\mathscr P)\sum_a |\partial_a\Psi\rangle q^a=0 .
    \label{eq:null_expanded}
\end{equation}
For a saturated 4D band, the null space is 3D over the quaternions. When the corresponding null directions can be chosen constant over the Brillouin zone, the band has a vortexability property analogous to that of ideal Chern bands~\cite{Wang2021exact,Wang2023origin,vortexability}. In suitable coordinates, if
\begin{equation}
    \Phi_\bk(t,x,y,z)=e^{i(k_tt+k_xx+k_yy+k_zz)}\Psi_\bk(t,x,y,z)
\end{equation}
is an occupied Bloch wave function, then the occupied subspace is closed under right multiplication by the degree-one polynomials (Proposition~\PropVortex)
\begin{equation}
    tI-x,\qquad tJ-y,\qquad tK-z .
\end{equation}
Iterating this closure generates wave functions multiplied by polynomials in three noncommuting variables. This parallels the quaternion-analytic structure of the quaternionic lowest Landau level (QLLL) in the 4D quantum Hall effect~\cite{Zhang2001,Elvang2003,LiWu2013,LiWu2013top}. 

We note that the QLLL have a restricted form with symmetric products of these variables, whereas the band-theoretic ideality condition permits more general noncommutative polynomials.
This distinction is intrinsically higher-dimensional: when restricted to a 2D subspace, only one coordinate, say \(Z_1=tI-x\), remains, and arbitrary polynomials in \(Z_1\) coincide with symmetric products. The 2D descendant of the quaternion-analyticity then makes direct contact with the quantum spin Hall construction: as the wave function $\ket{\Psi}\propto \phi_{\uparrow}-J\phi_{\downarrow}$ satisfies
\begin{gather}
    (\partial_t+I\partial_x) \phi_{\uparrow}=0,\qquad
    (\partial_t-I\partial_x)\phi_{\downarrow}=0,
\end{gather}
$\phi_{\uparrow}$ is holomorphic while $\phi_{\downarrow}$ is anti-holomorphic (Supplementary Text). This is precisely the wave function structure of the continuum quantum spin Hall Hamiltonian~\cite{BernevigZhang2006QSH}. In the present crystalline setting, saturation of the quantum geometric bound provides a quantum geometric criterion for when Bloch wave functions of a crystal can emulate the spin-dependent (anti-)holomorphic structure. This suggests a route to fractional class AII phases in nearly flat crystalline bands~\cite{Mera2021kahler,vortexability,LevinStern2009FTI,NeupertSantosRyuChamonMudry2011}.

\mbox{}\\
\textbf{Minimal four-band geometry.|}
The minimal realization occurs when both the occupied and unoccupied subspaces are a Kramers pair. The classifying space is then $\hp^1\simeq S^4$, and a representative Hamiltonian is the four-band Dirac form
\begin{equation}
    H(\bk)=\sum_{a=1}^{5}d_a(\bk)\Gamma^a,
    \qquad
    \{\Gamma^a,\Gamma^b\}=2\delta^{ab} .
    \label{eq:dirac_expanded}
\end{equation}
The topology depends on the map $\hat d:T^4\to S^4$, as long as the gap remains open. Because the target is already the minimal quaternionic projective space, Eq.~\eqref{eq:q_bound_expanded} is automatically saturated wherever the metric is non-degenerate (Proposition~\PropFour),
\begin{equation}
    \sqrt{\det g}\,\dd^4k=
    \pm\frac{1}{24}\sum_{A=I,J,K}\mathscr{F}^A\wedge \mathscr{F}^A .
\label{eq:fourband_saturation_expanded}
\end{equation}
This is the 4D analogue of the two-band Chern Hamiltonian, which maps into $\cp^1\simeq S^2$ and automatically saturates the Abelian metric--curvature inequality.
Thus the familiar relation between a two-level Hamiltonian and the Bloch sphere has a direct quaternionic analogue: a four-band Kramers-degenerate Hamiltonian maps to the quaternionic Bloch sphere $S^4$. In both cases, the minimal target has no room for extrinsic deformation of the occupied line. The quantum volume is therefore locked to the pullback of the target-space volume form, which is why the metric--curvature inequality becomes an identity in the minimal model.
In Dirac notation, the right-hand side of Eq.~\eqref{eq:fourband_saturation_expanded} is proportional to the pullback of the volume form on $S^4$, or equivalently to $\epsilon^{abcde}\hat d_a \dd\hat d_b\wedge \dd\hat d_c\wedge \dd\hat d_d\wedge \dd\hat d_e$~\cite{Murakami2004su2}.

The converse is equally important (Proposition~\PropRigid). A saturated, non-singular 4D image in $\hp^{n}$ must be a quaternionic submanifold. Such submanifolds are rigid: locally, up to a momentum-independent quaternion-unitary rotation, the image lies in an embedded $\hp^1\subset\hp^{n}$. 
Consequently, the family of occupied projectors $\{\mathscr{P}(\bk): \bk\in\bz{4}\}$ can be defined within a fixed $\mathbb H^2\subset\mathbb H^{n+1}$ subspace,
\begin{equation}
    U\mathscr P(\bk)U^\dagger=\mathscr P_{4b}(\bk)\oplus0,
\end{equation}
where $\mathscr{P}_{4b}$ is the projector matrix of a 4-band system.
Equivalently, on any connected region where $\det g>0$ and the inequality is saturated, the quaternionic Bloch state can be written using two fixed orthonormal quaternionic basis vectors $\mathbf{q}_1,\mathbf{q}_2\in\bb{H}^{n+1}$ and quaternion-valued functions $u_1,u_2$,
\begin{equation}
    \Psi(\bk)=\mathbf q_1u_1(\bk)+\mathbf q_2u_2(\bk),
\end{equation}
with all remaining quaternionic components absent after a global $\mathrm{Sp}(n+1)$ rotation. 
After spectral flattening, the Hamiltonian reduces to an ideal four-band block plus spectator bands. Generic momentum-dependent coupling to additional bands destroys this fixed $\hp^1$ embedding and makes the band non-ideal. The four-band Dirac Hamiltonian is therefore not merely a useful model; it captures the universal local form of ideal four-dimensional Kramers pair geometry.

\mbox{}\\
\textbf{Descendants in lower dimensions and class-AII bands.|}
We now consider lower-dimensional slices of four-dimensional time reversal symmetric bands. 2D and 3D Brillouin zones do not possess a four-form, but the same QQGT still yields local bounds. On a two-dimensional slice with coordinates $k_1,k_2$, non-negativity of the restricted $2\times2$ quaternionic Hermitian matrix gives (Proposition~\PropLowdim)
\begin{equation}
    \sqrt{\det g_{\mathrm{2D}}}\geq
    \frac{1}{2}\sqrt{(\mathscr{F}^I_{\mathrm{2D}})^2+(\mathscr{F}^J_{\mathrm{2D}})^2+(\mathscr{F}^K_{\mathrm{2D}})^2}
    \equiv |\Omega|_{\mathrm{2D}} .
\label{eq:2d_bound_expanded}
\end{equation}
\eq{eq:2d_bound_expanded} defines the canonical function $|\Omega|_{\mathrm{2D}}$  which is invariant under the local rotation of $\mathrm{SU}(2)$ Berry curvature.
At the same time, it is the non-Abelian extension of the Chern-band inequality. If a two-dimensional time reversal symmetric Hamiltonian decomposes into two time-reversed blocks,
\begin{equation}
    H(\bk) = \begin{pmatrix}h(\bk)&0\\0&h^*(\bk)\end{pmatrix},
\end{equation}
the $\mathrm{SU}(2)$ curvature has only one nonzero component. Equation~\eqref{eq:2d_bound_expanded} then reduces to the Abelian metric--curvature inequality for one block, and the integral of that curvature gives the spin-Chern representative of the Kane--Mele invariant modulo two (Proposition~\PropTwo). 
For a generic two-dimensional Kramers pair, $|\Omega|_{\mathrm{2D}}$ remains a well-defined function, but a universal lower bound by the $\mathbb Z_2$ index~\cite{yu2025universal} is not implied without additional structure.
This distinction is useful experimentally and conceptually. The canonical function is defined pointwise from wave-function geometry and is insensitive to the choice of local $\mathrm{SU}(2)$ frame, whereas the conventional $\mathbb Z_2$ invariant is a global topological quantity. The local bound therefore constrains geometric response even when no simple curvature integral represents the topology.

On a three-dimensional slice, the restricted QQGT gives (Proposition~\PropLowdim)
\begin{equation}
    \sqrt{\det g_{\mathrm{3D}}}\geq |\Omega|_{\mathrm{3D}},
    \label{eq:3d_bound_expanded}
\end{equation}
where $|\Omega|_{\mathrm{3D}}$ is another canonical function built from the restricted metric and the $\mathrm{SU}(2)$ curvature whose explicit form is shown in Eq.~(\EqOmegaThree). In general it is a local geometric quantity invariant under the rotation of $\mathrm{SU}(2)$ Berry curvature, but not a topological density. However, in four-term Dirac Hamiltonians,
\begin{equation}
    H(\bk) = \sum_{a=1}^{4}d_a(\bk)\Gamma^a,
\end{equation}
the normalized vector $\mathbf n=\mathbf d/|\mathbf d|$ defines a map $T^3\to S^3$. In this special class, $|\Omega|_{\mathrm{3D}}$ is proportional to the absolute winding density, while the parity of the winding number is the Fu--Kane--Mele invariant. This establishes a quantum volume bound by the Fu--Kane--Mele invariant (Proposition~\PropThree). 

Thus, the lower-dimensional inequalities are best understood as geometric descendants of the four-dimensional bound; their relation to $\mathbb Z_2$ topology depends on the Hamiltonian representation.

\mbox{}\\
\textbf{Inversion symmetry breaking.|}
The construction is not limited to inversion-symmetric crystals. If only time reversal symmetry is present, $\mathcal T$ maps $\bk$ to $-\bk$ rather than acting locally. One can fold the band structure by pairing a state $\psi_L(\bk)$ with its time-reversed partner $\mathcal T\psi_L(\bk)=\psi_R(-\bk)$ and treating the two as a local Kramers pair in a doubled description,
\begin{equation}
    \mathcal H^{\mathrm{fold}}_\bk=\mathbb C|\psi_L(\bk)\rangle+\mathbb C|\psi_R(-\bk)\rangle .
\end{equation}
Interchanging the two members of the pair changes the local Kramers frame by an $\mathrm{SU}(2)$ rotation together with momentum inversion. Therefore the folded metric and four-form are independent of the folding choice, while the three curvature components transform covariantly. Quaternionic quantum geometry therefore applies to class-AII bands generally, not only to systems with $\mathcal{PT}$ symmetry.

\begin{figure*}
    \centering
    \includegraphics [width=0.95\linewidth]{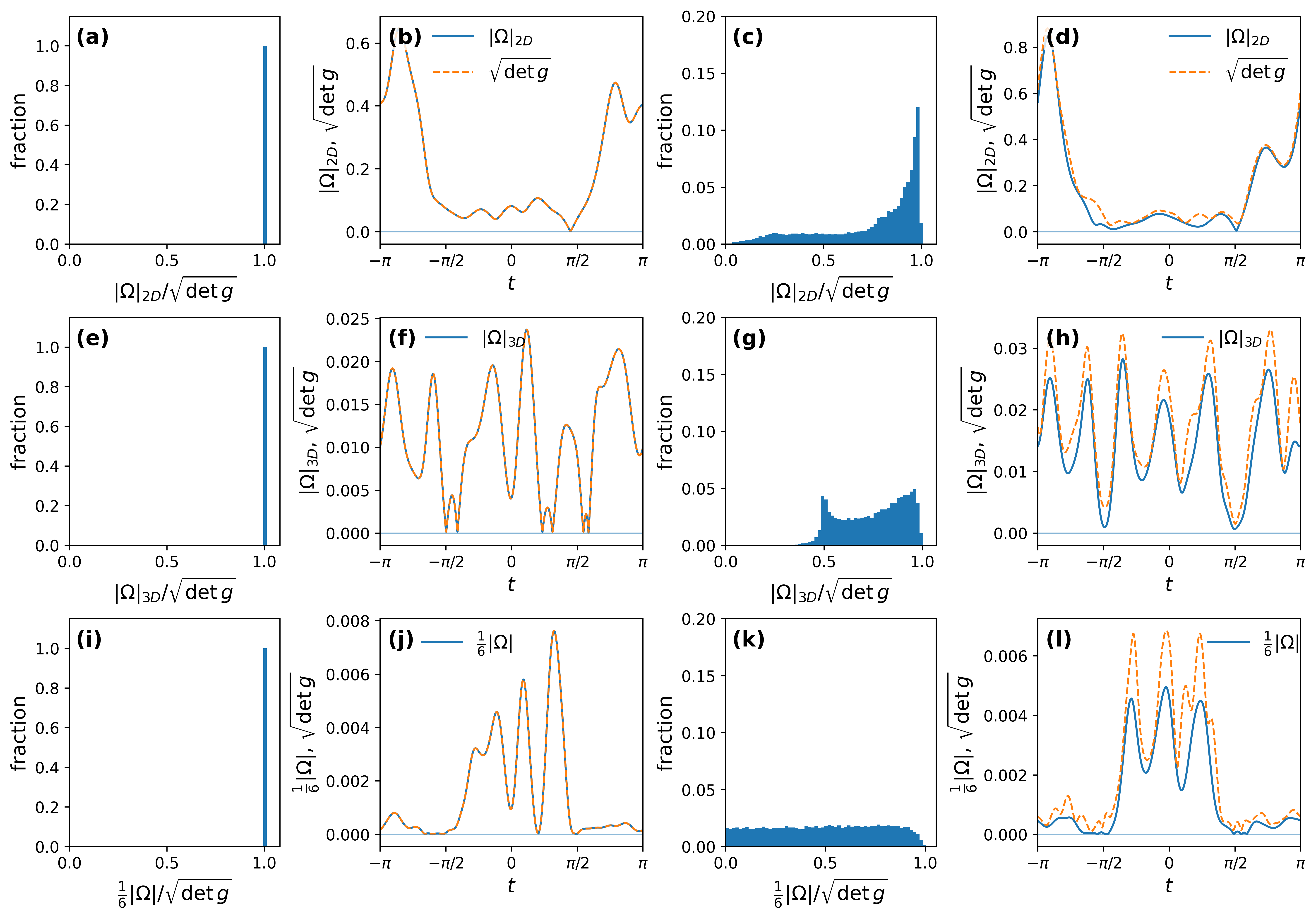}
\caption{
\textbf{Induced metric--curvature inequalities in random Wilson--Dirac models.}
The figure compares the quantum geometric local bounds in two, three, and four dimensions. The rows correspond respectively to the two-dimensional slice $k_3=k_4=0$, the three-dimensional slice $k_4=0$, and the full four-dimensional Brillouin zone. In the histogram panels, the plotted target function is $A_d(\bk)/\sqrt{\det g_d(\bk)}$, where $A_2=|\Omega|_{\mathrm{2D}}$, $A_3=|\Omega|_{\mathrm{3D}}$, and $A_4=|\Omega|/6$. In the profile panels, the blue solid curve is $A_d$ and the orange dashed curve is $\sqrt{\det g_d}$ along the momentum $\bk=(t,t,0,0)$ in the two-dimensional row, $(t,t,t,0)$ in the three-dimensional row, and $(t,t,t,t)$ in the four-dimensional row. \textbf{(a,b,e,f,i,j)} A four-band Wilson--Dirac model with a periodic $\mathcal{PT}$-preserving random perturbation in the Dirac channels. Since the Hamiltonian remains a map into $\hp^1$, the inequalities are saturated pointwise in all three dimensions. \textbf{(c,d,g,h,k,l)} An eight-band extension with periodic $\mathcal{PT}$-preserving random couplings to additional trivial bands. The occupied Kramers pair is no longer constrained to an $\hp^1$ subspace; the inequalities remain valid, but the histograms broaden below one and the profile curves separate.
}
\label{fig:numerics_wd_lowdim}
\end{figure*}

\mbox{}\\
\textbf{Lattice model.|}
We test the inequality in \eq{eq:q_bound_expanded} in lattice Hamiltonians derived from a four-dimensional Wilson--Dirac parent model. The first model is
\begin{equation}
    H_4(\bk) = \sum_{a=1}^{4}\sin k_a\, \Gamma^a+
    \left[m+\sum_{a=1}^{4}(1-\cos k_a)\right] \Gamma^5+V_4(\bk),
\label{eq:model4_expanded}
\end{equation}
where $V_4$ is a random perturbation that is periodic and preserves the local antiunitary symmetry. We compare the full four-dimensional geometry to its two- and three-dimensional restrictions by setting either $k_3=k_4=0$ or $k_4=0$. When $V_4$ is generated by the five Dirac matrices, the Hamiltonian still defines a map into $\hp^1$. The local density profiles can be strongly modulated by the perturbation, but the occupied Kramers pair remains an ideal quaternionic line. Consequently, the inequalities are saturated pointwise in all three dimensions.

The second model couples the four-band block to four additional trivial bands through a generic periodic perturbation preserving the same local $\mathcal{PT}$ symmetry,
\begin{equation}
    H_8(\bk)=
    \begin{pmatrix}
    H_4(\bk) & 0 \\ 0 & \Delta\mathbf 1_4\end{pmatrix}+V_8(\bk).
\end{equation}
The occupied Kramers pair remains isolated, but it is no longer constrained to lie in a fixed $\hp^1$ subspace; instead it maps into a larger $\hp^{n}$.
This second model is designed to separate symmetry from ideality. Both Hamiltonians have the same local antiunitary symmetry and the same Kramers pair structure, so both must obey the QQGT non-negativity bounds. Only the first, however, retains the minimal four-band geometry. The comparison therefore tests the geometric statement that saturation is controlled by the embedding of the projector, not by the presence of time reversal symmetry alone.

\begin{figure*}
    \centering
    \includegraphics [width=0.95\linewidth]{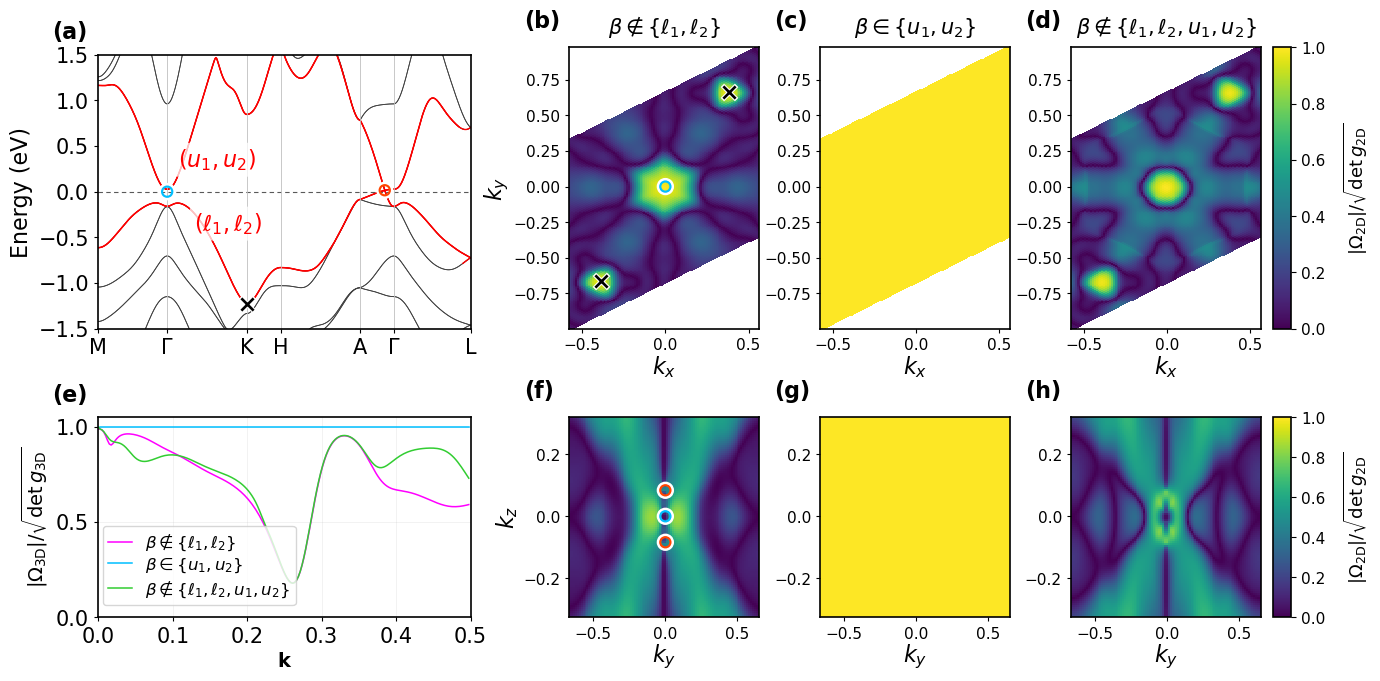}
\caption{
\textbf{Metric--curvature inequalities in $\mathrm{Na_3Bi}$.}
Numerical evaluation of the induced quaternionic metric--curvature bounds for the
$\mathrm{Na}_3\mathrm{Bi}$ band structure.
\textbf{(a)} Band dispersion along high-symmetry lines. The target lower
Kramers pair $\{\ell_1,\ell_2\}$ and the upper low-energy pair
$\{u_1,u_2\}$ are shown in red, and the dashed line denotes the Fermi level
$E_F$.
\textbf{(b--d)} Two-dimensional bound
$\sqrt{\det g_{\mathrm{2D}}} \geq |\Omega|_{\mathrm{2D}}$ on the displayed $(k_x,k_y)$ slice, represented by the saturation ratio $|\Omega|_{\mathrm{2D}}/\sqrt{\det g_{\mathrm{2D}}}$.
\textbf{(e--h)} Corresponding three-dimensional bound $\sqrt{\det g_{\mathrm{3D}}}\geq|\Omega|_{\mathrm{3D}}$.
Panel \textbf{(e)} shows a line cut along
$(k_1,k_2,k_3)=(t,t,t)$, whereas panels \textbf{(f--h)} show maps on the
displayed $(k_y,k_z)$ plane.
In the spectral decomposition of the quantum geometric tensor,
Eq.~\eqref{eq:spectral_QGT}, $\beta$ denotes the external-band index.
The last three columns show contributions from all bands external to the target pair, $\beta\notin\{\ell_1,\ell_2\}$; from the upper low-energy pair only, $\beta\in\{u_1,u_2\}$; and from all remaining external bands,
$\beta\notin\{\ell_1,\ell_2,u_1,u_2\}$, respectively. The blue circles in panels \textbf{(a,b,f)} mark the $\Gamma$ point, the orange circles mark the Dirac points along the $\Gamma$--$A$ direction, and the black crosses mark
the K points. Near $\Gamma$ and the Dirac points, saturation is governed by the four-band sector formed by $\{\ell_1, \ell_2, u_1, u_2\}$. At the K points, by contrast, the target pair $\{\ell_1, \ell_2\}$ forms an effective ideal four-band sector with a
lower-lying external Kramers pair rather than with $\{u_1,u_2\}$. This is reflected by the strong K-point saturation in \textbf{(b)}.
Away from these locally four-band regions, hybridization with
additional bands drives the projector away from the minimal
$\mathbb{H}P^1$ geometry, so the metric--curvature inequality remains valid
but is no longer saturated.
}
\label{fig:numerics_na3bi}
\end{figure*}

We compute, in each dimension, the canonical functions $A_d$: $A_2=|\Omega|_{\mathrm{2D}}$, $A_3=|\Omega|_{\mathrm{3D}}$, and $A_4=|\Omega|/6$ and compare it with the quantum-volume density $\sqrt{\det g_d}$. Figure~\ref{fig:numerics_wd_lowdim} shows the result. For the four-band model, the ratio $A_d/\sqrt{\det g_d}$ is pinned to one throughout the sampled Brillouin zone. For the eight-band model, the same inequalities hold but the ratio is broadly distributed below one. These calculations confirm the central distinction: time-reversal symmetry enforces quaternionic metric--curvature bounds, whereas ideality requires the stronger geometric condition that the occupied pair remain confined to an embedded $\hp^1$ sector.

\mbox{}\newline
\textbf{First principles calculation.|}
We next apply the same local bounds to the realistic band structure of $\mathrm{Na}_3 \mathrm{Bi}$, a Dirac semimetal whose low-energy bands are often
described by a four-band model. For the target lower Kramers pair
$\{|\ell_1\rangle, |\ell_2\rangle\}$, the QGT admits the
spectral decomposition
\begin{gather}
    Q_{ab}^{ij}
    =
    \sum_{\beta\notin\{\ell_1,\ell_2\}}
    \langle \partial_a \ell_i|\beta\rangle
    \langle\beta |\partial_b\ell_j\rangle, \qquad i,j=1,2,
  \label{eq:spectral_QGT}
\end{gather}
where $\beta$ labels bands external to the target pair. We separate the
contribution from the upper Kramers pair
$\beta\in\{u_1,u_2\}$ from
that of the remote bands
$\beta\notin\{\ell_1, \ell_2,u_1,u_2\}$.
Figure~\ref{fig:numerics_na3bi} compares the saturation ratio for the 2D and 3D metric--curvature inequalities after resolving the spectral contributions to the quantum geometry. The contribution from the active four bands saturates
the bound throughout the plotted region, as expected for a minimal four-band geometry. The full contribution from all external bands, however, is
qualitatively different. The saturation ratio approaches unity in the regions surrounding the Dirac points and in
separate regions near the $\Gamma$ and K points, where the active four-band
description accurately captures the wave-function texture.
Away from these regions, remote bands hybridize with the active states and change the occupied projector in a way that is not representable by the
minimal four-band model. The last column in Fig.~\ref{fig:numerics_na3bi} isolates this effect and show that these additional bands are responsible for the loss of saturation.

\mbox{}\newline
\textbf{\large Discussion}

\noindent
We have shown that time-reversal symmetry changes the algebraic type of quantum geometry. The familiar Abelian setting
\[
    \text{single band}/\mathrm{U}(1)/\text{complex}/\text{K\"ahler}
\]
is replaced, for Kramers pairs, by
\[
    \text{Kramers pair}/\mathrm{SU}(2)/\text{quaternionic}/\text{quaternion-K\"ahler}.
\]
In this framework, the saturation of a local non-Abelian metric--curvature inequality produces a second-Chern quantum volume bound and defines ideality. The bound-saturating geometry is automatic in minimal four-band Dirac models but fragile against generic coupling to spectator bands.

These results identify the ideal (bound-saturating) quaternionic geometry as a possible design principle for interaction-driven topological phases in crystalline Kramers-pair bands. The logic parallels the class A case: saturation of the K\"ahler metric--curvature bound allows Chern-band Bloch functions to emulate the holomorphic structure of the lowest Landau level, thereby providing a geometric foundation for fractional Chern insulators~\cite{Roy2014,Mera2021kahler,vortexability}. Here the saturated QQGT bound yields a non-Abelian vortexability condition, which identifies a concrete band-geometric criterion under which crystalline Bloch functions begin to emulate the QLLL wave functions~\cite{LiWu2013}. Since the QLLL is tied to the 4D quantum Hall effect and admits a Laughlin-type many-body state~\cite{Zhang2001,Elvang2003,LiWu2013,LiWu2013top}, our framework suggests a route to fractional class AII phases in crystals~\cite{LevinStern2009FTI,NeupertSantosRyuChamonMudry2011,MaciejkoQiKarchZhang2010FTI3D}.

\textbf{Note added.|}
Shortly after the appearance of the present work on arXiv, several independent studies addressing closely related questions were posted~\cite{ZhaoWu2026,hwang2026}.

\mbox{}\\
\begin{center}
\textbf{ACKNOWLEDGEMENTS}
\end{center}

\paragraph*{Funding:}
H.L., J.J., Y.Q., and B.-J.Y. were supported by the Samsung Science and Technology Foundation under Project No. SSTF-BA2601-02; by the
National Research Foundation of Korea (NRF), funded by the Korean government (MSIT), under Grants No.
RS-2021-NR060087 and No. RS-2025-00562579; by the
Global Research Development Center (GRDC) Cooperative Hub Program through the NRF, funded by the MSIT, under Grant No. RS-2023-00258359; and by the Global-LAMP program of the NRF, funded by the Ministry of Education, under Grant No. RS-2023-00301976.

\paragraph*{Author contributions:}
H.L. and B.-J.Y. conceived the project. 
H.L. developed the mathematical framework. 
H.L. and J.J. performed theoretical analysis and Y.Q. performed first-principles calculations. B.-J.Y. supervised the project.
The manuscript was written by H.L., J.J. and B.-J.Y. with input and comments from all authors.

\paragraph*{Competing interests:}
There are no competing interests to declare.


\nocite{Sudbery1979,HarveyLawson1982,Salamon1982,alesker20031,KaneMele2005,FuKaneMele2007,QiHughesZhang2008}

\makeatletter
\let\QK@originalbibsection\bibsection
\def\bibsection{%
  \let\QK@savedaddcontentsline\addcontentsline
  \let\addcontentsline\@gobblethree
  \QK@originalbibsection
  \let\addcontentsline\QK@savedaddcontentsline
}
\makeatother

\bibliography{refs}

\setcounter{enumiv}{55}
\clearpage

\ifcsname b@Simon1983\endcsname\else
\makeatletter
\bibcite{Simon1983}{{1}{1983}{{Simon}}{{}}}
\bibcite{Berry1984}{{2}{1984}{{Berry}}{{}}}
\bibcite{WilczekZee1984}{{3}{1984}{{Wilczek\ and\ Zee}}{{}}}
\bibcite{Zak1989}{{4}{1989}{{Zak}}{{}}}
\bibcite{Provost1980}{{5}{1980}{{Provost\ and\ Vallee}}{{}}}
\bibcite{KingSmithVanderbilt1993}{{6}{1993}{{King-Smith\ and\ Vanderbilt}}{{}}}
\bibcite{Resta1994}{{7}{1994}{{Resta}}{{}}}
\bibcite{Thonhauser2005}{{8}{2005}{{Thonhauser\ \emph  {et~al.}}}{{Thonhauser, Ceresoli, Vanderbilt,\ and\ Resta}}}
\bibcite{Ceresoli2006}{{9}{2006}{{Ceresoli\ \emph  {et~al.}}}{{Ceresoli, Thonhauser, Vanderbilt,\ and\ Resta}}}
\bibcite{EssinMooreVanderbilt2009}{{10}{2009}{{Essin\ \emph  {et~al.}}}{{Essin, Moore,\ and\ Vanderbilt}}}
\bibcite{GaoYangNiu2015}{{11}{2015}{{Gao\ \emph  {et~al.}}}{{Gao, Yang,\ and\ Niu}}}
\bibcite{GaoYangNiu2014}{{12}{2014}{{Gao\ \emph  {et~al.}}}{{Gao, Yang,\ and\ Niu}}}
\bibcite{SodemannFu2015}{{13}{2015}{{Sodemann\ and\ Fu}}{{}}}
\bibcite{Das2023}{{14}{2023}{{Das\ \emph  {et~al.}}}{{Das, Lahiri, Atencia, Culcer,\ and\ Agarwal}}}
\bibcite{Wang2023QuantumMetricTransport}{{15}{2023{}}{{Wang\ \emph  {et~al.}}}{{Wang \emph  {et~al.}}}}
\bibcite{Gao2023QuantumMetricNLHE}{{16}{2023}{{Gao\ \emph  {et~al.}}}{{Gao \emph  {et~al.}}}}
\bibcite{PeottaTorma2015}{{17}{2015}{{Peotta\ and\ T\"{o}rm\"{a}}}{{}}}
\bibcite{Julku2016}{{18}{2016}{{Julku\ \emph  {et~al.}}}{{Julku, Peotta, Vanhala, Kim,\ and\ T\"orm\"a}}}
\bibcite{Liang2017}{{19}{2017}{{Liang\ \emph  {et~al.}}}{{Liang \emph  {et~al.}}}}
\bibcite{Rhim2020}{{20}{2020}{{Rhim\ \emph  {et~al.}}}{{Rhim, Kim,\ and\ Yang}}}
\bibcite{Hwang2021}{{21}{2021}{{Hwang\ \emph  {et~al.}}}{{Hwang, Rhim,\ and\ Yang}}}
\bibcite{Jung2024}{{22}{2024}{{Jung\ \emph  {et~al.}}}{{Jung, Lim,\ and\ Yang}}}
\bibcite{JacksonMollerRoy2015}{{23}{2015}{{Jackson\ \emph  {et~al.}}}{{Jackson, M{\"o}ller,\ and\ Roy}}}
\bibcite{Roy2014}{{24}{2014}{{Roy}}{{}}}
\bibcite{Bauer2016}{{25}{2016}{{Bauer\ \emph  {et~al.}}}{{Bauer, Jackson,\ and\ Roy}}}
\bibcite{Yu2025QuantumGeometry}{{26}{2025{}}{{Yu\ \emph  {et~al.}}}{{Yu \emph  {et~al.}}}}
\bibcite{Kang2024}{{27}{2024}{{Kang\ \emph  {et~al.}}}{{Kang \emph  {et~al.}}}}
\bibcite{Kim2025}{{28}{2025}{{Kim\ \emph  {et~al.}}}{{Kim \emph  {et~al.}}}}
\bibcite{Yang2026}{{29}{2026}{{Yang}}{{}}}
\bibcite{Ledwith2020}{{30}{2020}{{Ledwith\ \emph  {et~al.}}}{{Ledwith, Tarnopolsky, Khalaf,\ and\ Vishwanath}}}
\bibcite{Mera2021kahler}{{31}{2021{}}{{Mera\ and\ Ozawa}}{{}}}
\bibcite{MeraOzawa2021Engineering}{{32}{2021{}}{{Mera\ and\ Ozawa}}{{}}}
\bibcite{Wang2021exact}{{33}{2021}{{Wang\ \emph  {et~al.}}}{{Wang, Cano, Millis, Liu,\ and\ Yang}}}
\bibcite{Wang2022hierarchy}{{34}{2022}{{Wang\ and\ Liu}}{{}}}
\bibcite{Wang2023origin}{{35}{2023{}}{{Wang\ \emph  {et~al.}}}{{Wang, Klevtsov,\ and\ Liu}}}
\bibcite{Ledwith2022}{{36}{2022}{{Ledwith\ \emph  {et~al.}}}{{Ledwith, Vishwanath,\ and\ Khalaf}}}
\bibcite{Hatsugai2010}{{37}{2010}{{Hatsugai}}{{}}}
\bibcite{Swann1991}{{38}{1991}{{Swann}}{{}}}
\bibcite{Freed1995}{{39}{1995}{{Freed\ and\ Uhlenbeck}}{{}}}
\bibcite{Zhang2001}{{40}{2001}{{Zhang\ and\ Hu}}{{}}}
\bibcite{Elvang2003}{{41}{2003}{{Elvang\ and\ Polchinski}}{{}}}
\bibcite{LiWu2013}{{42}{2013}{{Li\ and\ Wu}}{{}}}
\bibcite{LiWu2013top}{{43}{2013}{{Li\ \emph  {et~al.}}}{{Li, Zhang,\ and\ Wu}}}
\bibcite{gmp1986}{{44}{1986}{{Girvin\ \emph  {et~al.}}}{{Girvin, MacDonald,\ and\ Platzman}}}
\bibcite{Lee2017}{{45}{2017}{{Lee\ \emph  {et~al.}}}{{Lee, Claassen,\ and\ Thomale}}}
\bibcite{zhang2022relating}{{46}{2022}{{Mera\ \emph  {et~al.}}}{{Mera, Zhang,\ and\ Goldman}}}
\bibcite{karabali2022}{{47}{2002}{{Karabali\ and\ Nair}}{{}}}
\bibcite{Wang2023cp2}{{48}{2023{}}{{Wang\ \emph  {et~al.}}}{{Wang, Klevtsov,\ and\ Douglas}}}
\bibcite{vortexability}{{49}{2023}{{Ledwith\ \emph  {et~al.}}}{{Ledwith, Vishwanath,\ and\ Parker}}}
\bibcite{BernevigZhang2006QSH}{{50}{2006}{{Bernevig\ and\ Zhang}}{{}}}
\bibcite{LevinStern2009FTI}{{51}{2009}{{Levin\ and\ Stern}}{{}}}
\bibcite{NeupertSantosRyuChamonMudry2011}{{52}{2011}{{Neupert\ \emph  {et~al.}}}{{Neupert, Santos, Ryu, Chamon,\ and\ Mudry}}}
\bibcite{Murakami2004su2}{{53}{2004}{{Murakami\ \emph  {et~al.}}}{{Murakami, Nagaosa,\ and\ Zhang}}}
\bibcite{yu2025universal}{{54}{2025{}}{{Yu\ \emph  {et~al.}}}{{Yu, Herzog-Arbeitman,\ and\ Bernevig}}}
\bibcite{MaciejkoQiKarchZhang2010FTI3D}{{55}{2010}{{Maciejko\ \emph  {et~al.}}}{{Maciejko, Qi, Karch,\ and\ Zhang}}}
\bibcite{ZhaoWu2026}{{56}{2026}{{Zhao\ \emph  {et~al.}}}{{Zhao, Pan, Yang,\ and\ Wu}}}
\bibcite{hwang2026}{{57}{2026}{{Hwang\ \emph  {et~al.}}}{{Hwang, Yu,\ and\ Cho}}}
\bibcite{Sudbery1979}{{58}{1979}{{Sudbery}}{{}}}
\bibcite{HarveyLawson1982}{{59}{1982}{{Harvey\ and\ Lawson}}{{}}}
\bibcite{Salamon1982}{{60}{1982}{{Salamon}}{{}}}
\bibcite{alesker20031}{{61}{2003}{{Alesker}}{{}}}
\bibcite{KaneMele2005}{{62}{2005}{{Kane\ and\ Mele}}{{}}}
\bibcite{FuKaneMele2007}{{63}{2007}{{Fu\ \emph  {et~al.}}}{{Fu, Kane,\ and\ Mele}}}
\bibcite{QiHughesZhang2008}{{64}{2008}{{Qi\ \emph  {et~al.}}}{{Qi, Hughes,\ and\ Zhang}}}
\makeatother

\fi

\onecolumngrid

\renewcommand{\thefigure}{S\arabic{figure}}
\renewcommand{\thetable}{S\arabic{table}}
\renewcommand{\theequation}{S\arabic{equation}}
\renewcommand{\thepage}{S\arabic{page}}
\setcounter{figure}{0}
\setcounter{table}{0}
\setcounter{equation}{0}
\setcounter{page}{1}

\setcounter{definition}{0}
\setcounter{proposition}{0}
\setcounter{theorem}{0}
\setcounter{corollary}{0}
\setcounter{lemma}{0}
\renewcommand{\thedefinition}{S\arabic{definition}}
\renewcommand{\theproposition}{S\arabic{proposition}}
\renewcommand{\thetheorem}{S\arabic{theorem}}
\renewcommand{\thecorollary}{S\arabic{corollary}}
\renewcommand{\thelemma}{S\arabic{lemma}}

\begin{center}
\centering
\textbf{Supplementary Information for}

\textbf{``\scititle''}

\mbox{}\newline
Hyeongmuk~Lim$^{1,2,3}$,
Junseo~Jung$^{1,2,3}$,
Yuting~Qian$^{1,2,3}$,
Bohm-Jung~Yang$^{1,2,3,\ast}$\\ 
\small$^1$Department of Physics and Astronomy, Seoul National University, Seoul 08826, Korea.\\
\small$^2$Center for Theoretical Physics (CTP), Seoul National University, Seoul 08826, Korea.\\
\small$^3$Institute of Applied Physics, Seoul National University, Seoul 08826, Korea.\\
\small$^\ast$Corresponding author. Email: bjyang@snu.ac.kr\\
\end{center}


\tableofcontents


\clearpage


\subsection*{Supplementary Note 1.
Conventions and intuition for quaternionic linear algebra}

This subsection fixes the algebraic conventions used throughout the manuscript. Its main purpose is to explain why a Kramers pair is naturally treated as one quaternionic state. The point is not to introduce a new physical degree of freedom, but to use an algebra whose internal symmetry matches the symmetry already present in a time reversal symmetric spinful band.

The guiding analogy is
\begin{gather}
\begin{array}{c|c|c|c}
\text{band structure} & \text{state} & \text{gauge freedom} & \text{projective space} \\
\hline
\text{single complex band} & \text{complex line} & U(1) & \cp^n \\
\text{Kramers pair} & \text{quaternionic line} & \mathrm{Sp}(1)\simeq SU(2) & \hp^n .
\end{array}
\end{gather}
Thus, just as complex numbers are the natural language for the phase freedom of a single isolated band, quaternions are the natural language for the internal $\mathrm{SU}(2)$ freedom of a Kramers pair.

The quaternion algebra $\bb{H}$ is a four-dimensional real division algebra. A quaternion is written as
\begin{gather}
    q=a+bI+cJ+dK,\qquad a,b,c,d\in\bb{R}.
\end{gather}
The imaginary units obey
\begin{gather}
    I^2=J^2=K^2=-1,\qquad IJ=K,\qquad JK=I,\qquad KI=J,
\end{gather}
together with anticommutation, for example $JI=-K$, $KJ=-I$, and $IK=-J$. The order of multiplication is therefore part of the structure.

Choosing one imaginary unit, say $I$, gives a copy of the complex numbers inside $\bb{H}$:
\begin{gather}
    \bb{C}_I=\{a+bI\,|\,a,b\in\bb{R}\}\subset\bb{H}.
  \label{sm:ci}
\end{gather}
With this choice, the usual complex number $a+bi$ is identified with $a+bI$. The genuinely quaternionic structure appears only after adding another imaginary unit $J$. Then $K=IJ$ is forced to appear. If $I$ and $J$ were taken to commute, one would obtain two independent complex sectors $\bb{C}_I$ and $\bb{C}_J$, defined as in \eq{sm:ci}. The nontrivial extension is obtained by imposing $IJ=-JI$.

Quaternion conjugation extends complex conjugation:
\begin{gather}
    \overline{q}=a-bI-cJ-dK.
\end{gather}
It reverses the order of multiplication,
\begin{gather}
    \overline{q_1q_2} = \overline{q_2} \cdot \overline{q_1},
\end{gather}
and gives a positive norm,
\begin{gather}
    |q|^2=\overline{q}q = q\overline{q}=a^2+b^2+c^2+d^2.
\end{gather}
Hence every nonzero quaternion has an inverse,
\begin{gather}
    q^{-1}=\frac{\overline q}{|q|^2}.
\end{gather}

The unit quaternions form the group
\begin{gather}
    \mathrm{Sp}(1)=\{u\in\bb{H}\,|\,\overline{u}u=1\}\simeq \mathrm{SU}(2).
\end{gather}
A useful parametrization is
\begin{gather}
    u=\alpha-J\beta,\qquad \alpha,\beta\in\bb{C}_I,\qquad |\alpha|^2+|\beta|^2=1.
\end{gather}
This is the algebraic reason that quaternions are adapted to Kramers pairs: the allowed change of basis within a Kramers pair is precisely an $\mathrm{SU}(2)$ rotation.

Because $\bb{H}$ is noncommutative, one must specify whether scalars act from the left or from the right. We use right quaternionic vector spaces for kets. Thus quaternionic scalars multiply ket vectors from the right:
\begin{gather}
    \ket{\Phi q+\Xi r}=\ket{\Phi}q+\ket{\Xi}r,\qquad q,r\in\bb{H}.
\end{gather}
The inner product is quaternion-valued and satisfies
\begin{gather}
    \langle \Phi|\Xi q\rangle=\langle\Phi|\Xi\rangle q,\qquad
    \langle \Phi q|\Xi\rangle=\overline q\,\langle\Phi|\Xi\rangle,
\end{gather}
together with
\begin{gather}
    \langle \Phi|\Xi\rangle=\overline{\langle \Xi|\Phi\rangle}.
\end{gather}
Equivalently, taking the adjoint of a right multiplication turns it into a left multiplication by the conjugate quaternion:
\begin{gather}
    \big(\ket{\Phi}q\big)^\dagger=\overline q\,\bra{\Phi}.
\end{gather}

For a change of quaternionic frame by an invertible matrix $A$, our convention is
\begin{gather}
    \ket{\Psi'_i}=\sum_j\ket{\Psi_j}A_{ji},\qquad
    \bra{\Psi'_i}=\sum_j(A^\dagger)_{ij}\bra{\Psi_j},
\end{gather}
where $A^\dagger$ is the conjugate transpose. This is the same rule as in complex linear algebra, except that quaternionic conjugation reverses the order of products.

We now explain how a Kramers pair is repackaged as a quaternionic vector. Let
\begin{gather}
    \{\ket{\psi^1},\ket{\psi^2}\},\qquad \ket{\psi^2}=\Theta\ket{\psi^1},
\end{gather}
be a Kramers pair. As a complex vector space, the span of this pair is two-dimensional. As a quaternionic vector space, it is one-dimensional.

To make this statement concrete, identify $\bb{C}^2$ with $\bb{H}$ by
\begin{equation}
    F:
\begin{array}{ccc}
    \bb{C}^2 &\to& \bb{H} \\
    (c_{\uparrow},c_{\downarrow}) &\mapsto& c_{\uparrow}-Jc_{\downarrow},
\end{array}
\end{equation}
where the complex unit $i$ is identified with the quaternionic unit $I$. For example, if
\begin{gather}
    c_{\uparrow}=a-bi,\qquad c_{\downarrow}=c-di,
\end{gather}
then
\begin{gather}
    F(c_{\uparrow},c_{\downarrow})
    =(a-bI)-J(c-dI)
    =a-bI-cJ-dK.
\end{gather}

The quantities $c_{\uparrow}$ and $c_{\downarrow}$ are linear functions that return the two complex components of a state (ket vector). Therefore the map above identifies the pair of bras with the quaternion-valued bra
\begin{gather}
    \bra{\Psi}:=\frac{1}{\sqrt{2}}
    \big(\bra{\psi^1}-J\bra{\psi^2}\big).
\end{gather}
Taking the quaternionic conjugate transpose fixes the corresponding ket:
\begin{gather}
    \ket{\Psi}
    =\frac{1}{\sqrt{2}}
    \big(\ket{\psi^1}+\ket{\psi^2}J\big).
\end{gather}
The relative signs are not arbitrary; they are forced by quaternionic conjugation and by the convention that kets carry scalars from the right.

This convention also makes the $\mathrm{SU}(2)$ freedom of the Kramers pair transparent. A symmetry-compatible change of Kramers frame has the form
\begin{gather}
    \ket{\psi^1}\mapsto \alpha\ket{\psi^1}+\beta\ket{\psi^2},\qquad
    \ket{\psi^2}\mapsto -\beta^*\ket{\psi^1}+\alpha^*\ket{\psi^2},
    \qquad |\alpha|^2+|\beta|^2=1.
\end{gather}
In quaternionic notation this becomes
\begin{gather}
    \ket{\Psi}\mapsto \ket{\Psi}u,\qquad
    u=\alpha-J\beta,\qquad \overline u u=1.
\end{gather}
Thus the whole $\mathrm{SU}(2)$ rotation of a Kramers pair is represented by right multiplication by a unit quaternion.

The physical object is not the particular normalized vector $\ket{\Psi}$, but the quaternionic line it spans:
\begin{gather}
    \ket{\Psi}\sim \ket{\Psi}u,\qquad u\in\mathrm{Sp}(1).
\end{gather}
This is the direct analogue of the phase equivalence
\begin{gather}
    \ket{\psi}\sim e^{i\theta}\ket{\psi}
\end{gather}
for a single complex band. Accordingly, a Kramers pair defines a map into quaternion projective space,
\begin{gather}
    \hp^n
    =
    \big(\bb{H}^{n+1}\setminus\{0\}\big)/\bb{H}^{\times},
\end{gather}
the space of quaternionic lines in $\bb{H}^{n+1}$. This is why $\hp^n$ replaces $\cp^n$ in the geometry of a minimal Kramers pair.

The quaternionic projector onto the line is
\begin{gather}
    \mathscr{P} = \ket{\Psi}\bra{\Psi}.
\end{gather}
It is invariant under the unit-quaternion gauge transformation:
\begin{gather}
    \ket{\Psi}\mapsto \ket{\Psi}u,\qquad
    \bra{\Psi}\mapsto \overline u\,\bra{\Psi}
    \quad\Rightarrow\quad
    \mathscr{P}\mapsto \ket{\Psi}u\overline u\bra{\Psi}=\mathscr{P}.
\end{gather}
The ordinary complex projector onto the same Kramers pair is
\begin{gather}
    P = \ket{\psi^1}\bra{\psi^1}+\ket{\psi^2}\bra{\psi^2}.
\end{gather}
The two projectors encode the same subspace, but $\mathscr{P}$ keeps track of the quaternionic, or $\mathrm{SU}(2)$, structure in a compact form.

This notation is especially useful for the quantum geometric tensor. For a single complex band,
\begin{gather}
    Q^{\bb{C}}_{ab}
    =
    \bra{\partial_a\psi}(1-P)\ket{\partial_b\psi}
    =
    g_{ab}+\frac{i}{2}F_{ab},
\end{gather}
so the real part is the quantum metric and the imaginary part is the Berry curvature. For a Kramers pair, the analogous object is
\begin{gather}
    \mathscr{Q}_{ab}
    =
    \bra{\partial_a\Psi}(1-\mathscr{P})\ket{\partial_b\Psi}.
\end{gather}
Its quaternionic decomposition is
\begin{gather}
    \mathscr{Q}_{ab}
    =
    g_{ab}\,1
    +
    \frac{1}{2}
    \left(
        \mathscr{F}^{I}_{ab}I
        +
        \mathscr{F}^{J}_{ab}J
        +
        \mathscr{F}^{K}_{ab}K
    \right).
\end{gather}
Thus the real component is the quantum metric of the Kramers pair, while the three imaginary components are the three components of the non-Abelian $\mathrm{SU}(2)$ Berry curvature, written in a quaternionic basis.

Under a local change of Kramers frame,
\begin{gather}
    \ket{\Psi(\bk)}\mapsto \ket{\Psi(\bk)}u(\bk),
\end{gather}
the QQGT transforms covariantly:
\begin{gather}
    \mathscr{Q}_{ab}(\bk)
    \mapsto
    \overline{u(\bk)}\,\mathscr{Q}_{ab}(\bk)\,u(\bk).
\end{gather}
Consequently, $g_{ab}$ is gauge invariant, while the three curvature components
\begin{gather}
    \big(\mathscr{F}^{I},\mathscr{F}^{J},\mathscr{F}^{K}\big)
\end{gather}
rotate among themselves by an $SO(3)$ transformation. This is the quaternionic form of the usual statement that the Berry curvature of a Kramers pair is an $\mathrm{SU}(2)$ gauge field. Individual components depend on the local Kramers frame, but their span is intrinsic.

This point is also the origin of the quaternion-K\"ahler language used in the main text. In ordinary K\"ahler geometry there is one complex structure, compatible with the metric and the Berry curvature. In the quaternionic case there is instead a local triple of complex structures, corresponding to the three imaginary quaternionic directions. Because a change of Kramers frame rotates these three directions, the individual complex structures are generally local, while the rank-three bundle that they span is the globally meaningful object.

Finally, the non-negativity of the QQGT is the algebraic source of the metric--curvature inequalities discussed in the main text. For any quaternionic vector $q=(q^a)$, define
\begin{gather}
    X(q) = \sum_{a=1}^{4}
    (1-\mathscr{P}) \ket{\partial_a\Psi}\,q^a .
\end{gather}
Then
\begin{gather}
    \sum_{a,b}\overline{q^a}\,\mathscr{Q}_{ab}\,q^b
    =
    \langle X(q)|X(q)\rangle
    \geq 0.
\end{gather}
Equality holds precisely when
\begin{gather}
    X(q)=0.
\end{gather}
Therefore a null vector of $\mathscr{Q}$ is a derivative direction, or a quaternionic linear combination of derivative directions, whose component outside the target subspace vanishes. In the ideal limit studied in this work, such null directions acquire the additional algebraic structure responsible for the quaternionic analogue of Landau-level vortexability.

For systems with time reversal symmetry but without inversion, the two Kramers partners live at $\bk$ and $-\bk$ rather than at the same momentum. The same algebra still applies after the folding construction used in the main text, which pairs $\psi_L(\bk)$ with its time reversal partner $\psi_R(-\bk)$. Different folding choices are related by the same local $\mathrm{SU}(2)$, or unit-quaternion, freedom described above. Hence the quaternionic formalism is a symmetry-adapted way to describe Kramers pair geometry, not an additional assumption about the microscopic Hilbert space.

\subsection*{Supplementary Note 2.
The K\"ahler structure of \texorpdfstring{$\cp^n$}{complex projective space}
}

The complex projective space $\cp^n$ is the space of complex lines in a fixed complex vector space $\bb{C}^{n+1}$. Accordingly, for a system with $n+1$ bands, it parametrizes the rank-one projector associated with a chosen nondegenerate target band, while the remaining $n$ bands span its orthogonal complement.
$\cp^n$ carries a natural K\"ahler structure consisting of a Riemannian metric (the Fubini--Study metric $g_{\mathrm{FS}}$), a complex structure $\bb{J}$, and the associated K\"ahler 2-form $\omega_{\mathrm{FS}}$. Since points of $\cp^n$ are parameterized by the projector $P$ onto a line in $\bb{C}^{n+1}$, the triple ($g_{\mathrm{FS}},\omega_{\mathrm{FS}},\bb{J})$ can be written explicitly in terms of $P$.
In the following, we shall denote the unoccupied space projector as $P_{\perp}:=1-P$.

The Riemannian metric, K\"ahler form, and complex structure are tensors which act on tangent vectors. To evaluate them, we therefore need a concrete description of tangent vectors on $\cp^n$.
A tangent vector $X$ at $P\in\cp^n$ can be represented as the derivative of a curve $P(s)$, where $P(0)=P$ and $\dot{P}(0):= \partial_{s}P(s)|_{s=0}=X$.
Differentiating both sides of $P^2=P$ gives $P\dot{P} + \dot{P}P = \dot{P}$ and hence
\begin{gather}
    P\dot{P} = \dot{P} P_{\perp} \quad \Rightarrow\quad 
    P\dot{P}P = P_{\perp}\dot{P} P_{\perp} = 0,
\\
    \dot{P} = P\dot{P}P + P\dot{P}P_{\perp} + P_{\perp}\dot{P}P + P_{\perp}\dot{P}P_{\perp} = P\dot{P}P_{\perp} + P_{\perp}\dot{P} P.
  \label{sm:tangent_eq}
\end{gather}
However, since both $P$ and $P_{\perp}$ are projections, they are hermitian operators so that
\begin{gather}
    P_{\perp}\dot{P} P = \left(P\dot{P}P_{\perp} \right)^{\dagger}.
\end{gather}
Thus, $P\dot{P}P_{\perp}$ alone completely determines $\dot{P}$. These objects can be concretely described by square matrices as follows. Let $U\subset\cp^n$ be a local chart, and $\psi: U\to \bb{C}^{n+1}$ be a local function such that 
\begin{gather}
    \psi(P)^{\dagger}\psi(P) = \braket{\psi(P)}{\psi(P)} = 1, \qquad \psi(P)\psi^{\dagger}(P) = \ket{\psi(P)}\bra{\psi(P)} = P.
\end{gather}
Geometrically, $\psi$ is a local section of the fiber bundle $\bb{C}^{n+1}\to \cp^n$. Moreover, the section $\psi$ can be extended to a local orthonormal frame
\begin{gather}
    \{\psi,\phi_1,\cdots,\phi_n\}: U\to \bb{C}^{n+1}.
\end{gather}
Denoting $\phi_0:=\psi$, we have $\braket{\phi_i}{\phi_j} = \delta_{ij}\; (i,j\in\{0,1,\dots,n\})$.

Having chosen a local frame, we can represent the tangent vector $\dot{P}$ as a matrix with respect to this frame.
\begin{gather}
    \bra{\phi_i}X\ket{\phi_j} = \bra{\phi_i}\dot{P}(0)\ket{\phi_j} =
    \left[
    \begin{array}{c|ccc}
    0 & c_1 & \cdots & c_n \\
    \hline
    c_1^{*} & 0 & \cdots & 0 \\
    \vdots & \vdots & \ddots & \vdots \\
    c_n^{*} & 0 & \cdots & 0
    \end{array}
    \right]
    = 
    \begin{bmatrix}
        0 & \mathbf{c} \\ \mathbf{c}^{\dagger} & 0
    \end{bmatrix}
  \label{cpx_tangent}
\end{gather}
The matrix is manifestly determined by the row vector $\mathbf{c}=(c_1,\dots,c_n)$, which is precisely the matrix representation of $P\dot{P}P_{\perp}$ at $s=0$. We shall call $\mathbf{c}$ the \textit{directional vector} of $X$. Using the directional vector, a natural complex structure is defined by multiplying the imaginary unit $i$ to the vector $\mathbf{c}$:
\begin{gather}
    \bra{\phi_i}\,\bb{J}(X)\,\ket{\phi_j} \equiv 
    \begin{bmatrix}
        0 & i\mathbf{c} \\ -i\mathbf{c}^{\dagger} & 0
    \end{bmatrix}.
  \label{cpx_str}
\end{gather}
To check if $\bb{J}$ can be defined \textit{globally}, it suffices to check if it depends on the choice of local section $\psi$. For every smooth function $\theta:U\to\bb{R}$, $\psi'\equiv e^{i\theta}\psi$ qualifies as an equally good local section as $\psi$. With respect to the new local frame $\{\psi',\phi_1,\cdots,\phi_n\}$, the vector $\mathbf{c}$ transforms as $\mathbf{c}'=e^{-i\theta}\mathbf{c}$, which is multiplication by a phase. This obviously commutes with multiplication by $i$, hence the definition in \eq{cpx_str} is sound globally.

We turn to the Fubini--Study metric $g_{\mathrm{FS}}$ and the associated K\"ahler form $\omega_{\mathrm{FS}}$ on $\cp^n$. Let $X, Y$ be two tangent vectors at $P\in\cp^n$ with corresponding directional vectors $\mathbf{c,d}$. Then
\begin{align}
    g_{\mathrm{FS}}(X,Y) &\equiv \Tr[P\dd P\odot\dd P](X,Y)
    = \frac{1}{2} \Tr[P(XY + YX)] 
    = \frac{1}{2}(\mathbf{cd}^{\dagger} + \mathbf{dc}^{\dagger}),
\\
    \omega_{\mathrm{FS}}(X,Y) &\equiv \frac{-i}{2}\Tr[P\dd P\wedge\dd P](X,Y) = \frac{-i}{2} \Tr[P(XY - YX)] 
    = \frac{-i}{2}(\mathbf{cd}^{\dagger}-\mathbf{dc}^{\dagger}).
\end{align}
It is manifest that $g_{\mathrm{FS}}$ is real-valued, symmetric, bilinear, and positive-definite, whereas $\omega_{\mathrm{FS}}$ is real-valued, antisymmetric, bilinear, and nondegenerate. They satisfy the compatibility relation
\begin{gather}
    \omega_{\mathrm{FS}}(X,Y) = g_{\mathrm{FS}}(X,\bb{J}Y).
  \label{sm:hallmark_k}
\end{gather}
\eq{sm:hallmark_k} is a hallmark of K\"ahler geometry.

\subsection*{Supplementary Note 3.
The quaternion-K\"ahler structure of \texorpdfstring{$\hp^n$}{quaternion projective space}
}

We use the following operational definition of a quaternion-K\"ahler manifold.
\begin{definition}
  \label{sm:def_qk}
    A quaternion-K\"ahler manifold is a Riemannian manifold $(\mathcal{M},g)$ of dimension $4n$, further equipped with the following geometric structures. There is a suitable open cover $\{U_j\}$ of $\mathcal{M}$ such that
\begin{itemize}
    \item For each $U_j$, there is a triple of local sections of the endomorphism bundle
\begin{gather}
    \bb{I}_j,\bb{J}_j,\bb{K}_j \in \Gamma(\op{End}(T\mathcal{M}), U_j),
\end{gather}
    where $\Gamma(-, U_j)$ is the global section functor on $U_j$, such that
\begin{gather}
    \bb{I}_j^2 = \bb{J}_j^2 = \bb{K}_j^2 = \bb{I}_j\,\bb{J}_j\,\bb{K}_j = -1.
\end{gather}
    \item For each $U_j$, there is a triple of differential 2-forms
\begin{gather}
    \omega_j^I,\omega_j^J, \omega_j^K \in \Gamma(T^{*}\mathcal{M}\wedge T^{*}\mathcal{M}, U_j),
\end{gather}
    satisfying
\begin{gather}
    \omega_j^I(v_1,v_2) = g(v_1,\bb{I}_jv_2), \quad \omega_j^J(v_1,v_2) = g(v_1,\bb{J}_jv_2), \quad \omega_j^K(v_1,v_2) = g(v_1,\bb{K}_jv_2).
  \label{sm:qK_omgea_g}
\end{gather}
    \item It is further required that whenever $U_i\cap U_j\neq\emptyset$, the two sets of local endomorphisms 
\begin{gather}
    \{\bb{I}_i,\bb{J}_i, \bb{K}_i\}, \quad \{\bb{I}_j,\bb{J}_j,\bb{K}_j\}
\end{gather}
    are related by an orthogonal transformation in the sense that
\begin{subequations}
\begin{align}
    \bb{I}_i &= \bb{I}_jS^I{}_I + \bb{J}_jS^J{}_I + \bb{K}_jS^K{}_I
\\ 
    \bb{J}_i &= \bb{I}_jS^I{}_J + \bb{J}_jS^J{}_J + \bb{K}_jS^K{}_J
\\ 
    \bb{K}_i &= \bb{I}_jS^I{}_K + \bb{J}_jS^J{}_K + \bb{K}_jS^K{}_K
\end{align}
\end{subequations}
    for some matrix-valued function $S:U_i\cap U_j\to \mathrm{SO}(3)$.
    By \eq{sm:qK_omgea_g}, the same is true for the local 2-forms.
\end{itemize}
\end{definition}
Definition~\ref{sm:def_qk} is the operational notion used in this work. In standard differential geometry, quaternion-K\"ahler usually further requires compatibility between the endomorphisms $\{\bb{I}_j,\bb{J}_j,\bb{K}_j\}$ and the Levi--Civita connection. 
More precisely, there exists a skew-symmetric $3\times3$ matrix of one-forms,
$\theta_{ab}=-\theta_{ba}$, such that
\begin{gather}
    \nabla J_a = \sum_{b=1}^{3} \theta_{ab}\otimes J_b,
  \label{nablaQinQ}
\end{gather}
where $\nabla$ denotes the covariant derivative associated with the Levi--Civita connection. For our purposes, however, Eq.~\eqref{nablaQinQ} need not be verified directly. First, quaternionic projective space $\hp^n$ is known to satisfy this condition and provides a canonical example of a quaternion--K\"ahler manifold. Second, suppose that a target Kramers pair induces a quaternion--K\"ahler geometry on the Brillouin zone in the operational sense adopted here. On any momentum-space region where $\det g(\bk)\neq0$, the projector map $\mathscr{P}(\bk)$ is an immersion whose tangent image is preserved by the local complex structures $\bb{I,J,K}$. Its image therefore forms a quaternionic submanifold of $\hp^n$, and the induced geometry automatically satisfies Eq.~\eqref{nablaQinQ}.

As in the K\"ahler case, we can find the ingredients of the quaternion-K\"ahler structure of $\hp^n$ using a local frame.
With the unoccupied space projector $\mathscr{P}_{\perp}:=1-\mathscr{P}$, we have an equality for tangent vectors analogous to \eq{sm:tangent_eq}:
\begin{gather}
    \dot{\mathscr{P}} = \mathscr{P}\dot{\mathscr{P}}\mathscr{P}_{\perp} + \mathscr{P}_{\perp} \dot{\mathscr{P}}\mathscr{P}, \qquad 
    \dot{\mathscr{P}}:= \partial_{s}\mathscr{P}(s)\in T\hp^n.
\end{gather}
Thus an arbitrary tangent vector is represented by a matrix analogous to \eq{cpx_tangent}.

Let $U\subset\hp^n$ be an open chart, and $\Psi:U\to\bb{H}^{n+1}$ be a local section of the fibration $\bb{H}^{n+1}\to\hp^n$. We extend $\Psi$ to a (quaternionic) orthonormal frame 
\begin{gather}
    \{\Psi,\Phi_1,\cdots,\Phi_n\}: U\to \bb{H}^{n+1}.
\end{gather}
Denoting $\Phi_0:=\Psi$, we have $\braket{\Phi_i}{\Phi_j} = \delta_{ij}\;(i,j\in\{0,1,\dots,n\})$. With respect to this local frame, a tangent vector $X=\dot{\mathscr{P}}(0)$ is represented by the matrix
\begin{gather}
    \bra{\Phi_i}X\ket{\Phi_j} = \bra{\Phi_i}\dot{\mathscr{P}}(0)\ket{\Phi_j} =
    \left[
    \begin{array}{c|ccc}
    0 & q_1 & \cdots & q_n \\
    \hline
    \overline{q}_1 & 0 & \cdots & 0 \\
    \vdots & \vdots & \ddots & \vdots \\
    \overline{q}_n & 0 & \cdots & 0
    \end{array}
    \right]
    =
    \begin{bmatrix}
        0 & \mathbf{q} \\ \mathbf{q}^{\dagger} & 0
    \end{bmatrix}
\end{gather}
where $\mathbf{q}=(q_1,\cdots,q_n)$ represents $\mathscr{P}\dot{\mathscr{P}}\mathscr{P}_{\perp}$. A natural quaternionic structure is defined by multiplying the imaginary unit quaternions $I,J,K$ on the left of $\mathbf{q}$:
\begin{gather}
    \bra{\Phi_i}\bb{I}(X)\ket{\Phi_j} \equiv 
    \begin{bmatrix}
        0 & I\mathbf{q} \\
        -\mathbf{q}^{\dagger}I & 0
    \end{bmatrix}, \; 
    \bra{\Phi_i}\bb{J}(X)\ket{\Phi_j} \equiv 
    \begin{bmatrix}
        0 & J\mathbf{q} \\
        -\mathbf{q}^{\dagger}J & 0
    \end{bmatrix}, \; 
    \bra{\Phi_i}\bb{K}(X)\ket{\Phi_j} \equiv 
    \begin{bmatrix}
        0 & K\mathbf{q} \\
        -\mathbf{q}^{\dagger}K & 0
    \end{bmatrix}
\end{gather}
Unlike in the K\"ahler case, however, the global well-definedness of $\{\bb{I,J,K}\}$ is not guaranteed. If the local section is changed from $\Psi$ to $\Psi'=\Psi u$ for some function $u:U\to\bb{H}^{\times}$, the vector $\mathbf{q}$ transforms as $\mathbf{q}'=\overline{u}\mathbf{q}$. Multiplication by $\overline{u}$ on the left does not commute with multiplication by $\{I,J,K\}$ on the same side, because the quaternions are non-Abelian. Thus, the quaternionic structure is not a global invariant object; it is \textit{covariant} with respect to the frame change $\Psi'=\Psi u$ and transforms as
\begin{gather}
    \{\bb{I}',\bb{J}',\bb{K}'\} = \text{multiplication by } \{\overline{u}Iu,\,\overline{u}Ju,\,\overline{u}Ku\} \text{ on the left of } \mathbf{q}'.
  \label{quat_so3}
\end{gather}
It is not an error, but a feature of quaternion-K\"ahler geometry: What can be globally defined is not the quaternionic structure $\{\bb{I},\bb{J},\bb{K}\}$ itself, but only the rank 3 subbundle of the endomorphism bundle $\op{End}(T\hp^n)$ generated by the triple $\{\bb{I},\bb{J},\bb{K}\}$. We note that the transformation $X\mapsto \overline{u}Xu$ for $X\in\{aI+bJ+cK\,|\,a,b,c\in\bb{R}\}\simeq \bb{R}^{3}$ defines a rotation in $\mathrm{SO}(3)$, so that \eq{quat_so3} implies the global existence of the rank 3 vector bundle generated by $\{\bb{I},\bb{J},\bb{K}\}$.

We now turn to the Fubini--Study metric and the associated 2-forms. Let $X$ and $Y$ be tangent vectors at the same point $\mathscr{P}\in\hp^n$, with directional vectors $\mathbf{q,p}$. Then
\begin{subequations}
\begin{align}
    g_{\mathrm{FS}}(X,Y) &\equiv \Tr[\mathscr{P}\dd\mathscr{P}\odot\dd\mathscr{P}](X,Y) = \frac{1}{2}\Tr[\mathscr{P}(XY+YX)] = \frac{1}{2}(\mathbf{qp}^{\dagger} + \mathbf{pq}^{\dagger}),
\\
    \omega_{\mathrm{FS}}^{I}(X,Y) &\equiv \frac{1}{2}\Tr\real[-I\mathscr{P}\dd\mathscr{P}\wedge\dd\mathscr{P}](X,Y) = \frac{1}{2}\Tr\real[-I\mathscr{P}(XY-YX)] 
\nonumber\\
    &= \frac{1}{2}\real[-I(\mathbf{qp}^{\dagger} - \mathbf{pq}^{\dagger})] = \frac{1}{2}\real[-\mathbf{qp}^{\dagger}I + I\mathbf{pq}^{\dagger}] = \frac{1}{2}(-\mathbf{qp}^{\dagger}I + I\mathbf{pq}^{\dagger}),
\\
    \omega_{\mathrm{FS}}^{J}(X,Y) &= \frac{1}{2}\Tr\real[-J\mathscr{P}(XY-YX)]  = \frac{1}{2}(-\mathbf{qp}^{\dagger}J + J\mathbf{pq}^{\dagger}),
\\
    \omega_{\mathrm{FS}}^{K}(X,Y) &= \frac{1}{2}\Tr\real[-K\mathscr{P}(XY-YX)]  = \frac{1}{2}(-\mathbf{qp}^{\dagger}K + K\mathbf{pq}^{\dagger}).
\end{align}
  \label{sm:qk_strs}
\end{subequations}
In the last step of computing $\omega_{\mathrm{FS}}^I(X,Y)$, we used the fact
\begin{gather}
    \real[Ix] = \real[xI] \quad\text{and}\quad \real[x+x^{\dagger}]=x+x^{\dagger} \quad\text{for all } x\in\bb{H}.
\end{gather}
Analogous equalities hold with $J$ or $K$ in place of $I$. It is manifest that all quantities in \eq{sm:qk_strs} are real and satisfy symmetry, bilinearity, and nondegeneracy (absence of null vectors). They are tied together by the compatibility relations
\begin{gather}
    \omega_{\mathrm{FS}}^I(X,Y) = g_{\mathrm{FS}}(X,\bb{I}Y), \quad 
    \omega_{\mathrm{FS}}^J(X,Y) = g_{\mathrm{FS}}(X,\bb{J}Y), \quad 
    \omega_{\mathrm{FS}}^K(X,Y) = g_{\mathrm{FS}}(X,\bb{K}Y).
\end{gather}
Since the quaternionic structure $\{\bb{I,J,K}\}$ is only locally defined and does not make sense globally, the 2-forms $\{\omega_{\mathrm{FS}}^I,\omega_{\mathrm{FS}}^J,\omega_{\mathrm{FS}}^K\}$ are not globally defined either. However, the canonical four-form
\begin{gather}
    \Omega_{\mathrm{FS}} := \omega_{\mathrm{FS}}^I\wedge\omega_{\mathrm{FS}}^I + \omega_{\mathrm{FS}}^J\wedge\omega_{\mathrm{FS}}^J + \omega_{\mathrm{FS}}^K\wedge\omega_{\mathrm{FS}}^K
  \label{sm:canonical_form}
\end{gather}
is globally well-defined. To this end, observe that due to \eq{quat_so3}, a local basis change $\Psi'=\Psi u$ results in $\mathrm{SO}(3)$ rotation of 2-forms
\begin{gather}
    \omega_{\mathrm{FS}}'{}^{\alpha} = \sum_{\beta=I,J,K} S^{\alpha}{}_{\beta} \omega_{\mathrm{FS}}^{\beta}\qquad (\alpha=I,J,K),
  \label{sup:2form_rot}
\end{gather}
where the rotation matrix $S$ is determined by 
\begin{gather}
    \overline{u}\,(n^II+n^JJ+n^KK)\,u = \sum_{\beta=I,J,K} (S^{I}{}_{\beta}\,n^{\beta})I + (S^{J}{}_{\beta}\,n^{\beta})J + (S^{K}{}_{\beta}\,n^{\beta})K.
  \label{sup:n_rotation}
\end{gather}
Both sides of \eq{sup:n_rotation} are representations of the $\mathrm{SO}(3)$ rotation of a real vector $\mathbf{n}=(n^I,n^J,n^K)\in\bb{R}^3$. The matrix $S\in\mathrm{SO}(3)$ is uniquely determined by the unit quaternion $u$.
However, the canonical four-form in \eq{sm:canonical_form} is rotation invariant.

We have proved the following.

\begin{lemma}
  \label{lemma:hp}
    The quaternion projective space $(\hp^n,g_{\mathrm{FS}},\omega_{\mathrm{FS}}^{I},\omega_{\mathrm{FS}}^{J},\omega_{\mathrm{FS}}^{K}, \bb{I},\bb{J},\bb{K})$ is quaternion-K\"ahler. Besides the Fubini--Study metric, it has a globally defined canonical four-form
\begin{gather}
    \Omega_{\mathrm{FS}} = \omega_{\mathrm{FS}}^I\wedge\omega_{\mathrm{FS}}^I + \omega_{\mathrm{FS}}^J\wedge\omega_{\mathrm{FS}}^J + \omega_{\mathrm{FS}}^K\wedge\omega_{\mathrm{FS}}^K.
\end{gather}
\end{lemma}
%

\subsection*{Supplementary Note 4.
Quaternionic Landau levels
}

We work on the Euclidean 4D space with coordinates $\mathbf{r}=(x^{\mu})=(t,x,y,z)\in \bb{R}^4$.
Define the operators $e_{\mu}$ and $\Gamma^{\mu\nu}\;(\mu,\nu\in\{0,1,2,3\})$ acting on the spin-1/2 space $\bb{C}^2$, such that for all $i,j,k\in\{1,2,3\}$,
\begin{gather}
    e_{0} = 1_{2\times2}, \quad 
    e_{i}e_{j}= -\delta_{ij} 1_{2\times2} + \epsilon^{ijk}e_k, \quad
    \Gamma^{ij} = i\epsilon^{ijk}e_k, \quad 
    \Gamma^{k0} = -\Gamma^{0k} = ie_k.
  \label{algebra}
\end{gather}
The algebraic rules in \eq{algebra} can be represented by using the Pauli matrices. For the moment, however, we do not choose any particular representation of these operators.

Using the operators in \eq{algebra}, the quaternionic Landau level (QLL) is characterized by the Hamiltonian
\begin{gather}
    \hat{H}^{\mathrm{QLL}} = \underbrace{\frac{p_{4D}^2}{2m} + \frac{1}{2}m\omega_0^2 r_{4D}^2}_{\hat{H}_{\mathrm{osc}}} 
    \underbrace{- \omega_0\sum_{0\leq a<b\leq 3} \Gamma^{ab}L_{ab}}_{\hat{H}_{\mathrm{spin}}},
    \qquad 
    \br = (t,x,y,z), \quad 
    r_{4D} = |\br|,\quad 
    p_{4D} = -i\hbar\nabla_{\br}.
  \label{QLL_ham}
\end{gather}
Using the magnetic length $\ell = \sqrt{\hbar/m\omega_0}$, we define the creation/annihilation operators
\begin{gather}
    a_{\mu} = \frac{1}{\sqrt{2}} \left( \frac{x_{\mu}}{\ell} + \ell \frac{\partial}{\partial x^{\mu}}\right), 
    \qquad 
    a_{\mu}^{\dagger} = \frac{1}{\sqrt{2}} \left( \frac{x_{\mu}}{\ell} - \ell \frac{\partial}{\partial x^{\mu}}\right).
  \label{ca_operators}
\end{gather}
Using these operators, the two parts of the Hamiltonian in \eq{QLL_ham} can be rewritten as
\begin{subequations}
\begin{align}
    \hat{H}_{\mathrm{osc}} &= \hbar\omega_0 \sum_{\mu=0}^{3}\left( a_{\mu}^{\dagger}a_{\mu} + \frac{1}{2}\right) = \hbar\omega_0 (N+2), \qquad 
    N = \sum_{\mu=0}^{3} a_{\mu}^{\dagger}a_{\mu},
\\
    \hat{H}_{\mathrm{spin}} &= i\hbar\omega_0 \sum_{\mu<\nu} \Gamma^{\mu\nu} (a_{\mu}^{\dagger} a_{\nu} - a_{\nu}^{\dagger} a_{\mu}).
\end{align}
  \label{opr_hams}
\end{subequations}
Creating a state with definite angular momentum increases the oscillator contribution to the energy while decreasing the spin contribution. Thus, \eq{opr_hams} suggests a large degeneracy of states: the Landau levels. This is most transparent after introducing the quaternionic annihilation operator $\mathcal{D}_a$:
\begin{gather}
    \mathcal{D}_a =  \sum_{\mu=0}^{3} e_{\mu}a_{\mu}, 
    \quad 
    \mathcal{D}_a^{\dagger} = \sum_{\mu=0}^{3} \bar{e}_{\mu}a_{\mu}^{\dagger},
    \qquad 
    \bar{e}_{0}=e_0, \quad 
    \bar{e}_{k}=-e_k\;(k=1,2,3).
\end{gather}
Then the Hamiltonian in \eq{QLL_ham} can be rewritten as
\begin{gather}
    \hat{H}^{\mathrm{QLL}} = \hbar\omega_0( \mathcal{D}_a^{\dagger}\mathcal{D}_a + 2).
\end{gather}
Since the differential operator $\mathcal{D}_a^{\dagger}\mathcal{D}_a$ is positive definite, the ground-state subspace is precisely the kernel
\begin{gather}
    \mathscr{H}_{g}^{ \mathrm{QLL}} = \ker \mathcal{D}_a^{\dagger}\mathcal{D}_a = \ker \mathcal{D}_a = \set{\psi:\bb{R}^4\to\bb{C}^2}{\mathcal{D}_a\psi=0}.
\end{gather}
Since there are infinitely many solutions to the equation $\mathcal{D}_a\psi = 0$ in the flat space $\bb{R}^4$, the ground states are infinitely degenerate. This solution space constitutes the quaternionic lowest Landau level (QLLL).

To describe the solution space of $\mathcal{D}_a$, first factor out the exponential part by setting
\begin{gather}
    \psi(\mathbf{r}) = f(\mathbf{r}) e^{-r^2/2\ell^2}.
\end{gather}
Then the equation $\mathcal{D}_a\psi=0$ is equivalent to
\begin{gather}
    \sum_{\mu=0}^{3} e_{\mu}\partial_{\mu}f = 0.
\end{gather}
Now, we observe that the algebra in \eq{algebra} could be realized by setting $(e_0,e_1,e_2,e_3) = (\sigma_0,i\sigma_3,i\sigma_2,i\sigma_1)$, using the Pauli matrices
\begin{gather}
    \sigma_0 = \begin{pmatrix}1&0\\0&1
    \end{pmatrix},
    \qquad 
    \sigma_1 = \begin{pmatrix}0&1\\1&0
    \end{pmatrix},
    \qquad 
    \sigma_2 = \begin{pmatrix}0&-i\\i&0
    \end{pmatrix},
    \qquad 
    \sigma_3 = \begin{pmatrix}1&0\\0&-1
    \end{pmatrix}.
\end{gather}
The connection to quaternion numbers is as follows. Identifying the spinor space $\bb{C}^2$ with the quaternion algebra $\bb{H}$ by the correspondence
\begin{gather}
    \begin{pmatrix}
        c_{\uparrow} \\ c_{\downarrow}
    \end{pmatrix}\in\bb{C}^2
    \quad\xrightarrow{\qquad F\qquad} \quad
    c_{\uparrow} - Jc_{\downarrow}\in\bb{H},
  \label{C2_to_H}
\end{gather}
Left multiplication of the vector $(c_{\uparrow},c_{\downarrow})^{\top}$ by the Pauli matrices corresponds precisely to multiplying $c_{\uparrow} - Jc_{\downarrow}$ by the imaginary unit quaternions on the left:
\begin{equation}
\begin{alignedat}{2}
    &F\left[\sigma_0 \begin{pmatrix}
        c_{\uparrow} \\ c_{\downarrow}
    \end{pmatrix}\right]
    = 1\cdot(c_{\uparrow} - Jc_{\downarrow}),
    &\qquad& 
    F\left[i\sigma_3 \begin{pmatrix}
        c_{\uparrow} \\ c_{\downarrow}
    \end{pmatrix}\right]
    = I\cdot(c_{\uparrow} - Jc_{\downarrow}),
\\
    &F\left[i\sigma_2 \begin{pmatrix}
        c_{\uparrow} \\ c_{\downarrow}
    \end{pmatrix}\right]
    = J\cdot(c_{\uparrow} - Jc_{\downarrow}),
    &\qquad&
    F\left[i\sigma_1 \begin{pmatrix}
        c_{\uparrow} \\ c_{\downarrow}
    \end{pmatrix}\right]
    = K\cdot(c_{\uparrow} - Jc_{\downarrow}).
\end{alignedat}
\end{equation}
Through the identification in \eq{C2_to_H}, we can identify the wave function $f(\mathbf{r})$ with $\tilde{f}(\mathbf{r}) = F[f(\mathbf{r})]$ which satisfies
\begin{gather}
    \sum_{\mu=0}^{3} \lambda(\tilde{e}_{\mu})\partial_{\mu} \tilde{f} = 0, 
    \qquad 
    (\tilde{e}_{\mu}) := (1,I,J,K), \qquad 
    \lambda(\tilde{e}_{\mu})x := \tilde{e}_{\mu}\,x.
  \label{sm:Fueter}
\end{gather}
\eq{sm:Fueter} is the Fueter equation~\cite{Sudbery1979,LiWu2013}. 
Its polynomial solution space is spanned by symmetrized products of the three quaternionic variables~\cite{Sudbery1979}
\begin{gather}
    Z_1 := tI-x,\qquad
    Z_2 := tJ-y,\qquad
    Z_3 := tK-z.
\end{gather}
Representative low-degree solutions are
\begin{gather}
    Z_1,\qquad Z_2,\qquad Z_3,
    \nonumber\\
    Z_1Z_2+Z_2Z_1,\qquad
    Z_2Z_3+Z_3Z_2,\qquad
    Z_1Z_3+Z_3Z_1,
    \nonumber\\
    Z_1Z_2Z_3+Z_1Z_3Z_2+
    Z_2Z_3Z_1+Z_2Z_1Z_3+ Z_3Z_1Z_2+Z_3Z_2Z_1,\qquad \ldots .
\end{gather}
When restricted to the two-dimensional $(t,x)$ plane, the Fueter equation reduces to the holomorphic and anti-holomorphic wave-function structure proposed for the quantum spin Hall effect~\cite{BernevigZhang2006QSH}. To see this, consider the quaternionic wave function
\begin{gather}
    \Psi(t,x)
    =
    \widetilde{f}(t,x)\,
    e^{-(t^2+x^2)/2\ell^2},
    \qquad
    \widetilde{f}(t,x)
    =
    F\left[
    \begin{pmatrix}
        \phi_{\uparrow}(t,x)\\
        \phi_{\downarrow}(t,x)
    \end{pmatrix}
    \right]
    =
    \phi_{\uparrow}(t,x)-J\phi_{\downarrow}(t,x).
\end{gather}
Setting $\partial_y=\partial_z=0$, Eq.~\eqref{sm:Fueter} becomes
\begin{align}
    0
    &=
    \left(
        \partial_t + I\partial_x
    \right)\widetilde{f}(t,x)=
    \left(\partial_t + I\partial_x\right)\phi_{\uparrow}
    -
    J\left(\partial_t - I\partial_x\right)\phi_{\downarrow},
\end{align}
and hence
\begin{gather}
    \left(\partial_t + I\partial_x\right)\phi_{\uparrow}=0,
    \qquad
    \left(\partial_t-I\partial_x\right)\phi_{\downarrow}=0.
\end{gather}
Thus, with respect to the complex coordinate $w=t+Ix=-IZ_1$, $\phi_{\uparrow}$ is holomorphic whereas $\phi_{\downarrow}$ is anti-holomorphic. In this two-dimensional restriction, only the variable $Z_1=tI-x$ remains, so the distinction between symmetrized and arbitrary products disappears.

As shown in Proposition~\ref{prop:closed}, ideal four-dimensional time reversal symmetric crystalline bands possess an analogous quaternion-polynomial structure. The key difference is that the Fueter equation governing the QLLL selects symmetrized polynomials, whereas the ideal band condition permits arbitrary polynomials in the noncommuting variables $Z_1$, $Z_2$, and $Z_3$. The ideal band polynomial space therefore contains a distinguished symmetrized sector with the same algebraic structure as the QLLL wave functions. If the band energetics favor this sector, the resulting low-energy states may inherit the algebraic structure of the QLLL. This observation provides a possible route toward understanding interacting time reversal symmetric phases through their connection to quaternionic Landau level physics.

\subsection*{Supplementary Note 5.
Formal statements and proofs corresponding to main-text results
}

The main text presents the central results in prose. For completeness, we state the corresponding formal results here, preserving the assumptions, equivalence conditions, and normalization conventions used in the main text. We also provide the proofs of the formal statements, together with auxiliary lemmas used in those proofs.

We first review the basic definitions. The quaternionic Bloch state is
\begin{gather}
    \Phi_{\bk}(\br) = e^{i(k_tt+k_xx+k_yy+k_zz)} \,\Psi_{\bk}(\br)
\end{gather}
where the periodic part is a combination of two ordinary, complex-valued periodic Bloch vectors:
\begin{gather}
    \Psi_{\bk} = \frac{1}{\sqrt{2}} \left( \psi^1_{\bk} + \psi^2_{\bk}J \right).
\end{gather}
Then the QQGT (quaternionic QGT) is defined by
\begin{gather}
    \mathscr{Q}_{ab}(\bk) := \bra{\partial_a \Psi_{\bk}}(1-\mathscr{P}_{\bk}) \ket{\partial_b \Psi_{\bk}}
  \label{sup:def_qqgt}
\end{gather}
where the quaternionic projector $\mathscr{P}_{\bk}$ is
\begin{align}
    \mathscr{P}_{\bk} :=&\; \ket{\Psi_{\bk}}\bra{\Psi_{\bk}} 
\nonumber\\
    =&\; \frac{1}{2}
    \left( 
    \ket{\psi^1_{\bk}} + \ket{\psi^2_{\bk}}J 
        \right)
    \left( 
    \bra{\psi^1_{\bk}} - J\bra{\psi^2_{\bk}} 
        \right)
\nonumber\\
    =&\; \frac{1}{2}
    \left( 
    \ket{\psi^1_{\bk}} \bra{\psi^1_{\bk}} + \ket{\psi^2_{\bk}} \bra{\psi^2_{\bk}} + \ket{\psi^2_{\bk}}J \bra{\psi^1_{\bk}} - \ket{\psi^1_{\bk}}J \bra{\psi^2_{\bk}}
    \right)
\nonumber\\
    =&\; \frac{1}{2}P_{\bk} + \frac{1}{2} \left( 
    \ket{\psi^2_{\bk}}J \bra{\psi^1_{\bk}} - \ket{\psi^1_{\bk}}J \bra{\psi^2_{\bk}}
    \right).
  \label{SM:qProj}
\end{align}
\eq{sup:def_qqgt} should be compared to the complex matrix-valued QGT,
\begin{equation}
\begin{aligned}
    {Q}_{ab}^{ij}(\bk) &= \bra{\partial_{a}\psi^i(\bk)} \,(1-P)\, \ket{\partial_{b}\psi^j(\bk)}, \qquad 
    P=\sum_{i=1,2} \ket{\psi^i}\bra{\psi^i}.
  \label{q-QGT}
\end{aligned}
\end{equation}
The formal statements and proofs of the propositions needed in the main text are in order.

\begin{proposition}[Kramers partners have the same quantum metric]
    \label{prop:isometry}
    Let $\Theta=\mathcal{PT}$ satisfy $\Theta^2=-1$, and let $\psi^2=\Theta\psi^1$. For the Kramers-pair projector $P=\ket{\psi^1}\bra{\psi^1}+\ket{\psi^2}\bra{\psi^2}$,
\begin{align}
    g_{ab} &\equiv \frac{1}{2} \Big[\bra{\partial_{a}\psi^1(\bk)} \,(1-P)\, \ket{\partial_{b}\psi^1(\bk)} + (a\leftrightarrow b) \Big]
\nonumber\\
    &= \frac{1}{2} \Big[\bra{\partial_{a}\psi^2(\bk)} \,(1-P)\, \ket{\partial_{b}\psi^2(\bk)} + (a\leftrightarrow b) \Big].
\end{align}
\end{proposition}
\paragraph*{Proof of Proposition~\ref{prop:isometry}.}
    Using the anti-unitarity of $\Theta$, namely the property $\braket{\Theta v_1}{\Theta v_2} = \braket{v_1}{v_2}^{*}$,
\begin{align}
    \bra{\partial_{a} \psi^2(\bk)} \,(1-P)\, \ket{\partial_{b}\psi^2(\bk)} &= \bra{\Theta \partial_{a} \psi^1(\bk)} \,(1-P)\, \ket{\Theta \partial_{b} \psi^1(\bk)}
\nonumber\\
    &= \bra{\partial_{a} \psi^1(\bk)} \,(1-P)\, \ket{\partial_{b} \psi^1(\bk)}^{*}
\nonumber\\
    &= \bra{\partial_{b} \psi^1(\bk)} \,(1-P)\, \ket{\partial_{a} \psi^1(\bk)}.
\end{align}
    Changing the roles of $a$ and $b$ and summing gives the desired result.
\paragraph*{(end of proof)}
\mbox{}\newline

\begin{proposition}[Symmetric and antisymmetric parts of the Kramers-pair QGT]
  \label{parts_qgt}
    For the matrix-valued Kramers-pair QGT
\begin{gather}
    {Q}_{ab}^{ij}(\bk)=\bra{\partial_a\psi^i(\bk)}(1-P)\ket{\partial_b\psi^j(\bk)},
  \label{sup:def_q}
\end{gather}
    the antisymmetric part in the lower indices is the non-Abelian Berry curvature,
\begin{gather}
    {Q}_{ab}^{ij}-{Q}_{ba}^{ij}=F_{ab}^{ij}=F_{ab}^x\sigma_x+F_{ab}^y\sigma_y+F_{ab}^z\sigma_z.
\end{gather}
    The symmetric part defines the quantum metric matrix,
\begin{gather}
    g_{ab}^{ij}:=\frac{1}{2}\left({Q}_{ab}^{ij}+{Q}_{ba}^{ij}\right),
\end{gather}
    and for a Kramers pair $g_{ab}^{ij}(\bk)=g_{ab}(\bk)\delta^{ij}$. Hence
\begin{gather}
    {Q}_{ab}=g_{ab}\sigma_0+\frac{1}{2}\left(F_{ab}^x\sigma_x+F_{ab}^y\sigma_y+F_{ab}^z\sigma_z\right).
\end{gather}
\end{proposition}
\paragraph*{Proof of Proposition~\ref{parts_qgt}.}
    A basic property of the complex QGT that directly follows from the definition in \eq{q-QGT} is that
\begin{gather}
    {Q}_{ab}^{ij} = ({Q}_{ba}^{ji})^{*} = ({Q}_{ba}^{\dagger})^{ij}, \quad\text{i.e.,}\quad {Q}_{ab} = {Q}_{ba}^{\dagger}.
\end{gather}
    As a result, the symmetric (anti-symmetric) part of the QGT in lower indices is a hermitian (anti-Hermitian) matrix. In particular, the quantum metric matrix takes the form
\begin{gather}
    g_{ab}^{ij} = 
    \begin{pmatrix}
        u_{ab} & v_{ab}-iw_{ab} \\ v_{ab}+iw_{ab} & u'_{ab}
    \end{pmatrix}
  \label{sm:qmm}
\end{gather}
    for some real-valued functions $u_{ab}, u'_{ab},v_{ab},w_{ab}$. But Proposition~\ref{prop:isometry} shows that $u_{ab}=u'_{ab}$. Furthermore, repeating the proof of Proposition~\ref{prop:isometry} for different upper indices shows
\begin{alignat}{2}
    & \bra{\partial_{a}\psi^2(\bk)}\,(1-P)\, \ket{\partial_{b}\psi^1(\bk)} &\;=\;& \bra{\Theta\partial_{a}\psi^1(\bk)}\,(1-P)\, \ket{\partial_{b}\psi^1(\bk)}
\nonumber\\
    =\;& \bra{\Theta^2\partial_{a}\psi^1(\bk)}\,(1-P)\, \ket{\Theta\partial_{b}\psi^1(\bk)}^{*}
    &\;=\;& - \bra{\partial_{a}\psi^1(\bk)}\,(1-P)\, \ket{\Theta\partial_{b}\psi^1(\bk)}^{*}
\nonumber\\
    =\;& - \bra{\partial_{a}\psi^1(\bk)}\,(1-P)\, \ket{\partial_{b}\psi^2(\bk)}^{*}
    &\;=\;& - \bra{\partial_{b}\psi^2(\bk)}\,(1-P)\, \ket{\partial_{a}\psi^1(\bk)}.
\end{alignat}
    Changing the role of $a,b$ and summing over, one obtains $g_{ab}^{21}=-g_{ab}^{21}$ and hence $g_{ab}^{21}=0$. This means $v_{ab}=w_{ab}=0$ in \eq{sm:qmm}, so the quantum metric matrix is diagonal in spin indices. The surviving factor $u_{ab}$ is $g_{ab}$ by definition.

    We conclude by considering the anti-symmetric part of the QGT. Using $A_{a}^{ij}:= \braket{\psi^i}{\partial_{a}\psi^j}$, direct computation shows that
\begin{align}
    {Q}_{ab}^{ij} &= \braket{\partial_a\psi^i}{\partial_b\psi^j} - \sum_{k=1,2}\braket{\partial_a\psi^i}{\psi^k} \braket{\psi^k}{\partial_b\psi^j}
\nonumber\\
    &= \braket{\partial_a\psi^i}{\partial_b\psi^j} + A_{a}^{ik}A_{b}^{kj} = \braket{\partial_a\psi^i}{\partial_b\psi^j} + (A_{a}A_{b})^{ij},
\\
    {Q}_{ab}^{ij} - {Q}_{ba}^{ij} &= \braket{\partial_a\psi^i}{\partial_b\psi^j} - \braket{\partial_b\psi^i}{\partial_a\psi^j} + (A_aA_b-A_bA_a)^{ij}
\nonumber\\
    &= (\partial_{a}A_b - \partial_{b}A_a + [A_a,A_b])^{ij} = F_{ab}^{ij}.
\end{align}
\paragraph*{(end of proof)}
\mbox{}\newline
\mbox{}\newline
Now we can prove the following lemma, which will be useful for the proof of Proposition~\ref{prop:4band}.
\begin{lemma}
  \label{SM:lemma_PdPdP}
      The tensor ${Q}_{ab}$ in Proposition~\ref{parts_qgt} is related to the projector $P(\bk)$ by
\begin{gather}
    \sum_{i,j=1}^{2} \ket{\psi^i}\,{Q}_{ab}^{ij}\,\bra{\psi^j} = P\partial_{a}P\partial_{b}P,
  \label{sm:Q_PdPdP}
\end{gather}
    which is equivalent to
\begin{subequations}
\begin{gather}
    g\,\ket{\psi^i}\,(\sigma_0)^{ij}\,\bra{\psi^j} = P\dd P\odot\dd P,
\\
    \ket{\psi^i}\, \big(F^x\,\sigma_x + F^y\,\sigma_y + F^z\,\sigma_z\big)^{ij}\,\bra{\psi^j} = P\dd P\wedge\dd P,
\end{gather}
  \label{sm:PdPdP_sym_anti}
\end{subequations}
    where $\odot, \wedge$ denote the symmetric and anti-symmetric product of tensors, respectively, defined by
\begin{subequations}
\begin{align}
    (\omega_1\odot\omega_2)(X,Y) &= \frac{1}{2} \big[\omega_1(X) \omega_2(Y) + \omega_1(Y) \omega_2(X)\big],
\\
    (\omega_1\wedge\omega_2)(X,Y) &= \omega_1(X) \omega_2(Y) - \omega_1(Y) \omega_2(X).
\end{align}
\end{subequations}
\end{lemma}
\paragraph*{Proof of Lemma~\ref{SM:lemma_PdPdP}.}
    To see \eq{sm:Q_PdPdP}, it suffices to show that
\begin{gather}
    {Q}_{ab}^{ij} = \bra{\psi^i}P\partial_aP\partial_bP\ket{\psi^j}.
\end{gather}
    Plugging $P=\sum_{k}\ket{\psi^k}\bra{\psi^k}$, direct computation shows that
\begin{align}
    \bra{\psi^i}P\partial_aP &= \bra{\psi^i}\partial_aP = \bra{\psi^i} \sum_{k}\Big\{ \ket{\partial_a\psi^k}\bra{\psi^k} + \ket{\psi^k}\bra{\partial_a\psi^k}\Big\} = \bra{\partial_a\psi^i} + \sum_{k} \braket{\psi^i}{\partial_a\psi^k} \bra{\psi^k}
\nonumber\\
    &= \bra{\partial_a\psi^i} - \sum_{k} \braket{\partial_a\psi^i}{\psi^k} \bra{\psi^k} = \bra{\partial_a\psi^i} (1-P).
\end{align}
    Likewise,
\begin{gather}
    (\partial_bP)\ket{\psi^j} = (1-P) \ket{\partial_b\psi^j}.
\end{gather}
    Taking the product, we obtain
\begin{gather}
    \bra{\psi^i}P\partial_aP\partial_bP\ket{\psi^j} = \bra{\partial_a\psi^i} (1-P)^2\ket{\partial_b\psi^j} = \bra{\partial_a\psi^i} (1-P)\ket{\partial_b\psi^j} = {Q}_{ab}^{ij}.
\end{gather}
    (Anti-)symmetrization of the lower indices yields \eq{sm:PdPdP_sym_anti}.
\paragraph*{(end of proof)}
\mbox{}\newline
\mbox{}\newline
The following lemma is needed for the proof of Propositions~\ref{prop:QQGT},~\ref{prop:closed}. It essentially says that the complex matrix-valued projector $P$ and the quaternion-valued projector $\mathscr{P}$ can be used interchangeably. For another instance where $P$ and $\mathscr{P}$ are interchangeable, see the proof of Proposition~\ref{prop:positive}.

\begin{lemma}
  \label{SM:lemma_Proj}
    We can replace the quaternionic projector $\mathscr{P}$ by the ordinary projector $P=\ket{\psi^1}\bra{\psi^1} + \ket{\psi^2}\bra{\psi^2}$:
\begin{align}
    \mathscr{Q}_{ab}(\bk) :=&\; \bra{\partial_a\Psi_{\bk}}(1-\mathscr{P}_{\bk}) \ket{\partial_b\Psi_{\bk}}
\nonumber\\
    =&\; \bra{\partial_a\Psi_{\bk}} (1-P_{\bk}) \ket{\partial_b\Psi_{\bk}}.
  \label{SM:QQGT}
\end{align}
    hence, the QQGT can be written as
\begin{gather}
    \mathscr{Q}_{ab} = \frac{1}{2} \Big( \bra{\partial_a\psi^{1}} - J\bra{\partial_a\psi^{2}} \Big)
     \Big(1-P\Big)\, 
     \Big( \ket{\partial_b\psi^1} +  \ket{\partial_b\psi^{2}}J \Big)
  \label{SM:QQGT_re}
\end{gather}
\end{lemma}
\paragraph*{Proof of Lemma~\ref{SM:lemma_Proj}.}
To avoid cluttering with bras and kets, we use the notation 
\begin{gather}
    \ket{\Psi}=\Psi = \frac{1}{\sqrt{2}} (\psi^1 + \psi^2J), \qquad 
    \bra{\Psi}=\Psi^{\dagger} = \frac{1}{\sqrt{2}} (\psi^{1\dagger} - J\psi^{2\dagger}).
\end{gather}
We need to show that
\begin{gather}
    (\partial_a\Psi^{\dagger}) P (\partial_b\Psi) = (\partial_a\Psi^{\dagger}) \mathscr{P} (\partial_b\Psi).
\end{gather}
According to \eq{SM:qProj}, this is equivalent to
\begin{gather}
    (\partial_a\psi^{1\dagger}-J\partial_a\psi^{2\dagger}) (\psi^{1}\psi^{1\dagger} + \psi^{2}\psi^{2\dagger}) (\partial_b\psi^1+\partial_b\psi^2J) 
\nonumber\\
    = (\partial_a \psi^{1\dagger}-J\partial_a \psi^{2\dagger})
    (\psi^{2}J \psi^{1\dagger} - \psi^{1}J \psi^{2\dagger})
    (\partial_b\psi^1+ \partial_b\psi^2J).
  \label{SM:WTS_1}
\end{gather}
Now, using the symbols $f_{a}, g_{a}, h_{a}$ defined by the $\mathrm{SU}(2)$ Berry connection matrix
\begin{gather}
    A_{a}^{ij} = \psi^{i\dagger}\partial_a \psi^{j} = 
    \begin{pmatrix}
        if_a & g_a + ih_a \\
        -g_a+ih_a & -if_a
    \end{pmatrix},
  \label{sm:bc}
\end{gather}
one can verify by straightforward computation that both sides of \eq{SM:WTS_1} are equal to
\begin{gather}
    2f_af_b + 2(g_a+ih_a)(g_b-ih_b) + 2J\, \big[ 
    if_a(g_b-ih_b) - if_b(g_a-ih_a)
    \big].
\end{gather}
\paragraph*{(end of proof)}
\mbox{}\newline

\begin{proposition}[Quaternionic form of the QGT]
\label{prop:QQGT}
    The QQGT
\begin{gather}
    \mathscr{Q}_{ab}(\bk)=\bra{\partial_a\Psi(\bk)}(1-\mathscr{P}(\bk))\ket{\partial_b\Psi(\bk)}
\end{gather}
    is identical to
\begin{subequations}
\begin{gather}
    \mathscr{Q}_{ab}=g_{ab}\,1+\frac{1}{2}\left(\mathscr{F}_{ab}^I I+\mathscr{F}_{ab}^J J+\mathscr{F}_{ab}^K K\right),
\\
    \mathscr{F}_{ab}^I=-iF_{ab}^z,\qquad
    \mathscr{F}_{ab}^J=-iF_{ab}^y,\qquad
    \mathscr{F}_{ab}^K=-iF_{ab}^x.
\end{gather}
\end{subequations}
    Moreover, the curvature two-forms are pullbacks of the local symplectic forms on $\hp^n$:
\begin{gather}
    \mathscr{F}^I=2\mathscr{P}^{*}\omega_I,
    \qquad
    \mathscr{F}^J=2\mathscr{P}^{*}\omega_J,
    \qquad
    \mathscr{F}^K=2\mathscr{P}^{*}\omega_K.
\end{gather}
\end{proposition}
\paragraph*{Proof of Proposition~\ref{prop:QQGT}.}
By Lemma~\ref{SM:lemma_Proj}, we can take the definition of QGT in \eq{SM:QQGT_re}.
For notational simplicity, let us denote $\phi_{a}^{i} := (1-P) \partial_{a}\psi^{i}$.
We start with
\begin{align}
    2\mathscr{Q}_{ab} &= \big( \partial_a\psi^{1\dagger} - J\partial_a\psi^{2\dagger} \big)
    \big( 1-P\big)
    \big(\partial_b\psi^1 + \partial_b\psi^{2}J \big)
\nonumber\\
    &= \big(\phi_a^{1\dagger} - J\phi_a^{2\dagger}\big) \big(\phi_b^1 + \phi_b^{2}J \big)
\nonumber\\
    &= \underbrace{ {\phi_a^{1\dagger}}{\phi_b^1} + 
    {\phi_a^{2\top}}{\phi_b^{2*}}}_{=:\;C_1} 
    + \Big[ 
    \underbrace{ {\phi_a^{1\dagger}}{\phi_b^2} - 
    {\phi_a^{2\top}}{\phi_b^{1*}}}_{=:\;C_2}
        \Big] J.
  \label{proof2_c1c2}
\end{align}
The first part $C_1$ contains the coefficients of $1$ and $I$, and decomposes into the symmetric and the anti-symmetric part,
\begin{alignat*}{2}
    C_1 = {\phi_a^{1\dagger}}{\phi_b^1} + 
    {\phi_a^{2\top}}{\phi_b^{2*}}
    &= &&\frac{1}{2} 
    \Big( 
        \phi_a^{1\dagger} \phi_b^1 + \phi_b^{1\dagger} \phi_a^1 + \phi_a^{2\top} \phi_b^{2*} + \phi_b^{2\top} \phi_a^{2*}
    \Big)
\nonumber\\
    & &+\;&\frac{1}{2} 
    \Big( 
        \phi_a^{1\dagger} \phi_b^1 - \phi_b^{1\dagger} \phi_a^1 + \phi_a^{2\top} \phi_b^{2*} - \phi_b^{2\top} \phi_a^{2*}
    \Big).
\end{alignat*}
The symmetric part is manifestly real, and equals $2g_{ab}1$. The anti-symmetric part is manifestly imaginary, hence equals $ci$ for some $c$. We need to show that $c = \mathscr{F}^{I}_{ab}$. This can be checked by computing
\begin{align*}
    c\,i &= \frac{1}{2} 
    \left\{ 
        \phi_a^{1\dagger} \phi_b^1 - \phi_b^{1\dagger} \phi_a^1 + \phi_a^{2\top} \phi_b^{2*} - \phi_b^{2\top} \phi_a^{2*} 
    \right\}
\nonumber\\
    &= i\op{Im} 
    \left\{ 
        \phi_a^{1\dagger} \phi_b^1 - \phi_b^{2\top} \phi_a^{2*} 
    \right\} \;
    = i\op{Im} 
    \left\{ 
        {Q}_{ab}^{11} - {Q}_{ab}^{22}
    \right\} \;
    = i\op{Im} \op{tr} \left[ {Q}_{ab}\, \sigma_z \right],
\\
    c &= \op{Im} \op{tr} \left[ {Q}_{ab}\, \sigma_z \right] \;
    = \op{Im} \underbrace{\frac{1}{2} \op{tr} \left[ F_{ab}\, \sigma_z \right]}_{=F^z_{ab}} = -iF^z_{ab} =  \mathscr{F}_{ab}^I.
\end{align*}
In the last equality, we used the fact that $\mathscr{F}_{ab}^I$ is real-valued.
Now we turn to $C_2$ in \eq{proof2_c1c2}, which contains the coefficients of $J$ and $K$. Direct computation shows that the $J$-component is
\begin{gather*}
    \op{Re}\left\{ \phi_a^{1\dagger} \phi_b^2 - \phi_a^{2\top} \phi_b^{1*} \right\}
    = \op{Re}\left\{ {Q}_{ab}^{12} - {Q}_{ba}^{12} \right\} 
    = \frac{1}{2}\big({Q}_{ab}^{12} - {Q}_{ba}^{12}+{Q}_{ba}^{21}-{Q}_{ab}^{21}\big) 
\nonumber\\
    = \op{Re} \left\{ {Q}_{ab}^{12}-{Q}_{ab}^{21} \right\} = \op{Re}\op{tr}\big[{Q}_{ab}(-i\sigma_y)\big]
    = -iF^y_{ab}= \mathscr{F}_{ab}^J,
\end{gather*}
while the $K$-component is
\begin{gather*}
    \op{Im}\left\{ \phi_a^{1\dagger} \phi_b^2 - \phi_a^{2\top} \phi_b^{1*} \right\}
    = \op{Im}\left\{ {Q}_{ab}^{12} - {Q}_{ba}^{12} \right\} 
    = \frac{1}{2i}\big({Q}_{ab}^{12} - {Q}_{ba}^{12}-{Q}_{ba}^{21}+{Q}_{ab}^{21}\big) 
\nonumber\\
    = \op{Im} \left\{ {Q}_{ab}^{12} + {Q}_{ab}^{21} \right\}
    = \op{Im}\op{tr} \big[{Q}_{ab}\,\sigma_x\big] = -iF^x_{ab}= \mathscr{F}_{ab}^K.
\end{gather*}
Thus, we have proved Eq.~(\MainQQGTEq) in the main text:
\begin{gather}
    \mathscr{Q}_{ab} = g_{ab}1 + \frac{1}{2} \Big[ \mathscr{F}^I_{ab}\,I + \mathscr{F}^J_{ab}\,J + \mathscr{F}^K_{ab}\,K \Big].
\end{gather}
It remains to show that the Berry curvatures are the pullback of the two-forms in \eq{sm:qk_strs}. By definition, the pullback of the forms $\omega_{I,J,K}$ are obtained by plugging $\mathscr{P}=\mathscr{P}(\bk)$ to \eq{sm:qk_strs}:
\begin{equation}
\begin{aligned}
    2\mathscr{P}^{*}\omega_{I} &= \Tr\real[-I\mathscr{P}(\bk) \partial_{a}\mathscr{P}(\bk) \partial_{b}\mathscr{P}(\bk)] dk_a\wedge dk_b
\\
    2\mathscr{P}^{*}\omega_{J} &= \Tr\real[-J\mathscr{P}(\bk) \partial_{a}\mathscr{P}(\bk) \partial_{b}\mathscr{P}(\bk)] dk_a\wedge dk_b
\\
    2\mathscr{P}^{*}\omega_{K} &= \Tr\real[-K\mathscr{P}(\bk) \partial_{a}\mathscr{P}(\bk) \partial_{b}\mathscr{P}(\bk)] dk_a\wedge dk_b
\end{aligned}
  \label{omegas_pullback}
\end{equation}
Now, using $\mathscr{P}(\bk)=\Psi_{\bk}\Psi^{\dagger}_{\bk}$, we compute
\begin{align}
    &\Tr \big[\mathscr{P}(\bk) \partial_{a}\mathscr{P}(\bk) \partial_{b}\mathscr{P}(\bk) \big]
\nonumber\\
    &= \Tr \Big[\Psi_{\bk}\Psi^{\dagger}_{\bk} \Big( \partial_a\Psi_{\bk} \Psi^{\dagger}_{\bk} + \Psi_{\bk} \partial_a\Psi^{\dagger}_{\bk} \Big)
    \Big( \partial_b\Psi_{\bk} \Psi^{\dagger}_{\bk} + \Psi_{\bk} \partial_b\Psi^{\dagger}_{\bk} \Big) \Big]
\nonumber\\
    &= (\Psi^{\dagger}_{\bk} \partial_a\Psi_{\bk}) (\Psi^{\dagger}_{\bk} \partial_b\Psi_{\bk}) + (\Psi^{\dagger}_{\bk} \partial_a\Psi_{\bk}) (\partial_b\Psi^{\dagger}_{\bk} \Psi_{\bk}) + \partial_a\Psi^{\dagger}_{\bk} \partial_b\Psi_{\bk} + (\partial_a\Psi^{\dagger}_{\bk} \Psi_{\bk}) (\partial_b\Psi^{\dagger}_{\bk} \Psi_{\bk})
\nonumber\\
    &= \partial_a\Psi^{\dagger}_{\bk} \partial_b\Psi_{\bk} + (\partial_a\Psi^{\dagger}_{\bk} \Psi_{\bk}) (\partial_b\Psi^{\dagger}_{\bk} \Psi_{\bk})
\nonumber\\
    &= \partial_a\Psi^{\dagger}_{\bk} (1-\Psi_{\bk} \Psi^{\dagger}_{\bk}) \partial_b\Psi_{\bk}
\nonumber\\
    &= \bra{\partial_a\Psi(\bk)} (1-\mathscr{P}(\bk)) \ket{\partial_b\Psi(\bk)}
    = \mathscr{Q}_{ab}
    = g_{ab}1 + \frac{1}{2} \Big[ \mathscr{F}^I_{ab}\,I + \mathscr{F}^J_{ab}\,J + \mathscr{F}^K_{ab}\,K \Big]
\end{align}
Plugging this result to \eq{omegas_pullback}, we get the desired result
\begin{gather}
    2\mathscr{P}^{*} \omega_{\mathrm{FS}}^{I} = \frac{1}{2}\mathscr{F}_{ab}^{I} \dd k_a \wedge \dd k_b = \mathscr{F}^{I}, \qquad 
    2\mathscr{P}^{*}\omega_{\mathrm{FS}}^{J} = \mathscr{F}^{J}, \qquad 
    2\mathscr{P}^{*}\omega_{\mathrm{FS}}^{K} = \mathscr{F}^{K}.
\end{gather}
As a consequence, the local complex structures $(\bb{I,J,K})$ of $\hp^n$ can be pulled back to the Brillouin zone torus $(A_{\bb{I}},A_{\bb{J}},A_{\bb{K}})$ by demanding the relations
\begin{gather}
    \frac{1}{2}\mathscr{F}^I(v_1, v_2) = g(v_1, A_{\bb{I}}v_2), \quad
    \frac{1}{2}\mathscr{F}^J(v_1, v_2) = g(v_1, A_{\bb{J}}v_2), \quad
    \frac{1}{2}\mathscr{F}^K(v_1, v_2) = g(v_1, A_{\bb{K}}v_2)
  \label{T4_trinity}
\end{gather}
for all tangent vectors $v_1,v_2$ on the Brillouin zone torus. \eq{T4_trinity} implies that the relevant tensors, when represented as a matrix in a local coordinate system, satisfy
\begin{gather}
    [A_{\bb{I}}]=\frac{1}{2}[g]^{-1}[\mathscr{F}^I],\quad
    [A_{\bb{J}}]=\frac{1}{2}[g]^{-1}[\mathscr{F}^J],\quad
    [A_{\bb{K}}]=\frac{1}{2}[g]^{-1}[\mathscr{F}^K],
\end{gather}
where $[\cdot]$ denotes the matrix representation.
\paragraph*{(end of proof)}
\mbox{}\newline

\begin{proposition}[Non-negativity and null vectors of the QQGT]
  \label{prop:positive}
    The QQGT is positive semidefinite. For every quaternion vector $\mathbf{q}=(q^a)$,
\begin{gather}
    \sum_{a,b}\overline{q}^a\mathscr{Q}_{ab}q^b\geq 0.
\end{gather}
    Equality holds if and only if
\begin{gather}
    X(\mathbf{q})\equiv\sum_a(1-P)\ket{\partial_a\Psi}\,q^a
    =\sum_a(1-\mathscr{P})\ket{\partial_a\Psi}\,q^a=0.
\end{gather}
    In particular, such a quaternion vector is a null vector of $\mathscr{Q}_{ab}$.
\end{proposition}
\paragraph*{Proof of Proposition~\ref{prop:positive}.}
    For given $\mathbf{q}$, define $X = (1-P)(\partial_a\psi^1 + \partial_a\psi^{2}J) q^a$. Then using \eq{SM:QQGT_re},
\begin{gather}
    \overline{q}^a \mathscr{Q}_{ab} q^b = \overline{q}^a \Big(\partial_a\psi^{1\dagger}-J\partial_a \psi^{2\dagger}\Big) \Big(1-P\Big) \Big(
    \partial_b\psi^1 + \partial_b\psi^{2}J\Big) q^b = \overline{X}X \geq 0.
\end{gather}
    Obviously, the equality holds if and only if $X=0$. The proof is complete if we show that
\begin{gather}
    P\ket{\partial_a\Psi} = \mathscr{P}\ket{\partial_a\Psi}.
  \label{sm:Pket}
\end{gather}
Using the Berry connection matrix in \eq{sm:bc}, straightforward computation shows that both sides of \eq{sm:Pket} equal $\ket{\Psi}\left\{if_a+(g_a+ih_a)J\right\}$.
\paragraph*{(end of proof)}
\mbox{}\newline
\mbox{}\newline
We turn to Theorem~\ref{prop:q-wirtinger}, which is a main result of this work. To this end, we begin by establishing a crucial lemma:
\begin{lemma}
  \label{SM:lemma_qk}
    Let $(\mathcal{M},g, \omega^I,\omega^J,\omega^K,\bb{I,J,K})$ be a quaternion-K\"ahler manifold, and consider an arbitrary rank-4 subspace $E\leq T_{x}\mathcal{M}$ of the tangent space of $\mathcal{M}$ at $x$. Then
\begin{gather}
    \sqrt{\det g_{E}} \geq \frac{1}{6}\,|\Omega_{E}|
    \quad\text{with}\quad \Omega:= \omega^I\wedge\omega^I + \omega^J\wedge\omega^J + \omega^K\wedge\omega^K,
  \label{sm:qwirtinger}
\end{gather}
    where $(g_E, \Omega_E)$ are restrictions of $(g, \Omega)$ to $E$.
    Furthermore, the equality holds in \eq{sm:qwirtinger} if and only if $E$ is stable under the quaternionic structure in the sense that
\begin{gather}
    \bb{I}(E)=\bb{J}(E)=\bb{K}(E)=E.
\end{gather}
\end{lemma}
\paragraph*{Proof of Lemma~\ref{SM:lemma_qk}.}
    Define the 4-form $\Omega:= \omega^I\wedge\omega^I + \omega^J\wedge\omega^J + \omega^K\wedge\omega^K$. Choose a set of vectors $S = \{X_1,X_2,X_3,X_4\}\subset E$. We want to show that
\begin{gather}
    \dd\op{Vol}(E):= \sqrt{\det [g(X_a,X_b)]_{a,b=1}^{4}} \geq \frac{1}{6} \big| \Omega(X_1,X_2,X_3,X_4) \big|.
  \label{sm_qk_wts}
\end{gather}
    If $S$ is linearly dependent, both sides of \eq{sm_qk_wts} vanish and there is nothing to prove. Assume therefore that the span $\langle S\rangle=E$ is four-dimensional. By the Gram--Schmidt process, choose a basis $\{e_1,e_2,e_3,e_4\}$ of $E$ that is orthonormal with respect to $g|_E$. Suppose $X_a = \sum_{i=1}^{4} e_i A^{i}_{a}$ with $A=(A^i_a)\in\mathrm{GL}(\bb{R}^4)$. By multilinearity of $\Omega$ and $\det$, we have
\begin{gather}
    \dd\op{Vol}(E) = |\det A|, \qquad \Omega(X_1,X_2,X_3,X_4) = \det A\cdot \Omega(e_1,e_2,e_3,e_4).
\end{gather}
    Thus, it is enough to show
\begin{gather}
    |\Omega(e_1,e_2,e_3,e_4)|\leq 6.
\end{gather}
    for every orthonormal frame $\{e_1,e_2,e_3,e_4\}$ of $E$.
    To this end, let $\alpha$ be an arbitrary 2-form on $E$ with $\alpha_{ij}:= \alpha(e_i,e_j)\in\bb{R}$. Then
\begin{gather}
    (\alpha\wedge\alpha)(e_1,e_2,e_3,e_4) = 2(\alpha_{12}\alpha_{34} - \alpha_{13}\alpha_{24} + \alpha_{14}\alpha_{23}).
\end{gather}
    Using the triangle inequality and $2|ab|\leq a^2+b^2$, we obtain
\begin{gather}
    |(\alpha\wedge\alpha)(e_1,e_2,e_3,e_4)| \leq 2 (|\alpha_{12}\alpha_{34}| + |\alpha_{13}\alpha_{24}| + |\alpha_{14}\alpha_{23}|)
    \leq \sum_{1\leq i<j\leq 4} \alpha_{ij}^2 =: \|\alpha\|^2.
  \label{sm:chain_ineq}
\end{gather}
    Applying \eq{sm:chain_ineq} to $\alpha=\omega^I,\omega^J,\omega^K$, we find
\begin{align}
    |\Omega(e_1,e_2,e_3,e_4)| &\leq \|\omega^I\|^2 + \|\omega^J\|^2 + \|\omega^K\|^2
\nonumber\\
    &= \sum_{1\leq i<j\leq 4} \left[ g(e_i,\bb{I}e_j)^2 + g(e_i,\bb{J}e_j)^2 + g(e_i,\bb{K}e_j)^2 \right].
\end{align}
    For each fixed pair $i<j$, consider the space $e_j^{\perp}:= \{v\in T_{x}\mathcal{M}\,|\,g(v,e_j)=0\}$. Since the vectors $\{\bb{I}e_j,\bb{J}e_j,\bb{K}e_j\} \subset e_j^{\perp}$ are orthonormal and $e_i\in e_j^{\perp}$,
\begin{gather}
    g(e_i,\bb{I}e_j)^2 + g(e_i,\bb{J}e_j)^2 + g(e_i,\bb{K}e_j)^2 \leq g(e_i,e_i)^2=1.
  \label{sm:ineq_proj}
\end{gather}
    There are $\binom{4}{2}=6$ such pairs, so
\begin{gather}
    |\Omega(e_1,e_2,e_3,e_4)|\leq6.
  \label{sm:lem2_conclusion}
\end{gather}
    Now, suppose $|\Omega(e_1,e_2,e_3,e_4)|=6$ holds. Then equality holds in \eq{sm:ineq_proj} for every pair $i<j$. This requires $e_i$ to be spanned by $\{\bb{I}e_j,\bb{J}e_j,\bb{K}e_j\}$. Setting $j=1$ and leaving $i$ arbitrary, it follows that all vectors in $E$ are generated by the basis $\{e_1,\bb{I}e_1,\bb{J}e_1,\bb{K}e_1\}$. Thus, $E$ is stable under the action of $\{\bb{I,J,K}\}$. Conversely, if $E$ is stable under the quaternionic structure, $\{e_1,\bb{I}e_1,\bb{J}e_1,\bb{K}e_1\}$ is an orthonormal basis of $E$. It follows that equality holds in \eq{sm:ineq_proj} and hence in \eq{sm:lem2_conclusion}.
\paragraph*{(end of proof)}
\mbox{}\newline

\begin{theorem}[Quaternionic metric--curvature inequality]
  \label{prop:q-wirtinger}
    At every point $\bk$ of the four-dimensional Brillouin zone,
\begin{gather}
    \sqrt{\det(g_{\mathrm{FS}})} \geq \frac{1}{24} \left|\mathscr{F}^I\wedge\mathscr{F}^I+\mathscr{F}^J\wedge\mathscr{F}^J+\mathscr{F}^K\wedge\mathscr{F}^K\right|.
  \label{sm:qkWirtinger}
\end{gather}
    At any point where the quantum metric is nondegenerate, the following conditions are equivalent:
\begin{itemize}
    \item[(a)] Equality holds in the inequality above.
    \item[(b)] The image of $T_{\bk}\bz{4}$ under the projector map is preserved by the local complex structures of $\hp^n$: $\bb{A}(\dd\mathscr{P}(T_{\bk}\bz{4}))=\dd\mathscr{P}(T_{\bk}\bz{4})$ for $\bb{A}=\bb{I},\bb{J},\bb{K}$.
    \item[(c)] The pullback operators $\{A_{\bb{I}},A_{\bb{J}},A_{\bb{K}}\}$ defined by Eq.~\eqref{T4_trinity} satisfy $A_{\bb{I}}^2=A_{\bb{J}}^2=A_{\bb{K}}^2=A_{\bb{I}}A_{\bb{J}}A_{\bb{K}}=-1$ at $\bk$.
    \item[(d)] The QQGT has three independent quaternionic null vectors: $\dim_{\bb{H}}\ker\mathscr{Q}(\bk)=3$.
\end{itemize}
\end{theorem}
\eq{sm:qkWirtinger} is the quaternionic analogue of the Wirtinger inequality, or equivalently the calibration inequality for the quaternionic 4-form~\cite{HarveyLawson1982,Salamon1982}.
\paragraph*{Proof of Theorem~\ref{prop:q-wirtinger}.}
    First, we prove the inequality. For arbitrary tuple of tangent vectors $\{V_1,V_2,V_3,V_4\}$ at $\bk\in\bz{4}$, we want to show
\begin{gather}
    \sqrt{\det [g(V_a,V_b)]_{a,b=1}^{4}} \geq \frac{1}{6} \big| \Omega(V_1,V_2,V_3,V_4) \big|
  \label{sm:main_ineq}
\\
    \text{with}\quad \Omega:= \frac{1}{4} (\mathscr{F}^I\wedge \mathscr{F}^I + \mathscr{F}^J\wedge \mathscr{F}^J + \mathscr{F}^K\wedge \mathscr{F}^K).
\end{gather}
    Since the tensors $(g,\Omega)$ in \eq{sm:main_ineq} are the pullbacks of the tensors $(g_{\mathrm{FS}},\Omega_{\mathrm{FS}})$ on $\hp^n$ along the quaternionic projector map $\mathscr{P}(\bk)$, the desired inequality follows by noticing that $\mathcal{M}=\hp^n$ is quaternion-K\"ahler by Lemma~\ref{lemma:hp} and applying Lemma~\ref{SM:lemma_qk} to the vectors
\begin{gather}
    X_i = \dd\mathscr{P}(V_i).
\end{gather}
    It remains to establish the equivalence of conditions (a-d).
    
    $(a\Leftrightarrow b)$ This follows from the equivalence statement of Lemma~\ref{SM:lemma_qk}.

    $(b\Rightarrow c)$
    Assume that $(b)$ is true at $\bk$. Since $g$ is non-degenerate (which we assume for generic $\bk$), the map $\dd\mathscr{P}$ is nonzero and hence $E:=\dd\mathscr{P}(T_{\bk}\bz{4})\neq\{0\}$. Choose a nonzero vector $X_0\in E$. Since $E$ is closed under $\bb{I,J,K}$, and since the vectors $\{X_0,\bb{I}X_0,\bb{J}X_0, \bb{K}X_0\}$ are orthogonal, $\dim E\geq 4$. However, $\dim E$ cannot exceed the dimension of $T_{\bk}\bz{4}$, which is 4. Therefore, $\dim E = 4$ and the map
\begin{equation}
    f:
\begin{array}{ccc}
    T_{\bk}\bz{4} &\to& E \\
    V &\mapsto& \dd\mathscr{P}(V)
\end{array}
\end{equation}
    is an isomorphism.
    Using this map, one can define the linear operators on $T_{\bk}\bz{4}$
\begin{gather}
    \widetilde{\bb{I}}:= f^{-1}\circ\bb{I}\circ f, \quad \widetilde{\bb{J}}:= f^{-1}\circ\bb{J}\circ f, \quad \widetilde{\bb{K}}:= f^{-1}\circ\bb{K}\circ f
  \label{sm:pullback_quat}
\end{gather}
    which obviously satisfy
\begin{gather}
    \mathscr{P}^{*} \omega_{\mathrm{FS}}^{I}(v_1,v_2) = \mathscr{P}^{*}g_{\mathrm{FS}}(v_1,\widetilde{\bb{I}}v_2)
\end{gather}
    and similar relations for $\widetilde{\bb{J}}, \widetilde{\bb{K}}$. This is equivalent to \eq{T4_trinity} with $A_{\bb{I}}$ replaced by $\widetilde{\bb{I}}$. Since \eq{T4_trinity} uniquely determines the operator $A_{\bb{I}}$ whenever the metric $g$ is non-degenerate, we have
\begin{gather}
    A_{\bb{I}} = \widetilde{\bb{I}}, \quad 
    A_{\bb{J}} = \widetilde{\bb{J}}, \quad 
    A_{\bb{K}} = \widetilde{\bb{K}}.
\end{gather}
    However, it is obvious from \eq{sm:pullback_quat} that
\begin{gather}
    \widetilde{\bb{I}}^2= \widetilde{\bb{J}}^2= \widetilde{\bb{K}}^2= \widetilde{\bb{I}}\;\widetilde{\bb{J}}\,\widetilde{\bb{K}}=-1.
\end{gather}
    
    $(d\Rightarrow b)$
    Choose a basis $e_a:=\partial/\partial k_a$ of $T_{\bk}\bz{4}$, and set
\begin{gather}
    X_a:= \dd\mathscr{P}(e_a) \in E.
\end{gather}
    While proving Lemma~\ref{lemma:hp}, we represented tangent vectors of $\hp^n$ by row vectors $\mathbf{q}\in\bb{H}^n$. Furthermore, from \eq{sm:qk_strs} we find that
\begin{gather}
    g_{\mathrm{FS}}(X,Y) + \omega_{\mathrm{FS}}^I(X,Y)I + \omega_{\mathrm{FS}}^J(X,Y)J + \omega_{\mathrm{FS}}^K(X,Y)K
\nonumber\\
    = \frac{1}{2}\Tr[\mathscr{P}(XY+YX)] + \frac{1}{2}\Tr[\mathscr{P}(XY-YX)]
    = \Tr[\mathscr{P}XY]
    = \mathbf{qp}^{\dagger}
  \label{sm:Q_in_HP}
\end{gather}
    where $\mathbf{q,p}$ are the directional vectors of $X,Y$. 
To see \eq{sm:Q_in_HP}, it suffices to show that
\begin{gather}
    \omega_{\mathrm{FS}}^I (X,Y)I + \omega_{\mathrm{FS}}^J (X,Y)J + \omega_{\mathrm{FS}}^K (X,Y)K = \frac{1}{2}\Tr[\mathscr{P}(XY-YX)].
\end{gather}
First, we observe that the quantity $W:=\frac{1}{2}\Tr[\mathscr{P}(XY-YX)]=\frac{1}{2}(\mathbf{qp}^{\dagger}-\mathbf{pq}^{\dagger})$ is purely imaginary:
\begin{gather}
    \overline{W} = \frac{1}{2}(\mathbf{pq}^{\dagger}-\mathbf{qp}^{\dagger})=-W.
\end{gather}
Therefore, $W=W_II+W_JJ+W_KK$ for some real numbers $W_I,W_J,W_K\in\bb{R}$. Then
\begin{subequations}
\begin{gather}
    \omega_{\mathrm{FS}}^I (X,Y) = \real[-IW] = W_I,
\\
    \omega_{\mathrm{FS}}^J (X,Y) = \real[-JW] = W_J,
\\
    \omega_{\mathrm{FS}}^K (X,Y) = \real[-KW] = W_K
\\ 
    \Rightarrow\quad \omega_{\mathrm{FS}}^I (X,Y)I + \omega_{\mathrm{FS}}^J (X,Y)J + \omega_{\mathrm{FS}}^K (X,Y)K = W_II+W_JJ+W_KK = W.
\end{gather}
\end{subequations}
    Since the QQGT is the pullback of the form in \eq{sm:Q_in_HP},
\begin{gather}
    \mathscr{Q}_{ab} = \mathscr{Q}(e_a,e_b) = \mathbf{q}_a\mathbf{q}_b^{\dagger}
\end{gather}
    where $\mathbf{q}_a$ is the directional vector of $\dd\mathscr{P}(e_a)=X_a$. Defining the 4-by-$n$ quaternion matrix $M$ by putting $\mathbf{q}_a$ in the $a$-th row of $M$,
\begin{gather}
    \mathscr{Q} = MM^{\dagger}.
\end{gather}
    Therefore
\begin{gather}
    \ker_{\bb{H}} \mathscr{Q} = \ker_{\bb{H}} M^{\dagger},
\end{gather}
    and the rank-nullity theorem gives
\begin{gather}
    \dim \ker_{\bb{H}}\mathscr{Q} = 4 - \op{rank}_{\bb{H}}M = 4 - \dim_{\bb{H}} \langle X_0,X_1,X_2,X_3\rangle.
\end{gather}
    If $\dim \ker_{\bb{H}}\mathscr{Q}=3$, then the quaternionic span of $\{X_a\}$ has quaternionic dimension 1. But the real span of the same vectors is the whole $E$, which has real dimension 4. Hence $E$ is exactly the quaternionic line
\begin{gather}
    E = \bb{H}\cdot X = \langle X,\bb{I}X,\bb{J}X,\bb{K}X\rangle
\end{gather}
    generated by a single nonzero vector $X$. It follows that $\bb{I}(E)=\bb{J}(E)=\bb{K}(E)=E$, which is condition (b).
    
    $(c\Rightarrow d)$ Let $\mathbf{q} = (q^a)\in\bb{H}^4$ be a fixed quaternion vector with four components. We shall investigate the condition for $\mathbf{q}$ to be a null vector of $\mathscr{Q}$, i.e.,
\begin{gather}
    \mathscr{Q}_{ab}q^b = 0.
  \label{methods:null}
\end{gather}
    One can explicitly write the real and imaginary parts of each quaternion number $q^a$ as 
\begin{gather}
    q^a = q_0^a + q_I^a I + q_J^a J + q_K^a K,
  \label{methods:q}
\end{gather}
whereby one obtains four distinct real vectors 
\begin{gather}
    \mathbf{q}_0=(q_0^a),\quad \mathbf{q}_I=(q_I^a),\quad \mathbf{q}_J=(q_J^a),\quad \mathbf{q}_K=(q_K^a),\quad 
    \text{all in } \bb{R}^4.
\end{gather}
Now, \eq{methods:null} is equivalent to
\begin{gather}
    \big\{ \delta_{ab}1 + (A_{\bb{I}})_{ab}I + (A_{\bb{J}})_{ab}J + (A_{\bb{K}})_{ab}K \big\}q^b = 0, \quad\text{where}
\\
    A_{\bb{I}} = (1/2) g^{-1}\mathscr{F}^I,\quad A_{\bb{J}} = (1/2) g^{-1}\mathscr{F}^J,\quad A_{\bb{K}} = (1/2) g^{-1}\mathscr{F}^K.
  \label{null_components}
\end{gather}
In \eq{null_components}, we are interpreting $g,\mathscr{F}^I,\mathscr{F}^J,\mathscr{F}^K$ as matrices which act on the coordinate vectors with respect to the basis
\begin{gather}
    \hat{e}_0 = \frac{\partial}{\partial k_t}, \quad
    \hat{e}_1 = \frac{\partial}{\partial k_x}, \quad
    \hat{e}_2 = \frac{\partial}{\partial k_y}, \quad
    \hat{e}_3 = \frac{\partial}{\partial k_z}
\end{gather}
of the tangent space of the Brillouin zone. The matrices $A_{\bb{I}}, A_{\bb{J}}, A_{\bb{K}}$ are precisely the matrix representation of ``pullback'' complex structures defined in \eq{T4_trinity}.

Expanding $q^b$ by \eq{methods:q}, one obtains
\begin{align}
    0 &= \big\{ \delta_{ab}1 + (A_{\bb{I}})_{ab}I + (A_{\bb{J}})_{ab}J + (A_{\bb{K}})_{ab}K \big\} (q_0^b + q_I^bI + q_J^bJ + q_K^bK)
\nonumber\\
    &= \big\{\delta_{ab}q_0^b - (A_{\bb{I}})_{ab} q_I^b - (A_{\bb{J}})_{ab} q_J^b - (A_{\bb{K}})_{ab} q_K^b \big\}\,1 
\nonumber\\
    &\quad + 
    \big\{ (A_{\bb{I}})_{ab}q_0^b + \delta_{ab}q_I^b - (A_{\bb{K}})_{ab}q_J^b + (A_{\bb{J}})_{ab}q_K^b \big\}\,I
\nonumber\\
    &\quad + \big\{ (A_{\bb{J}})_{ab}q_0^b + (A_{\bb{K}})_{ab}q_I^b + \delta_{ab}q_J^b - (A_{\bb{I}})_{ab}q_K^b \big\}\,J 
\nonumber\\
    &\quad + 
    \big\{ (A_{\bb{K}})_{ab}q_0^b - (A_{\bb{J}})_{ab}q_I^b + (A_{\bb{I}})_{ab}q_J^b + \delta_{ab}q_K^b \big\}\,K.
  \label{methods:big}
\end{align}
One can reorganize \eq{methods:big} as a 16-by-16 matrix equation
\begin{gather}
    \begin{pmatrix}
        \delta & -A_{\bb{I}} & -A_{\bb{J}} & -A_{\bb{K}} \\
        A_{\bb{I}} & \delta & -A_{\bb{K}} & A_{\bb{J}} \\
        A_{\bb{J}} & A_{\bb{K}} & \delta & -A_{\bb{I}} \\
        A_{\bb{K}} & -A_{\bb{J}} & A_{\bb{I}} & \delta
    \end{pmatrix}
    \begin{pmatrix}
        \mathbf{q}_0 \\ \mathbf{q}_I \\ \mathbf{q}_J \\ \mathbf{q}_K
    \end{pmatrix} = 0
  \label{methods:big_mat}
\end{gather}
where $\delta$ stands for the 4-by-4 identity matrix.
\eq{methods:big_mat} can be solved by multiplying an invertible matrix on the left:
\begin{gather}
    \begin{pmatrix}
        \delta & 0 & 0 & 0 \\
        -A_{\bb{I}} & \delta & 0 & 0 \\
        -A_{\bb{J}} & 0 & \delta & 0 \\
        -A_{\bb{K}} & 0 & 0 & \delta
    \end{pmatrix}
    \begin{pmatrix}
        \delta & -A_{\bb{I}} & -A_{\bb{J}} & -A_{\bb{K}} \\
        A_{\bb{I}} & \delta & -A_{\bb{K}} & A_{\bb{J}} \\
        A_{\bb{J}} & A_{\bb{K}} & \delta & -A_{\bb{I}} \\
        A_{\bb{K}} & -A_{\bb{J}} & A_{\bb{I}} & \delta
    \end{pmatrix}
    \begin{pmatrix}
        \mathbf{q}_0 \\ \mathbf{q}_I \\ \mathbf{q}_J \\ \mathbf{q}_K
    \end{pmatrix} = 
    \begin{pmatrix}
        \delta & -A_{\bb{I}} & -A_{\bb{J}} & -A_{\bb{K}} \\
        0 & 0 & 0 & 0 \\
        0 & 0 & 0 & 0 \\
        0 & 0 & 0 & 0 
    \end{pmatrix}
    \begin{pmatrix}
        \mathbf{q}_0 \\ \mathbf{q}_I \\ \mathbf{q}_J \\ \mathbf{q}_K
    \end{pmatrix} = 0.
\end{gather}
Thus, a basic set of solutions is given by
\begin{gather}
    \begin{pmatrix}
        \mathbf{q}_0 \\ \mathbf{q}_I \\ \mathbf{q}_J \\ \mathbf{q}_K
    \end{pmatrix} = 
    \begin{pmatrix}
        A_{\bb{I}}v_1\\ v_1\\ 0\\ 0
    \end{pmatrix}, \;\;
    \begin{pmatrix}
        A_{\bb{J}}v_2\\ 0\\ v_2\\ 0
    \end{pmatrix}, \;\;
    \begin{pmatrix}
        A_{\bb{K}}v_3\\ 0\\ 0\\ v_3
    \end{pmatrix}, \qquad 
    v_1,\, v_2,\, v_3\in\bb{R}^4.
  \label{methods:sol}
\end{gather}
The solution space to \eq{methods:null} is closed under the right multiplication of quaternions, i.e., $q^b\to q^bu$ for some fixed $u\in\bb{H}$. This transformation is equivalent to changing the vector $v_i\in\bb{R}^4$ in \eq{methods:sol} by an orthogonal transformation. The general solution to \eq{methods:sol} is
\begin{gather}
    (\mathbf{q}_0, \mathbf{q}_I, \mathbf{q}_J, \mathbf{q}_K)^{\top} = (A_{\bb{I}}v_1+A_{\bb{J}}v_2+A_{\bb{K}}v_3, v_1, v_2, v_3)^{\top} \in\bb{R}^{16}
  \label{sup:gen_sol}
\end{gather}
which has 12 free parameters. Since the transformation $q^b\to q^bu$ has 4 independent parameters, a single solution $(q^b)$ generates a four-dimensional equivalence class in the solution space. Therefore, the total number of equivalence classes of solutions is $12/4=3$.
\paragraph*{(end of proof)}
\mbox{}\newline
\mbox{}\newline
An immediate corollary is the quantum volume bound set by the second Chern number. It follows straightforwardly from integrating both sides of \eq{sm:qkWirtinger}, hence we omit the proof.

\begin{corollary}[Four-dimensional quantum-volume bound]
  \label{cor:qbound}
    Integrating the local inequality and using the second Chern number normalization gives
\begin{gather}
    \operatorname{Vol}_{\operatorname{4D}}(g)\geq\frac{2\pi^2}{3}|\mathcal{C}_2|,
\end{gather}
    with equality precisely when the induced structure on the four-dimensional Brillouin zone is quaternion-K\"ahler.
\end{corollary}
Next, we turn to the class AII analogue of vortexability~\cite{Mera2021kahler,vortexability,LiWu2013}.

\begin{proposition}[Vortexability of ideal quaternion-K\"ahler bands]
  \label{prop:closed}
    Under a suitable coordinate system on the four-dimensional Brillouin zone and on four-dimensional real space, the occupied subspace of an ideal quaternion-K\"ahler band is closed under right multiplication by the degree-1 polynomials $tI-x$, $tJ-y$, and $tK-z$. More precisely, let
\begin{gather}
    \Phi_{\bk}(t,x,y,z)=e^{i(k_tt+k_xx+k_yy+k_zz)}\Psi_{\bk}(t,x,y,z)
\end{gather}
    be an occupied Bloch wave function whose periodic part is $\Psi_{\bk}=\psi^1_{\bk}+\psi^2_{\bk}J$. Then
\begin{gather}
    \Phi_{\bk}(t,x,y,z)(tI-x),\qquad
    \Phi_{\bk}(t,x,y,z)(tJ-y),\qquad
    \Phi_{\bk}(t,x,y,z)(tK-z)
\end{gather}
    belong to the occupied subspace. Iterating this closure, any wave function of the form
\begin{gather}
    \Phi_{\bk}(t,x,y,z)\cdot P[tI-x,tJ-y,tK-z],
\end{gather}
    with $P[Z_1,Z_2,Z_3]$ a polynomial in three noncommuting variables, belongs to the occupied subspace.
\end{proposition}
\paragraph*{Proof of Proposition~\ref{prop:closed}.}
    The ideal quaternionic Bloch state is
\begin{gather}
    \Phi_{\bk}(\br) = e^{ik_{\nu} r^{\nu}}\Psi_{\bk}(\br).
  \label{sm:full_qbloch}
\end{gather} 
Fix a null vector $(q^a)$ of the QQGT. Denoting $\mathcal{D} = \sum_{a=0}^{3}\rho(q^a)\partial/\partial k_a$ with $\rho(q)$ being the right-multiplication operator $\rho(q)u=uq$,
\begin{gather}
    \sum_{\mu=0}^{3} \rho(q^\mu) \frac{\partial}{\partial k_\mu} e^{ik_\nu x^\nu} = \sum_{\mu=0}^{3} \rho(q^\mu)ix^\mu \quad\Longrightarrow\quad 
    \mathcal{D} \Phi_{\bk}(\br) = \sum_{\mu=0}^{3} \rho(q^\mu) ix^{\mu} \Phi_{\bk}(\br) + e^{ik_\nu r^\nu} \mathcal{D} \Psi_{\bk}(\br).
\end{gather}
Next, we define the total projection operator onto the target Kramers band by
\begin{gather}
    P_{\text{tot}} = \sum_{\bk} \ket{\Phi_{\bk}}\bra{\Phi_{\bk}}.
\end{gather}
Then $P_{\text{tot}}\ket{\mathcal{D}\Phi_{\bk}} = \ket{\mathcal{D}\Phi_{\bk}}$, due to the following reason. Notice that the derivative $\partial/\partial k_a$ is defined via infinitesimal difference, i.e.,
\begin{gather}
    \frac{\partial}{\partial k_a} \ket{\Phi_{\bk}} = \lim_{\epsilon\to 0} \frac{\ket{\Phi_{\bk+\epsilon \hat{e}_a}} - \ket{\Phi_{\bk}}}{\epsilon}
\end{gather}
where $\hat{e}_a$ is the unit vector in the $k_a$-direction.
Therefore, $\mathcal{D}\Phi_{\bk}$ is nothing but
\begin{gather}
    \ket{\mathcal{D} \Phi_{\bk}} = \lim_{\epsilon\to0} \frac{1}{\epsilon} \left[ \sum_{a=0}^{3} \Big(\ket{\Phi_{\bk+\epsilon \hat{e}_a}} -\ket{\Phi_{\bk}}\Big) q^a\right] \equiv \lim_{\epsilon\to0} \Delta(\epsilon).
\end{gather}
However, it is obvious that $P_{\text{tot}}\Delta(\epsilon)=\Delta(\epsilon)$ for all $\epsilon$. Taking the limit $\epsilon\to0$, we obtain $P_{\text{tot}}\ket{\mathcal{D}\Phi_{\bk}} = \ket{\mathcal{D}\Phi_{\bk}}$.

Thus, we write
\begin{align}
    0 &= (1-P_{\text{tot}}) \mathcal{D}\Phi_{\bk}(\br)
\nonumber\\
    &= (1-P_{\text{tot}}) (\sum_{\mu=0}^{3} \rho(q^\mu)ix^\mu) \Phi_{\bk}(\br) + e^{ik_\mu x^\mu} 
    \underbrace{ 
    (1-P_{\bk}) \mathcal{D} \Psi_{\bk}(\br)
    }_{=0\text{ by Proposition }\ref{prop:positive}}.
  \label{sm:proof_vortex}
\end{align}
Note that we can use either $P$ or $\mathscr{P}$ interchangeably in \eq{sm:proof_vortex}, invoking Lemma~\ref{SM:lemma_Proj} or \eq{sm:Pket}.
From \eq{sm:proof_vortex} it follows that the space of occupied Bloch states is closed under the right multiplication by a suitable ``vortex function'' $\mathscr{R}(\mathbf{q})$:
\begin{gather}
    (1-P_{\text{tot}}) \rho(\mathscr{R})\, \Phi_{\bk}(\br) = 0, \qquad 
    \mathscr{R}(\mathbf{q}) := \sum_{\mu=0}^{3} x^{\mu}q^{\mu}.
  \label{sm:vortex}
\end{gather}
    Now we apply \eq{sm:vortex} to specific null vectors $\mathbf{q}$. By a suitable coordinate transformation in 4D BZ, one can set
\begin{gather}
    A_{\bb{I}} = 
    \begin{pmatrix}
        0&1&0&0 \\
        -1&0&0&0 \\
        0&0&0&-1 \\
        0&0&1&0
    \end{pmatrix},
    \qquad 
    A_{\bb{J}} = 
    \begin{pmatrix}
        0&0&1&0 \\
        0&0&0&1 \\
        -1&0&0&0& \\
        0&-1&0&0
    \end{pmatrix},
    \qquad
    A_{\bb{K}} = 
    \begin{pmatrix}
        0&0&0&1 \\
        0&0&-1&0 \\
        0&1&0&0& \\
        -1&0&0&0
    \end{pmatrix}.
\end{gather}
    Choose $v_1=v_2=v_3=(1,0,0,0)^{\top}$ in \eq{methods:sol}. Then the corresponding null vectors $\mathbf{q}_1,\, \mathbf{q}_2,\, \mathbf{q}_3$ are
\begin{gather}
    \mathbf{q}_1 = (I,-1,0,0)^{\top}, \quad 
    \mathbf{q}_2 = (J,0,-1,0)^{\top}, \quad 
    \mathbf{q}_3 = (K,0,0,-1)^{\top},
\end{gather}
    which are independent over $\bb{H}$. 
For example, the first null vector $\mathbf{q}_1$ is produced by computing that
\begin{align}
    \mathbf{q}_1 &= (\mathbf{q}_1)_0 + (\mathbf{q}_1)_I\cdot I + (\mathbf{q}_1)_J\cdot J + (\mathbf{q}_1)_K\cdot K 
\nonumber\\
    &= (A_{\bb{I}}v_1) + (v_1)\cdot I
\nonumber\\
    &= (0,-1,0,0)^{\top} + (1,0,0,0)^{\top}\cdot I 
\nonumber\\
    &= (I,-1,0,0)^{\top}.
\end{align}
    The corresponding vortex functions are
\begin{gather}
    \mathscr{R}(\mathbf{q}_1) = tI-x, \quad 
    \mathscr{R}(\mathbf{q}_2) = tJ-y, \quad 
    \mathscr{R}(\mathbf{q}_3) = tK-z.
\end{gather}
\paragraph*{(end of proof)}
\mbox{}\newline
\mbox{}\newline
The following proposition shows that in the minimal 4-band systems, the metric--curvature inequality is automatically saturated~\cite{Murakami2004su2,zhang2022relating}.

\begin{proposition}[Four-band saturation]
  \label{prop:4band}
    In a $\mathcal{PT}$-symmetric four-band system with $(\mathcal{PT})^2=-1$,
\begin{gather}
    \sqrt{\det(g)}\, \dd^4k\equiv\pm\frac{1}{24}\left(\mathscr{F}^I\wedge\mathscr{F}^I+\mathscr{F}^J\wedge\mathscr{F}^J+\mathscr{F}^K\wedge\mathscr{F}^K\right).
  \label{sup:4band_sat}
\end{gather}
    If $\det g(\bk)\neq0$ for all $\bk$, the sign does not depend on $\bk$.
\end{proposition}
\paragraph*{Proof of Proposition~\ref{prop:4band}.}
    We can assume that the Hamiltonian has been flattened, i.e.
\begin{gather}
    \hat{H}(\bk) = \sum_{a=1}^{5} n_a(\bk)\Gamma^a = 1-2P(\bk), \qquad 
    P(\bk) = \frac{1}{2}\big\{1-n_a(\bk)\Gamma^a\big\}
\end{gather}
    for some unit vector $\mathbf{n}=(n_a)$. We may leave the gamma matrices arbitrary, but for computations it is convenient to set
\begin{gather}
    \Gamma^1 = \sigma_x\tau_0, \quad 
    \Gamma^2 = \sigma_y\tau_0, \quad 
    \Gamma^3 = \sigma_z\tau_x, \quad 
    \Gamma^4 = \sigma_z\tau_y, \quad 
    \Gamma^5 = \sigma_z\tau_z.
\end{gather}
    To express both sides of \eq{sup:4band_sat} in terms of $\mathbf{n}$, we begin with the quantum metric. By Lemma~\ref{SM:lemma_PdPdP},
\begin{gather}
    g\,\ket{\psi^i}\,(\sigma_0)^{ij}\,\bra{\psi^j} = P\dd P\odot\dd P = \frac{1}{8}\big(1-n_{\lambda}\Gamma^{\lambda}\big)\Gamma^{\mu}\Gamma^{\nu} \dd n_{\mu}\odot\dd n_{\nu}
\nonumber\\
    \underset{\text{trace}}{\Longrightarrow}\quad 2g = \op{tr}[g\sigma_0] = \frac{1}{8}\op{tr}[\Gamma^{\mu}\Gamma^{\nu}]\dd n_{\mu}\odot\dd n_{\nu}.
\end{gather}
    Since $\op{tr}[\Gamma^{\mu}\Gamma^{\nu}] = 4\delta^{\mu\nu}$ and the trace of the product of three gamma matrices vanishes, the metric is given by
\begin{gather}
    g = \frac{1}{4}\dd n^{\mu}\odot\dd n_{\mu} = \frac{1}{4} \sum_{a=1}^{5} (\dd n_a)^2 = \frac{1}{4} (f^{*}g_{1}),
  \label{sm:vol_prop}
\end{gather}
    where $g_1$ is the standard metric of the unit sphere and $f^{*}g_1$ is its pullback along the map
\begin{equation}
    f:
\begin{array}{rcl}
    \bz{4} &\to& \bb{S}^4 \\
    \bk &\mapsto& \mathbf{n}(\bk).
\end{array}
\end{equation}
    Introducing the totally anti-symmetric symbol on five indices $\epsilon$, the volume form of the unit sphere $(\bb{S}^4,g_1)$ is
\begin{gather}
    \dd\op{Vol}(\bb{S}^4,g_1) = \frac{1}{24}\epsilon^{\lambda\mu\nu\sigma\tau}n_{\lambda}\dd n_{\mu}\wedge \dd n_{\nu} \wedge \dd n_{\sigma} \wedge \dd n_{\tau}.
\end{gather}
    Therefore, the proportionality relation in \eq{sm:vol_prop} implies the quantum volume form
\begin{align}
    \dd\op{Vol}(\bz{4},g) &= \frac{1}{16} f^{*}\dd\op{Vol}(\bb{S}^4,g_1)
\nonumber\\
    &= \frac{1}{16\cdot24}\epsilon^{\lambda\mu\nu\sigma\tau} f^{*} \big[n_{\lambda}\dd n_{\mu}\wedge \dd n_{\nu} \wedge \dd n_{\sigma} \wedge \dd n_{\tau}\big]
\nonumber\\
    &= \frac{1}{16\cdot24}\epsilon^{\lambda\mu\nu\sigma\tau}n_{\lambda}\frac{\partial n_{\mu}}{\partial k_a} \frac{\partial n_{\nu}}{\partial k_b} \frac{\partial n_{\sigma}}{\partial k_c} \frac{\partial n_{\tau}}{\partial k_d}\,\dd k_a\wedge \dd k_b\wedge \dd k_c\wedge\dd k_d
\nonumber\\ 
    &= \frac{1}{16\cdot24}\epsilon^{\lambda\mu\nu\sigma\tau} \varepsilon^{abcd} n_{\lambda}\frac{\partial n_{\mu}}{\partial k_a} \frac{\partial n_{\nu}}{\partial k_b} \frac{\partial n_{\sigma}}{\partial k_c} \frac{\partial n_{\tau}}{\partial k_d}\,\dd^4k
  \label{sm:dvol}
\end{align}
    where we introduced another totally anti-symmetric symbol $\varepsilon$, this time on four momentum indices.
    Since \eq{sm:dvol} is a signed volume which can take both positive and negative values, it is equal to $\pm\sqrt{\det g}\,\dd^4k$. Thus,
\begin{gather}
    \sqrt{\det g} = \frac{\pm1}{16\cdot24}\epsilon^{\lambda\mu\nu\sigma\tau} \varepsilon^{abcd} n_{\lambda}\frac{\partial n_{\mu}}{\partial k_a} \frac{\partial n_{\nu}}{\partial k_b} \frac{\partial n_{\sigma}}{\partial k_c} \frac{\partial n_{\tau}}{\partial k_d}.
  \label{sm:pmvol}
\end{gather}
    We turn to the right-hand side of \eq{sup:4band_sat}. Let us define the matrix-valued 2-form (cf. Lemma~\ref{SM:lemma_PdPdP})
\begin{gather}
    T := \ket{\psi^i}\, \big(F^x\,\sigma_x + F^y\,\sigma_y + F^z\,\sigma_z\big)^{ij}\,\bra{\psi^j} = P\dd P\wedge\dd P.
\end{gather}
    Taking the wedge product with itself (we omit the pullback symbol $f^{*}$),
\begin{align}
    T\wedge T &= \sum_{i=1,2} \ket{\psi^i}\, \big(F^x\wedge F^x + F^y\wedge F^y + F^z\wedge F^z\big)\,\bra{\psi^i} 
\nonumber\\
    &= (P\dd P\wedge\dd P) \wedge (P\dd P\wedge\dd P)
\nonumber\\
    &= P\dd P\wedge\dd P\wedge\dd P\wedge\dd P 
\nonumber\\
    &= \frac{1}{32}\big(1-n_{\lambda}\Gamma^{\lambda}\big)\Gamma^{\mu}\Gamma^{\nu}\Gamma^{\sigma}\Gamma^{\tau} \dd n_{\mu}\wedge\dd n_{\nu}\wedge\dd n_{\sigma}\wedge\dd n_{\tau}
\nonumber\\
    &= \frac{1}{128}\big(1-n_{\lambda}\Gamma^{\lambda}\big)[\Gamma^{\mu},\Gamma^{\nu}][\Gamma^{\sigma},\Gamma^{\tau}] \dd n_{\mu}\wedge\dd n_{\nu}\wedge\dd n_{\sigma}\wedge\dd n_{\tau}.
\end{align}
    Taking the trace and using the formulae
\begin{align}
\op{tr}\big([\Gamma^{\mu},\Gamma^{\nu}]\,[\Gamma^{\sigma},\Gamma^{\tau}]\big)
&=-16\Big(\delta^{\mu\tau}\delta^{\nu\sigma}-\delta^{\mu\sigma}\delta^{\nu\tau}\Big),\\
\op{tr}\big(\Gamma^{a}\,[\Gamma^{\mu},\Gamma^{\nu}]\,[\Gamma^{\sigma},\Gamma^{\tau}]\big)
&=-16\,\varepsilon^{a\mu\nu\sigma\tau},
\end{align}
    we obtain
\begin{align}
    2\big(\mathscr{F}^I\wedge \mathscr{F}^I + \mathscr{F}^J\wedge \mathscr{F}^J + \mathscr{F}^K\wedge \mathscr{F}^K\big)
    &= -2\big(F^x\wedge F^x + F^y\wedge F^y + F^z\wedge F^z\big)
\nonumber\\
    &= \frac{-1}{8} \epsilon^{\lambda\mu\nu\sigma\tau}n_{\lambda}\dd n_{\mu}\wedge \dd n_{\nu} \wedge \dd n_{\sigma} \wedge \dd n_{\tau}
\\
    \Longrightarrow\quad \mathscr{F}^I\wedge \mathscr{F}^I + \mathscr{F}^J\wedge \mathscr{F}^J + \mathscr{F}^K\wedge \mathscr{F}^K &= \frac{-1}{16} \epsilon^{\lambda\mu\nu\sigma\tau}n_{\lambda}\dd n_{\mu}\wedge \dd n_{\nu} \wedge \dd n_{\sigma} \wedge \dd n_{\tau}
  \label{sm:canonical}
\end{align}
    Combining \eq{sm:pmvol} with \eq{sm:canonical} completes the proof.
\paragraph*{(end of proof)}
\mbox{}\newline
\mbox{}\newline
There exists a converse to Proposition~\ref{prop:4band}, which states that saturation of the inequality over the whole Brillouin zone implies a form of equivalence to a 4-band system:

\begin{proposition}[Rigidity of locally saturated four-dimensional images]
  \label{prop:fourband_unique}
Let $\mathscr{P}:U\to\hp^n$ be the quaternionic projector map of a Kramers pair on a connected open set $U\subset\bz{4}$. Suppose that the quantum metric is non-degenerate on $U$ and that equality holds in the four-dimensional metric--curvature inequality at every point of $U$. Then $\mathscr{P}(U)$ is contained in a totally geodesic quaternionic projective line
\begin{gather}
    \hp^1\subset\hp^n .
\end{gather}
Equivalently, there exists a $\bk$-independent quaternion-unitary transformation $U_0\in\mathrm{Sp}(n+1)$ such that
\begin{gather}
    U_0\Psi(\bk)
    =
    \begin{pmatrix}
        u_1(\bk)\\
        u_2(\bk)\\
        0\\
        \vdots\\
        0
    \end{pmatrix}
\end{gather}
for quaternion-valued functions $u_1,u_2$. Hence the spectrally flattened Hamiltonian is, up to spectator bands, a four-band Hamiltonian.

Moreover, any four-band Hamiltonian with a $\bk$-preserving anti-unitary symmetry $\Theta^2=-1$ can be written as
\begin{gather}
    h_{\mathrm{4b}}(\bk)
    =
    d_0(\bk)\mathbf{1}
    +
    \sum_{a=1}^{5}d_a(\bk)\Gamma^a,
\end{gather}
where the five matrices $\Gamma^a$ satisfy $\{\Gamma^a,\Gamma^b\}=2\delta^{ab}$.
\end{proposition}
\paragraph*{Proof of Proposition~\ref{prop:fourband_unique}.}
By Theorem~\ref{prop:q-wirtinger}, equality in the four-dimensional metric--curvature inequality, together with non-degeneracy of the quantum metric, implies that the tangent space
\begin{gather}
    \dd\mathscr{P}(T_{\bk}U)
    \subset
    T_{\mathscr{P}(\bk)}\hp^n
\end{gather}
is preserved by the local quaternionic structure of $\hp^n$. Thus $\mathscr{P}(U)$ is a quaternionic submanifold of $\hp^n$ of real dimension four.

We recall the standard rigidity argument in this special setting. Let $B$ be the second fundamental form of this submanifold, and let $\bb{I},\bb{J},\bb{K}$ denote the local quaternionic structures of $\hp^n$. Since the tangent bundle of the submanifold is preserved by $\bb{I},\bb{J},\bb{K}$, the normal bundle is also preserved by them. Taking the normal component of the covariant derivative of $\bb{A}Y$, with $\bb{A}=\bb{I},\bb{J},\bb{K}$, gives
\begin{gather}
    B(X,\bb{A}Y) = \bb{A}B(X,Y)
\end{gather}
for tangent vectors $X,Y$. By symmetry of the second fundamental form, this also implies
\begin{gather}
    B(\bb{A}X,Y) = \bb{A}B(X,Y).
\end{gather}
Using two different quaternionic structures, we find
\begin{align}
    B(\bb{I}X,\bb{J}Y)
    &=
    \bb{I}B(X,\bb{J}Y)
    =
    \bb{I}\bb{J}B(X,Y)
    =
    \bb{K}B(X,Y),
\\
    B(\bb{I}X,\bb{J}Y)
    &=
    B(\bb{J}Y,\bb{I}X)
    =
    \bb{J}B(Y,\bb{I}X)
    =
    \bb{J}\bb{I}B(Y,X)
    =
    -\bb{K}B(X,Y).
\end{align}
Therefore $B(X,Y)=0$ for all $X,Y$, so the submanifold is totally geodesic.

The totally geodesic quaternionic submanifolds of $\hp^n$ are quaternionic projective spaces associated with quaternionic linear subspaces of $\bb{H}^{n+1}$. Since the submanifold above has real dimension four, it is a quaternionic projective line. Hence there exist fixed orthonormal quaternionic vectors $\mathbf{q}_1,\mathbf{q}_2\in\bb{H}^{n+1}$ such that
\begin{gather}
    \Psi(\bk)
    =
    \mathbf{q}_1u_1(\bk)
    +
    \mathbf{q}_2u_2(\bk).
\end{gather}
Choosing $U_0\in\mathrm{Sp}(n+1)$ with $U_0\mathbf{q}_1=e_1$ and $U_0\mathbf{q}_2=e_2$ gives
\begin{gather}
    U_0\Psi(\bk)
    =
    \begin{pmatrix}
        u_1(\bk)\\
        u_2(\bk)\\
        0\\
        \vdots\\
        0
    \end{pmatrix}.
\end{gather}
Thus the projector is supported on a fixed copy of $\bb{H}^2\subset\bb{H}^{n+1}$, and the flattened Hamiltonian $1-2\mathscr{P}$ is a four-band block plus spectator bands.

It remains to identify the general form of the four-band block. To this end, note that any 4-band Hamiltonian is a linear combination of gamma matrices
\begin{gather}
    \mathbf{1}, \quad \Gamma^a\;(1\leq a\leq 5), \quad \Gamma^{ab}:= i\Gamma^a\Gamma^b \;(1\leq a<b\leq 5),
  \label{16_gammas}
\end{gather}
where $\{\Gamma^a,\Gamma^b\}=2\delta_{ab}$. Without loss of generality, one can choose
\begin{gather}
    \Gamma^1 = \tau_0\sigma_x,\quad \Gamma^2 = \tau_0\sigma_z,\quad \Gamma^3 = \tau_x\sigma_y,\quad 
    \Gamma^4 = \tau_y\sigma_y,\quad 
    \Gamma^5 = \tau_z\sigma_y.
\end{gather}
Now, we represent the operator $\Theta$ by
\begin{gather}
    \Theta = i\tau_y\sigma_0\, \mathcal{K}
\end{gather}
where $\mathcal{K}$ is the complex conjugation. Among the general gamma matrices in \eq{16_gammas}, only the five matrices $\Gamma^a$ and the identity $\mathbf{1}$  satisfy the symmetry condition
\begin{gather}
    \Theta \Gamma^a \Theta^{-1} = \Gamma^a.
\end{gather}
Thus, the $\bk$-local symmetry of the Hamiltonian, $\Theta H(\bk)\Theta^{-1} = H(\bk)$, forces that $H(\bk)$ is the Dirac Hamiltonian
\begin{gather}
    H(\bk) = d_0(\bk)\mathbf{1} + \sum_{a=1}^{5} d_{a}(\bk)\Gamma^{a}.
\end{gather}
\paragraph*{(end of proof)}
\mbox{}\newline
\mbox{}\newline
The four-dimensional bound discussed so far uses the canonical four-form on the Brillouin zone torus. Lower-dimensional time reversal symmetric band structures do not carry four-forms, but the same quaternionic quantum geometry still gives useful local constraints. The reason is that the QQGT is positive semidefinite on every tangent subspace, and hence on every two- or three-dimensional slice. 
After deriving these local inequalities, we explain how these functions connect, in special Hamiltonian classes, to two- and three-dimensional $\bb{Z}_2$ invariants.
\begin{proposition}[two- and three-dimensional metric--curvature bound]
  \label{prop:lowd_bound}
  Let $g_{d\mathrm{D}}\,(d=2,3)$ be the quantum metric of a four-dimensional $\mathcal{PT}$-symmetric band system $(\mathcal{PT}^2=-1)$ restricted to a $d$-dimensional subspace of the Brillouin zone. If $d=2$, let $(\mathscr{F}^{I}_{\mathrm {2D}}, \mathscr{F}^{J}_{\mathrm {2D}}, \mathscr{F}^{K}_{\mathrm {2D}})$ be the $\mathrm{SU}(2)$ Berry curvature restricted to the subspace. Then
\begin{gather}
    \det g_{\mathrm{2D}}(\bk) \geq |\Omega|^2_{\mathrm{2D}}(\bk):= \frac{1}{4} \left[(\mathscr{F}_{\mathrm {2D}}^{I})^2+(\mathscr{F}_{\mathrm {2D}}^{J})^2+(\mathscr{F}_{\mathrm {2D}}^{K})^2\right].
  \label{qWirtinger_2D}
\end{gather}
  If $d=3$, let $\mathscr{F}_{ab}=(\mathscr{F}_{ab}^I, \mathscr{F}_{ab}^J, \mathscr{F}_{ab}^K)\,((a,b)=(2,3),(3,1),(1,2))$ be the $\mathrm{SU}(2)$ Berry curvature along three independent directions of the subspace. Then \eq{qWirtinger_2D} holds for every two-dimensional slice of the submanifold. This guarantees that the function 
\begin{align}
    |\Omega|^2_{\mathrm{3D}}(\bk)&:= \frac{1}{4}\big[ g_{11}|\mathscr{F}_{23}|^2 + g_{22}|\mathscr{F}_{31}|^2 + g_{33}|\mathscr{F}_{12}|^2
    - (\mathscr{F}_{23}\cdot(\mathscr{F}_{31}\times\mathscr{F}_{12}))  
\nonumber\\
    &\quad + g_{12}(\mathscr{F}_{23}\cdot \mathscr{F}_{31}) + g_{23}(\mathscr{F}_{31}\cdot \mathscr{F}_{12}) + g_{31}(\mathscr{F}_{12}\cdot \mathscr{F}_{23}) \big]
  \label{sup:canfun_3d}
\end{align}
    is non-negative. Furthermore,
\begin{gather}
    \det g_{\mathrm{3D}}(\bk) \geq |\Omega|^2_{\mathrm{3D}}(\bk).
  \label{qWirtinger_3D}
\end{gather}
    Therefore, we have two- and three-dimensional descendants of the time reversal symmetric quantum geometric bound: $\sqrt{\det g_{d\mathrm{D}}}\geq |\Omega|_{d\mathrm{D}}\,(d=2,3)$.
\end{proposition}
\paragraph*{Proof of Proposition~\ref{prop:lowd_bound}.}
The proof uses only the non-negativity of the restricted QQGT. Since the full $\mathscr{Q}$ is positive semidefinite, its restriction to any two-dimensional tangent plane is also positive semidefinite:
\begin{gather}
    \mathscr{Q}_{\mathrm{2D}} = 
\begin{pmatrix}
    g_{11} & g_{12}+\omega_{12}\\
    g_{12}-\omega_{12} & g_{22}
\end{pmatrix}, \qquad
    \omega_{12} = \frac{1}{2} (\mathscr{F}_{12}^{I}I + \mathscr{F}_{12}^{J}J + \mathscr{F}_{12}^{K}K).
\end{gather}
For this $2\times2$ quaternionic Hermitian matrix, positive-semidefinite property implies non-negativity of the determinant,
\begin{align}
    & \det \mathscr{Q}_{\mathrm{2D}} = \det g_{\mathrm{2D}} + \omega_{\mathrm{2D}}^2 = \det g_{\mathrm{2D}} - \|\omega_{\mathrm{2D}}\|^2 \geq 0
\\
    \Rightarrow\quad &\det g_{\mathrm{2D}} \geq \|\omega_{12}\|^2 = \frac{1}{4}[(\mathscr{F}_{12}^{I})^2+(\mathscr{F}_{12}^{J})^2+(\mathscr{F}_{12}^{K})^2].
  \label{sup:pf_qWirtinger_2d}
\end{align}
This proves \eq{qWirtinger_2D}. Now we turn to the 3D case. First, we show that the right-hand side of \eq{sup:canfun_3d} is non-negative. By changing the coordinates, one may assume that the metric tensor $g_{\mathrm{3D}}$ is the identity matrix: $(g_{\mathrm{3D}})_{ab}=\delta_{ab}$. Then the function in question becomes
\begin{gather}
    |\mathscr{F}_{23}|^2 + |\mathscr{F}_{31}|^2 + |\mathscr{F}_{12}|^2
    - (\mathscr{F}_{23} \cdot(\mathscr{F}_{31} \times\mathscr{F}_{12})) . 
\end{gather} 
Using the simplified notation
\begin{gather}
    x=|\mathscr{F}_{23}|,\quad y=|\mathscr{F}_{31}|,\quad z=|\mathscr{F}_{12}|,
\end{gather}
it suffices to show that $x^2+y^2+z^2-xyz\geq 0$, because $(\mathscr{F}_{23} \cdot(\mathscr{F}_{31} \times\mathscr{F}_{12}))\leq |\mathscr{F}_{23}| |\mathscr{F}_{31}| |\mathscr{F}_{12}|.$ To this end, we use the non-negativity of the QQGT restricted to the 3D submanifold:
\begin{gather}
    \mathscr{Q}_{\mathrm{3D}} = 
    \begin{pmatrix}
        1 & \omega_{12} & \omega_{13} \\
        - \omega_{12} & 1 & \omega_{23} \\
        - \omega_{13} & - \omega_{23} & 1
    \end{pmatrix},\quad
    \omega_{ab}=\frac{1}{2} (\mathscr{F}_{ab}^{I}I + \mathscr{F}_{ab}^{J}J + \mathscr{F}_{ab}^{K}K).
\end{gather}
For $\mathscr{Q}_{\mathrm{3D}}$ to be positive-semidefinite, the 2-by-2 minors should be non-negative. This gives \eq{sup:pf_qWirtinger_2d} for the 2D slices:
\begin{gather}
    x^2\leq4,\quad y^2\leq4,\quad z^2\leq4.
  \label{sup:xyz_bound}
\end{gather}
Given \eq{sup:xyz_bound}, we conclude
\begin{gather}
    x^2+y^2+z^2-xyz = (z-\frac{xy}{2})^2 + x^2(1-\frac{y^2}{4}) + y^2 \geq 0.
\end{gather}
Now we can denote the right-hand side of \eq{sup:canfun_3d}, which has been shown to be non-negative, by $|\Omega|_{\mathrm{3D}}^2$. Finally, the matrix $\mathscr{Q}_{\mathrm{3D}}$ being positive semi-definite implies that its Moore determinant is non-negative~\cite{alesker20031}. But its Moore determinant is precisely
\begin{gather}
    \op{Mdet} \mathscr{Q}_{\mathrm{3D}} = \det g_{\mathrm{3D}} - |\Omega|_{\mathrm{3D}}^2.
\end{gather}
\paragraph*{(end of proof)}
\mbox{}\newline
\mbox{}\newline
In some special cases, the low-dimensional metric--curvature inequalities in Proposition~\ref{prop:lowd_bound} yield a quantum volume bound set by the 2D Kane--Mele~\cite{KaneMele2005,yu2025universal} and the 3D Fu--Kane--Mele $\bb{Z}_2$ invariants~\cite{FuKaneMele2007,QiHughesZhang2008}:
\begin{proposition}[Quantum volume bound for 2D spin $S_z$-preserving Hamiltonians]
  \label{prop:2d_qvbound}
    Suppose that the Kramers pair of Bloch vectors can be written as
\begin{gather}
    \psi^1(\bk)=(u(\bk),0)^{\top}, \quad \psi^2(\bk)=(0,u^{*}(\bk))^{\top}.
\end{gather}
    Equivalently, the 2-by-2 Hamiltonian relevant to a Kramers pair is unitarily equivalent to
\begin{gather}
    H(\bk) = 
    \begin{pmatrix}
        h(\bk)&0\\0&h^{*}(\bk)
    \end{pmatrix},
  \label{block_pt}
\end{gather}
    where the upper (lower) block describes the evolution of spin-up (spin-down) component of the Bloch vector. In two spatial dimensions, the topology of this Hamiltonian is described by the Kane--Mele $\bb{Z}_2$ invariant $\nu_{\mathrm{2D}}\in\{0,1\}$. In this setting, we have the quantum volume bound
\begin{gather}
    \op{Vol}_{\mathrm{2D}}(g) \geq \int\dd^2k\, |\Omega|_{\mathrm{2D}}(\bk) \geq \pi \,\nu_{\mathrm{2D}} \quad (\nu_{\mathrm{2D}} \in\{0,1\}).
\end{gather}
\end{proposition}
\paragraph*{Proof of Proposition~\ref{prop:2d_qvbound}.}
The matrix-valued $\mathrm{SU}(2)$ Berry curvature is
\begin{gather}
    F_{ab}^{ij}(\bk) = \braket{\partial_a\psi^i}{\partial_b\psi^j}-(a\leftrightarrow b) = 
    \begin{pmatrix}
        F_{ab}^{11}(\bk)& F_{ab}^{12}(\bk)\\
        F_{ab}^{21}(\bk)& F_{ab}^{22}(\bk)
    \end{pmatrix}.
\end{gather}
To determine the off-diagonal components, define
\begin{gather}
    \eta_{ab}:= \braket{\partial_a u^{*}(\bk)}{\partial_b u(\bk)}.
\end{gather}
The complex conjugate of $\eta_{ab}$ can be evaluated in two equivalent ways:
\begin{subequations}
\begin{align}
    \eta_{ab}^{*} &= \braket{\partial_b u(\bk)}{\partial_a u^{*}(\bk)},
  \label{eta12_1}
\\
    \eta_{ab}^{*} &= \braket{\partial_a u(\bk)}{\partial_b u^{*}(\bk)} = \eta^{*}_{ba}.
  \label{eta12_2}
\end{align}
\end{subequations}
In \eq{eta12_1}, complex conjugation exchanges bra and ket; in \eq{eta12_2}, it acts on the Bloch vector itself. The two results imply that the off-diagonal Berry curvature vanishes:
\begin{gather}
    F_{ab}^{12} = \eta^{*}_{ab}-\eta^{*}_{ba} = 0, \qquad F_{ab}^{21} = \eta_{ab}-\eta_{ba} = 0.
\end{gather}
The same argument gives $F_{ab}^{11}+F_{ab}^{22}=0$. Therefore, with $f(\bk):= F_{ab}^{11}(\bk)$,
\begin{align}
    &F_{\mathrm{2D}}(\bk) = 
    \begin{pmatrix}
        f(\bk)&0\\0&-f(\bk)
    \end{pmatrix} = f(\bk)\,\sigma_z
\\
    &\Rightarrow\quad (F_{\mathrm{2D}}^{x},F_{\mathrm{2D}}^{y},F_{\mathrm{2D}}^{z}) = (0,0,f(\bk)).
\end{align}
Equivalently, since $\mathscr{F}^{I}= -iF^z,\, \mathscr{F}^{J}= -iF^y,\, \mathscr{F}^{K}= -iF^x$,
\begin{gather}
    (\mathscr{F}_{\mathrm{2D}}^{I}, \mathscr{F}_{\mathrm{2D}}^{J}, \mathscr{F}_{\mathrm{2D}}^{K}) = (-if(\bk),0,0).
  \label{curv_align}
\end{gather}
Thus the curvature vector has only one component, and \eq{qWirtinger_2D} reduces to $\sqrt{\det g_{\mathrm{2D}}}\geq |f(\bk)|/2$, the familiar K\"ahler metric--curvature inequality for the block Hamiltonian $h(\bk)$.
In two dimensions, if the Hamiltonian decomposes as in \eq{block_pt} globally over the BZ, the Kane--Mele invariant is represented, modulo two, by the spin Chern number of one block. In the block-diagonal basis this representative is the integral of the $z$-component of the $\mathrm{SU}(2)$ Berry curvature (or the $I$-component in the quaternion notation), which is the only nonzero component by \eq{curv_align}:
\begin{gather}
    \nu_{\mathrm{2D}} = \frac{1}{2\pi} \int\dd^2k\, \mathscr{F}_{\mathrm{2D}}^{I} \quad \mod 2.
\end{gather}
For this block-diagonal class, \eq{curv_align} shows that the two-dimensional canonical function becomes
\begin{gather}
    |\Omega|_{\mathrm{2D}}(\bk) = \frac{1}{2} \sqrt{(\mathscr{F}_{\mathrm{2D}}^{I})^2+(\mathscr{F}_{\mathrm{2D}}^{J})^2+(\mathscr{F}_{\mathrm{2D}}^{K})^2} = \frac{1}{2} |\mathscr{F}_{\mathrm{2D}}^{I}|,
\end{gather}
and the local inequality \eq{qWirtinger_2D} implies the desired bound
\begin{gather}
    \op{Vol}(g) \geq \int\dd^2k\, |\Omega|_{\mathrm{2D}}(\bk) \geq \pi \,\nu_{\mathrm{2D}}.
\end{gather}
\paragraph*{(end of proof)}
\mbox{}\newline
\begin{proposition}[3D inequality and the 3D winding density]
  \label{prop:3d_qvbound}
Let
\begin{gather}
    H(\bk)=
    \sum_{\mu=1}^{4}d_\mu(\bk)\Gamma^\mu,
    \qquad
    \{\Gamma^\mu,\Gamma^\nu\}=2\delta_{\mu\nu},
  \label{sup:4gammaham}
\end{gather}
be a gapped four-band Dirac Hamiltonian with four gamma matrices, defined on a 3D Brillouin zone. Let
\begin{gather}
    n_\mu(\bk) = \frac{d_\mu(\bk)}{|d(\bk)|},
    \qquad
    P(\bk)=\frac{1-n_\mu(\bk)\Gamma^\mu}{2}
\end{gather}
be the occupied projector. Let $g_{ij}$ be the quantum metric of the occupied Kramers pair, and let $|\Omega|_{\mathrm{3D}}$ be the canonical function defined by \eq{qWirtinger_3D}. Then, everywhere in momentum space,
\begin{gather}
    \sqrt{\det g(\bk)}
    =
    |\Omega|_{\mathrm{3D}}(\bk)
    =
    \frac{\pi^2}{\sqrt{2}}
    |q_{\mathrm{wind}}(\bk)|,
  \label{sup:3d_z2bound}
\end{gather}
where
\begin{gather}
    q_{\mathrm{wind}}(\bk)
    =
    \frac{1}{2\pi^2}
    \det
    \left(
    \mathbf n,
    \frac{\partial\mathbf n}{\partial k_1},
    \frac{\partial\mathbf n}{\partial k_2},
    \frac{\partial\mathbf n}{\partial k_3}
    \right).
\end{gather}
Furthermore, for Dirac representatives of three-dimensional class AII insulators, the parity of the Dirac-map degree
\begin{gather}
    N_3=
    \int_{\mathrm{BZ}} \dd^3k\,q_{\mathrm{wind}}(\bk)
\end{gather}
is the strong Fu--Kane--Mele invariant $\nu_{\mathrm{3D}}\in\{0,1\}$.
\end{proposition}
\paragraph*{Proof of Proposition~\ref{prop:3d_qvbound}.}
All statements are pointwise, so fix a momentum $\bk$ and write
\begin{gather}
    v_i:=\partial_i\mathbf n,
    \qquad
    i=1,2,3 .
\end{gather}
Since $\mathbf n\cdot\mathbf n=1$, the vectors $v_i$ lie in the tangent space $T_{\mathbf n}\bb{S}^3$.
The projector satisfies
\begin{gather}
    \partial_iP
    =
    -\frac{1}{2}
    (\partial_i n_\mu)\Gamma^\mu
    =
    -\frac{1}{2}v_{i,\mu}\Gamma^\mu .
\end{gather}
Using $P=(1-n_\mu\Gamma^\mu)/2$, $\Tr(\Gamma^\mu\Gamma^\nu)=4\delta_{\mu\nu}$, and the vanishing of the trace of an odd product of gamma matrices, the quantum metric is
\begin{align}
    g_{ij}
    &=
    \real\Tr[P\partial_i P\partial_jP]
      =\frac{1}{2}v_i\cdot v_j .
\end{align}
Let $G_{ij}=v_i\cdot v_j$. Then $g=G/2$ and
\begin{gather}
    \det g=\frac{1}{8}\det G .
  \label{sm:3d_metric_gram}
\end{gather}
If
\begin{gather}
    D:=\det \left( \mathbf n, v_1,v_2,v_3 \right),
\end{gather}
where the determinant is taken in the ambient $\bb{R}^4$, then the Gram determinant identity gives
\begin{gather}
    D^2=\det G .
  \label{sm:3d_gram_identity}
\end{gather}
Combining \eq{sm:3d_metric_gram} and \eq{sm:3d_gram_identity} with the normalization $q_{\mathrm{wind}}=D/(2\pi^2)$ gives
\begin{gather}
    \sqrt{\det g}
    =
    \frac{|D|}{2\sqrt{2}}
    =
    \frac{\pi^2}{\sqrt{2}}
    |q_{\mathrm{wind}}| .
  \label{sm:3d_metric_winding}
\end{gather}
It remains to identify the three-dimensional canonical function. In a quaternionic frame of the occupied two-plane, the traceless non-Abelian curvature of the Dirac projector is the pullback of the natural cross product on $T_{\mathbf n}\bb{S}^3$:
\begin{gather}
    2\omega_{ij}
    =
    v_i\times_{\mathbf n}v_j,
    \qquad
    (u\times_{\mathbf n}v)_\mu
    =
    \epsilon_{\mu\nu\rho\sigma}
    n_\nu u_\rho v_\sigma,
\end{gather}
up to the fixed quaternionic orientation. This orientation only changes the simultaneous sign of the three imaginary curvature components and therefore does not affect $|\Omega|_{\mathrm{3D}}$.
The identity $|\Omega|_{\mathrm{3D}}^2=\det g$ is invariant under a change of local coordinates on the Brillouin zone. We may therefore choose, at the fixed point $\bk$, an oriented orthonormal frame $e_1,e_2,e_3$ of $T_{\mathbf n}\bb{S}^3$ and local coordinates such that
\begin{gather}
    v_1=L_1e_1,
    \qquad
    v_2=L_2e_2,
    \qquad
    v_3=L_3e_3 .
\end{gather}
Then
\begin{gather}
    g_{ii}=\frac{1}{2}L_i^2,
    \qquad
    g_{ij}=0\quad (i\neq j),
  \label{sm:3d_diag_metric}
\end{gather}
and, with the quaternionic orientation chosen so that $e_1\times_{\mathbf{n}}e_2=e_3$, the curvature variables entering \eq{qWirtinger_3D} are
\begin{gather}
    \omega_{23}
    =
    \frac{1}{2}L_2L_3e_1,
    \qquad
    \omega_{13}
    =
    -\frac{1}{2}L_1L_3e_2,
    \qquad
    \omega_{12}
    =
    \frac{1}{2}L_1L_2e_3 .
  \label{sm:3d_curvature_diag}
\end{gather}
Thus
\begin{gather}
    |\omega_{23}|^2=\frac{1}{4}L_2^2L_3^2,
    \qquad
    |\omega_{13}|^2=\frac{1}{4}L_1^2L_3^2,
    \qquad
    |\omega_{12}|^2=\frac{1}{4}L_1^2L_2^2,
\end{gather}
while the pairwise real parts vanish if we identify $(e_1,e_2,e_3)$ with the imaginary quaternions $(I,J,K)$:
\begin{gather}
    \real[\omega_{23} \omega_{13}] =
    \real[\omega_{12} \omega_{23}] =
    \real[\omega_{12} \omega_{13}] =
    0 .
\end{gather}
The triple product is
\begin{gather}
    \real[\omega_{23}\omega_{13}\omega_{12}]
    =
    \frac{1}{8}L_1^2L_2^2L_3^2 .
  \label{sm:3d_triple_product}
\end{gather}
Substituting \eq{sm:3d_diag_metric}--\eq{sm:3d_triple_product} into the definition of $|\Omega|_{\mathrm{3D}}^2$ in \eq{qWirtinger_3D}, we find
\begin{align}
    |\Omega|_{\mathrm{3D}}^2
    &=
    \frac{1}{8}L_1^2L_2^2L_3^2
    +
    \frac{1}{8}L_1^2L_2^2L_3^2
    +
    \frac{1}{8}L_1^2L_2^2L_3^2
    -
    2\left(
    \frac{1}{8}L_1^2L_2^2L_3^2
    \right)
\nonumber\\
    &=
    \frac{1}{8}L_1^2L_2^2L_3^2
    =
    \det g .
\end{align}
Together with \eq{sm:3d_metric_winding}, this proves
\begin{gather}
    \sqrt{\det g}
    =
    |\Omega|_{\mathrm{3D}}
    =
    \frac{\pi^2}{\sqrt{2}}|q_{\mathrm{wind}}| .
\end{gather}
Finally, for a four-gamma Dirac representative of a class AII insulator, one may choose a chiral representation in which
\begin{gather}
    U(\mathbf n)
    =
    n_4\mathbf 1_2
    +
    i\sum_{a=1}^{3}n_a\sigma_a
    \in
    \mathrm{SU}(2)
\end{gather}
and the occupied frame is $W=2^{-1/2}(-U,\mathbf 1_2)^{\top}$. In this Dirac frame the non-Abelian Berry connection is
\begin{gather}
    A=W^\dagger \dd W=\frac{1}{2}U^\dagger\dd U .
\end{gather}
Consequently the Chern--Simons invariant obeys $2\op{CS}(A)=\pm N_3$, with the sign fixed by the orientation convention. Therefore $2\op{CS}(A)$ modulo two, which is the strong Fu--Kane--Mele invariant for this representative, equals $N_3$ modulo two.
\paragraph*{(end of proof)}
\mbox{}\newline
\mbox{}\newline
Integrating both sides of the local inequality \eq{sup:3d_z2bound}, we obtain the global quantum-volume bound for the special type of Hamiltonian in \eq{sup:4gammaham}:
\begin{corollary}[Quantum volume bound for the 3D four-term Dirac Hamiltonians]
  \label{cor:3d_z2bound}
\begin{gather}
    \op{Vol}_{\mathrm{3D}}(g) = \int\dd^3k\,|\Omega|_{\mathrm{3D}}(\bk) \geq \frac{\pi^2}{\sqrt{2}}\, \nu_{\mathrm{3D}} \quad (\nu_{\mathrm{3D}} \in\{0,1\}).
\end{gather}
\end{corollary}
%

\subsection*{Supplementary Note 6.
Removing space inversion symmetry
}

\begin{figure}[ht]
    \centering
    \includegraphics [width=1\linewidth]{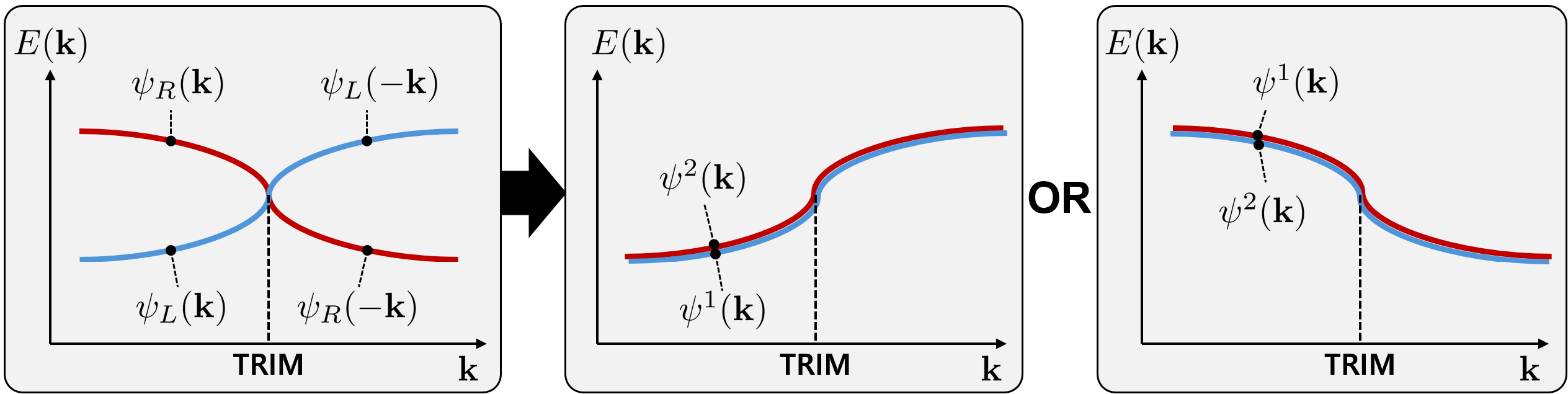}
\caption{\textbf{Band structure folding.}
Without inversion symmetry, time reversal symmetry maps $\bk$ to $-\bk$ rather than acting locally at a fixed momentum. By pairing time-reversal-related states at $\bk$ and $-\bk$, one obtains a folded description with a local Kramers pair at each momentum. The QQGT can then be defined as in the $\mathcal{PT}$-symmetric case. Gauge-invariant quantities, including the quantum metric and canonical four-form, do not depend on the folding choice.
}
  \label{fig:folding}
\end{figure}

The construction so far assumed that the Kramers pair is local in momentum. For time reversal symmetry without inversion, $\mathcal{T}$ maps $\bk$ to $-\bk$, so the two partners do not sit at the same momentum. The same quaternionic geometry can nevertheless be applied by folding the pair into a local doubled description; see Fig.~\ref{fig:folding}.

Let $\mathcal{T}$ be the time reversal operator with $\mathcal{T}^2=-1$. Choose two bands that are isolated from the remaining bands and mapped into each other by $\mathcal{T}$. Label them $\psi_L(\bk)$ and $\psi_R(\bk)$, with $\mathcal{T}\psi_L=\psi_R$, and locally define
\begin{gather}
    \psi^1(\bk) \equiv \psi_L(\bk),
    \qquad
    \psi^2(\bk) \equiv \mathcal{T}[\psi_L(\bk)] = \psi_R(-\bk).
  \label{kramers_partner2}
\end{gather}
The framework above can then be applied to $(\psi^1,\psi^2)$. Equivalently, the original local Hilbert space $\mathcal{H}_{\bk}=\bb{C}\ket{\psi_L(\bk)}+\bb{C}\ket{\psi_R(\bk)}$ is reorganized as the folded space
\begin{gather}
    \mathcal{H}_{\bk}^{\text{new}} = \bb{C}\ket{\psi^1(\bk)} +  \bb{C}\ket{\psi^2(\bk)} = \bb{C}\ket{\psi_L(\bk)} +  \bb{C}\ket{\psi_R(-\bk)}.
    \label{Kramers_double}
\end{gather}

After folding, the quaternion-valued projector $\mathscr{P}(\bk)$, the quantum metric, and the canonical four-form are independent of which member is chosen as $\psi_L$. The $\mathrm{SU}(2)$ Berry curvature depends on that choice only covariantly, as in \eq{sup:2form_rot}. Interchanging the roles of $\psi_L$ and $\psi_R$ gives
\begin{subequations}
\begin{gather}
    (\psi_{L,\text{new}},\psi_{R,\text{new}}) = (\psi_{R,\text{old}},-\psi_{L,\text{old}}),
\\
    \psi^1_{\text{new}}(\bk) = \psi_{L,\text{new}}(\bk) = \psi_{R,\text{old}} = \psi^2_{\text{old}}(-\bk),
\\
    \psi^2_{\text{new}}(\bk) = \psi_{R,\text{new}}(-\bk) = -\psi_{L,\text{old}}(-\bk) = -\psi^1_{\text{old}}(-\bk).
\end{gather}
  \label{old_new}
\end{subequations}
The local part of this transformation is the $\mathrm{SU}(2)$ rotation $(\psi_{\bk}^1,\psi_{\bk}^2)\mapsto(\psi_{\bk}^2,-\psi_{\bk}^1)$, and the remaining operation is momentum inversion. Because the QGT is built from the projector, which is invariant under this change at each momentum, the folded quantum geometry of a class AII band is independent of the folding scheme.

\clearpage
\subsection*{
Supplementary Figures
}

\clearpage 



\end{document}